\documentclass[3p,fleqn]{elsarticle}

\usepackage{hyperref}
\usepackage{tabularx}
\usepackage[subrefformat=parens]{subcaption}
\usepackage{bm}
\usepackage{amsmath,amssymb}

\journal{arXiv}

\begin{document}

\begin{frontmatter}

  \title{Topology-free immersed boundary method for incompressible turbulence flows: An aerodynamic simulation for ''dirty'' CAD geometry}


\author[mymainaddress]{Keiji Onishi\corref{mycorrespondingauthor}}
\ead[url]{www.r-ccs.riken.jp}
\cortext[mycorrespondingauthor]{Corresponding author}
\ead{keiji.onishi@riken.jp}



\author[mymainaddress,myfifthaddress]{Makoto Tsubokura}

\address[mymainaddress]{RIKEN, Center for Computational Science, 7-1-26, Minatojima-minami-machi, Chuo-ku, Kobe, Hyogo, 650-0047, Japan}
\address[myfifthaddress]{Kobe University, 1-1 Rokkodai-cho, Nada-ku, Kobe, Hyogo, 657-8501, Japan}

\begin{abstract}
 To design a method to solve the issues of handling ''dirty'' and highly complex geometries, the topology-free method combined with the immersed boundary method is presented for viscous and incompressible flows at a high Reynolds number. The method simultaneously employs a ghost-cell technique and distributed forcing technique to impose the boundary conditions. An axis-projected interpolation scheme is used to avoid searching failures during fluid and solid identification. This method yields a topology-free immersed boundary, which particularly suits flow simulations of highly complex geometries. Difficulties generally arise when generating the calculation grid for these scenarios. This method allows dirty data to be handled without any preparatory treatment work to simplify or clean-up the geometry. This method is also applicable to the coherent structural turbulence model employed in this study. The verification cases, used in conjunction with the second-order central-difference scheme, resulted in first-order accuracy at finer resolution, although the coarser resolution retained second-order accuracy. This method is fully parallelized for distributed memory platforms. In this study, the accuracy and fidelity of this method were examined by simulating the flow around the bluff body, past a flat plate, and past dirty spheres. These simulations were compared with experimental data and other established results. Finally, results from the simulation of practical applications demonstrate the ability of the method to model highly complex, non-canonical three-dimensional flows. The countermeasure based on the accurate classification of geometric features has provided a robust and reasonable solution.
\end{abstract}


\end{frontmatter}


\section{Introduction}

Advances in computational fluid dynamics (CFD) in engineering are often derived from the improvement of shape reproducibility for complex geometries. This is largely because of the development of grid generation methods. Over the past two decades, commercial software has been widely used in this field \cite{Boysan2009}, and considerable progress has been made in its utilization in technology such as easy-to-use grid generation software, the unstructured grid (tetrahedron, hexahedron, polyhedron, etc.) or cut-cell technology, and surface wrapping techniques \cite{Edelsbrunner2003}. Consequently, commercial software is now an indispensable technology in engineering product design. 

The graphical user interface (GUI) has aided the dissemination of CFD technology by providing intuitive operation to non-specialized engineers; it visualizes complex geometries to suit the human eye and simplifies operations, and has become an indispensable factor in CFD. Unstructured grids and cut cells, introduced in the early 2000s, make it easy to create computational grids for arbitrary complex three-dimensional geometries by filling the space with body-fitting polygon elements. This is only possible with an intuitive GUI, and in fact, shape reproducibility has dramatically improved as grid resolution has increased during this time. Furthermore, surface wrapping, which was popularized in the 2010s, has made it easy to convert complex surfaces to a suitable closed volume grid, thus reducing workloads and improving shape reproducibility. However, in many cases, it is still difficult to complete the grid generation without encountering failures. To process a spatial grid generation, it is necessary to complete a preparatory work, thereby correcting the geometry to be ''watertight.'' The geometry should consist of closed volumes without cracks between faces to form fluid regions. This process generally takes several days to several weeks \cite{furukawa2004application}. It is performed using manual operations that require expert knowledge of the software operations. The difficulty of this operation is firstly due to the nonconformity included in the geometry data, such as gaps, overlaps, and defects in surfaces, inhibiting the formation of a closed volume. These difficulties are mainly owing to errors in data conversion processes of computer-aided-design (CAD) data, which occur during information exchange between CAD and CFD software. In the early design stages, the CAD data itself may not have been created cleanly. Secondly, intersections of thin-plate structures or nested surfaces that exist inside of other parts make it difficult to judge whether the inside/outside of the region should be recognized as a flow field. Surfaces without thicknesses also cause the same difficulty. Thirdly, tiny flow paths with sizes below the grid resolution should be eliminated, as they disturb the grid generation process itself. Geometry data that features these types of features are called ''dirty'' CAD data. Preparing clean geometry from dirty geometry data is both costly and requires a large manual operation workload, even with an excellent GUI. This sometimes influences the results depending on the operator's skill. The burdensome nature of this preparatory work is still one of the factors hindering the utilization of CFD in industrial fields \cite{Michal2009}. Moreover, optimization technologies such as big data mining or machine learning are making significant progress. Automation in these massively parallel computing environments is a key factor. If this preparatory workload issue is not addressed, it will become a bottleneck on progress; thus, it is necessary to solve this issue.

The surface wrapping technique \cite{Vollmer2001} originates from the image scanning technology of three-dimensional geometries. It regenerates a new shape covering an existing shape as a surface grid composing a single closed volume. This makes it possible to separate the fluid and solid regions as a unique ''watertight'' volume. This in turn makes it easy to judge the inside and outside of the region. Examples of full-vehicle aerodynamic analyses featuring dirty geometries have been successfully reported \cite{Kotapati2009,Kandasamy2012,Regin2013,Nakamura2015} with surface wrapping, using either the unstructured grid method or the lattice Boltzmann method. The full-vehicle has parts ranging from thousands to tens of thousands, such as the engine bay, underfloor, suspension modules, inner frames, wire-harnesses, and bolt/nuts. There are several defects in the geometry data; thus, surface wrapping was necessary to correct geometrical problems. However, the shapes were often simplified at low resolution (on the order of 10 to 100 mm), as the surface wrapping software performance was fundamentally limited due to the serial operation process. A high surface resolution that satisfies the requirement of CFD on the order of approximately 1 mm \cite{Kandasamy2012,Regin2013} is not always obtained. In addition, as the spatial resolution increases following advances in parallel computing platforms, the serial computing paradigm will not scale up. Consequently, this preparatory work becomes difficult, and the number of operations needed to ensure further detail increases. There is concern, therefore, that the burden on CFD workers will increase in the near future.

The isogeometric analysis proposed by Hughes et al.\cite{Cottrell2009} is expected to provide a solution to this problem. As this method directly uses non-uniform rational B-spline data, which are the basis of most CAD packages, it is considered effective for dirty CAD. The parametric surface shape is treated as finite elements, in which the degree of freedom is increased, then the high-order of discretization is applicable. However, this method requires there to be a guaranteed connection between faces. It cannot be classified as dirty CAD capable if the geometry data needs to be modified in advance to satisfy the connectivity between the surface elements. The immersed body methods of Dettmer et al. \cite{Dettmer2016} and Wu et al. \cite{Wu2017} are in the same category. Even if the mesh-less method is combined \cite{Bazilevs2017}, this is the same if it requires to judge the inside/outside of the fluid regions beforehand. The difficulty of region judgment is fundamentally the same as the one of grid generation.

An analysis method that uses a Cartesian grid makes it easy to create a computational grid. However, when considering its application to industrial fields, we infer that many industrial products are made on the basis of thin plates. Apparently, in this case, the number of calculation grids required would significantly increase. To represent the shape of thin plates as a stack of the cubic grid, the number of grids required would increase at a scale of $(L/h)^3$, where $L$ is the characteristic length of the model and $h$ is the plate thickness on the order of sub-millimeters. Therefore, a method for handling the surface shape, regardless of the grid resolution, has been proposed for this purpose. The immersed boundary method (IBM) proposed by Peskin \cite{Peskin1972} is a method of expressing the object as an external force term; the dispersion of the continuous force is introduced in the momentum equation, replacing the object boundary grid in space. Peskin successfully solved the moving boundary problems of a thin elastic wall, such as a heart valve pulsation or linear object locomotion. Other methods proposed by Saiki and Biringen \cite{Saiki1996}, Mohd-Yusof et al. \cite{Mohd-Yusof1997}, Hu et al. \cite{Hu2001}, Shirgaonkar et al. \cite{Shirgaonkar2009}, and others \cite{Fauci1988,Peskin1989,Tu1992,Goldstein1993,Mohd-Yusof1997,Fadlun2000,Peskin2003,Bao2003,Newren2008,Borazjani2013} are in the same category. However, if there is an intersection of a thin wall and a thick body, for example a heart valve and cardiovascular vessel (as shown in \cite{Borazjani2013}), or a full-vehicle model (as shown in \cite{Kandasamy2012,Regin2013}), it is necessary to cancel the calculation in the inside of the solid object. That is, this requires inside/outside identification. Furthermore, these methods have suffered from the severe timestep restrictions that are needed to maintain stability for rigid bodies, as has been well documented in the literature \cite{Newren2007}. Ye, Mittal et al. \cite{Ye1999}, Ghias, Mittal et al. \cite{Mittal2007} proposed an IBM that could be applied to complex geometries that contain sharp shapes, with the elimination of associated stability constraints on the forcing term. This method satisfies the boundary condition by using a cell located at the solid--fluid interface, termed a ghost cell. However, as this method requires at least one cell to express the boundary, it is difficult to deal with the shape without thickness, and it is necessary to identify the inside and outside of the solid. Many similar methods have been proposed \cite{Kim2001,Tseng2003,Iaccarino2003,Nishida2009,Ono2002,Ochi2005,hattori2006large,Kang2009,Mittal2008,Bergmann2013}, all of which can be regarded as belonging to the same category as Mittal et al. \cite{Mittal2007} -- refer to the review \cite{Mittal2005} for the details. Furthermore, many other methods have been proposed that improve the accuracy and stability of the flow solution in the region of complex shapes, for example, a method that solves a simplified vehicle geometry \cite{Jindal2005,Iaccarino2003} by applying a Reynolds averaged turbulence model and a wall model based on IBM. Further examples include a method that replaces computational grids near the object using a level-set function or mesh-less method on IBM \cite{Wang2004,Choi2007}, and a method of interpolating a sharp edge with a ghost cell with several grid widths near a wall on IBM \cite{Nakahashi2011}. However, to the authors' knowledge, all these methods still implicitly require closed volume geometry, leading to difficulties in handling dirty geometries. In other words, there have not yet been reported successful cases of industrial products, such as a full-vehicle aerodynamics model, which has numerous dirty surfaces.

The objective of this study is to design a method to solve this issue of dirty CAD handling on IBM: the handling of dirty geometry is a long-standing challenge in CFD, however it may be possible to solve it by applying simple ideas. In this paper, a topology-free method for IBM is formulated, and the results of flow around a dirty geometry are presented. This method is based on Mittal's boundary condition \cite{Mittal2007}, coupled with Peskin's \cite{Peskin1972} distributed forcing method. The dummy cell definition was applied to avoid searching fluid and solid regions; this also works for the zero-thickness treatment. The formulation of the IBM was modified by projecting the interpolation onto the axial direction, to avoid failures when searching for interpolation points. This model exhibited a high affinity with the solution to the problem caused by the narrow flow path. The method was verified on the Taylor-Green vortex decay problem, and validated for three-dimensional flow around the Ahmed body, flow past a flat plate for a wide range of angles of attack, and flow past a sphere for various geometries that have dirty topology. The method was then applied to full-vehicle aerodynamic analysis and city area wind environment analysis to demonstrate its ability to automatically and robustly simulate highly complex geometry flows with dirty topologies.

\section{Numerical methods}

The governing equations are the spatially filtered incompressible Navier--Stokes equations and the continuity equation, including the external forcing term of the IBM, denoted here by $\bm{f}$. This is a formulation for large eddy simulation (LES). The fundamental equations are non-dimensionalized using Reynolds number $Re$, and are described as:
\begin{align}
  \label{eq:mass}
  & \nabla \cdot \bar{\bm{u}} = 0 \quad in \quad \Omega, & \\
  \label{eq:momentum}
  & \frac{\partial \bar{\bm{u}}}{\partial t} + \bar{\bm{u}} \cdot \nabla \bar{\bm{u}} = - \nabla p + \frac{1}{Re} \nabla^2 \bar{\bm{u}} - \nabla \cdot \bm{\tau} + \bm{f} \quad in \quad \Omega, & \\
\label{eq:stress}
& \tau_{ij} = \overline{u_{i} u_{j}} - \bar{u_{i}} \bar{u_{j}}, &
\end{align}
where $(\;\bar{ }\;)$ denotes the grid-filtering operator, $\bar{\bm{u}}$ is the filtered fluid velocity, $p$ is the pressure, and $t$ is the time. The $\bm{\tau}$ is the subgrid-scale(SGS) stress tensor, which is generally modeled with $\bar{\bm{u}}$ to close the equation. The SGS models used in this study are based on the eddy viscosity concept:
\begin{align}
  \label{eq:SGSstress}
  & R_{ij} = \tau_{ij} - \frac{1}{3} \delta_{ij} \tau_{kk} = - \frac{2}{Re_{t}} \bar{\bm{S}}, \quad \bar{\bm{S}}=\frac{1}{2} \left( \nabla \bar{\bm{u}} + \left( \nabla \bar{\bm{u}}\right)^{T} \right), &
\end{align}
where $Re_{t}$ is the Reynolds number based on the eddy viscosity $\nu_{t}$, $\bar{\bm{S}}$ is the velocity strain tensor, and $\delta_{ij}$ is the Kronecker's delta. The definition of the SGS model is described in section \ref{turbulence_model}. The immersed body (IB) boundary is denoted by $\Gamma$, as shown in Fig. \ref{fig:grid_system}\subref{fig:schematic_body}. All the regular fluid regions separated by $\Gamma$ are denoted $\Omega$ without specifying explicit solid or fluid regions in this study.

\subsection{Temporal and spatial discretization}

For the discretization of the basic equations, a Cartesian coordinate system and finite volume method with collocated variables are used. As the velocity and pressure are defined at the same point, that is, at the center of a cell, interpolation on the cell face is not necessary when modeling the IBM forcing term.

The scheme is based on a fractional step \cite{Chorin1969}, using the predicted intermediate velocity, which is corrected with the pressure gradient obtained from the Poisson equation for pressure. The advection term and diffusion term are expressed in the form of the flux and the divergence of velocity gradient, respectively, and are discretized using the interface velocity $U$ defined at the center of the face. The discretized equation of momentum and continuity equation is thus obtained using the Crank-Nicolson time integration scheme; it is written in a non-conservative form using discrete-time $n$, as follows:

\begin{align}
  \label{eq:crank-nicolson}
  & \frac {\bm{u}^{n+1/2} - \bm{u}^n} {\Delta t} =
  \frac{1}{2} \left[ -\bm{U}^n \cdot \nabla \bm{u}^{n+1/2} - \bm{U}^n \cdot \nabla \bm{u}^n \right]
  + \frac{1}{2} \left[ \frac{1}{Re} \left( \nabla^2 \bm{u}^{n+1/2} + \nabla^2 \bm{u}^n \right) \right] \nonumber & \\
  & \hspace{2cm} + \frac{1}{2} \left[ \frac{2}{Re_{t}} \left( \nabla \bm{S}^{n+1/2} + \nabla \bm{S}^n \right) \right]
  + \bm{f}^{n+1/2}, & \\
  \label{eq:correction}
  & \frac{\bm{u}^{n+1} - \bm{u}^{n+1/2}}{\Delta t} = -\nabla p^{n+1}, & \\
  \label{eq:poisson}
  & \nabla^2 p^{n+1} = \frac{1}{\Delta t} \left( \nabla \cdot \bm{U}^{n+1/2} \right), &
\end{align}
where the intermediate velocity is denoted by $n+1/2$ and the operator $(\;\bar{ }\;)$ and volume integrals are abbreviated. In this equation, advection-diffusion hybrid discretization is adapted to the Crank-Nicolson scheme. This hybrid form is applied to make the solution more stable than when it is applied only to the diffusion term. The Poisson equation \eqref{eq:poisson} and implicit time integration are solved using either the bi-conjugate gradient stabilized method (BiCGStab) based on Krylov subspace theory or the red-black colored successive over-relaxation method (SOR). This system of equation is solved for the entire domain of $\Omega$.

\begin{figure}[htb]
  \centering
  \begin{minipage}{0.3\hsize}
      \includegraphics[keepaspectratio,width=\textwidth]{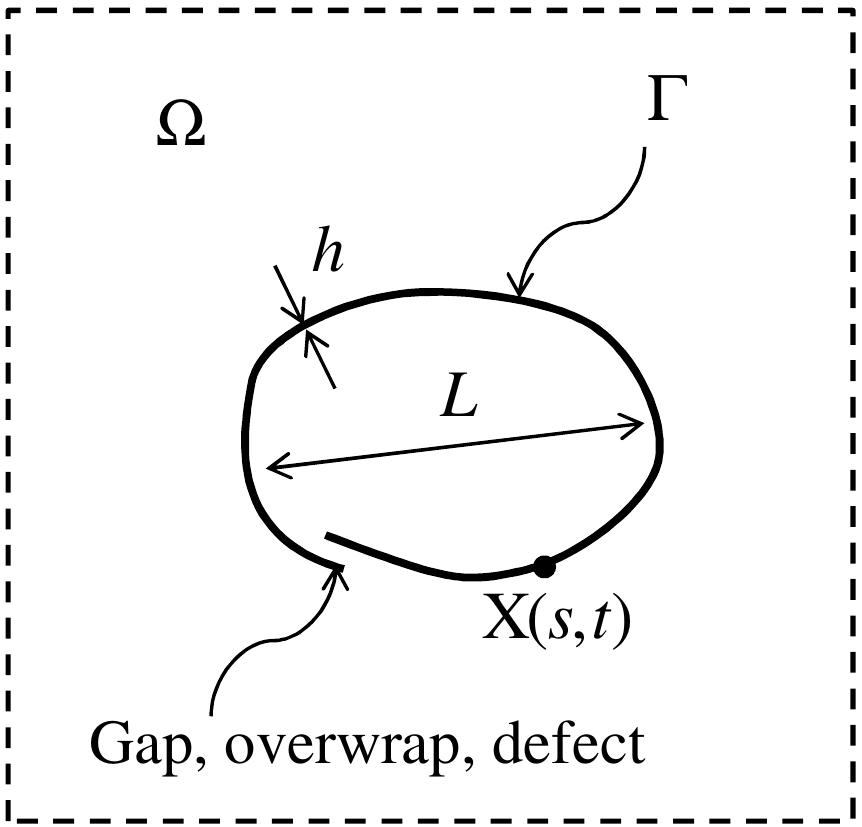}
    \subcaption{}
    \label{fig:schematic_body}
  \end{minipage}
  \hspace{0.4cm}
  \begin{minipage}{0.3\hsize}
      \vspace{0.2cm}
      \includegraphics[keepaspectratio,width=\textwidth]{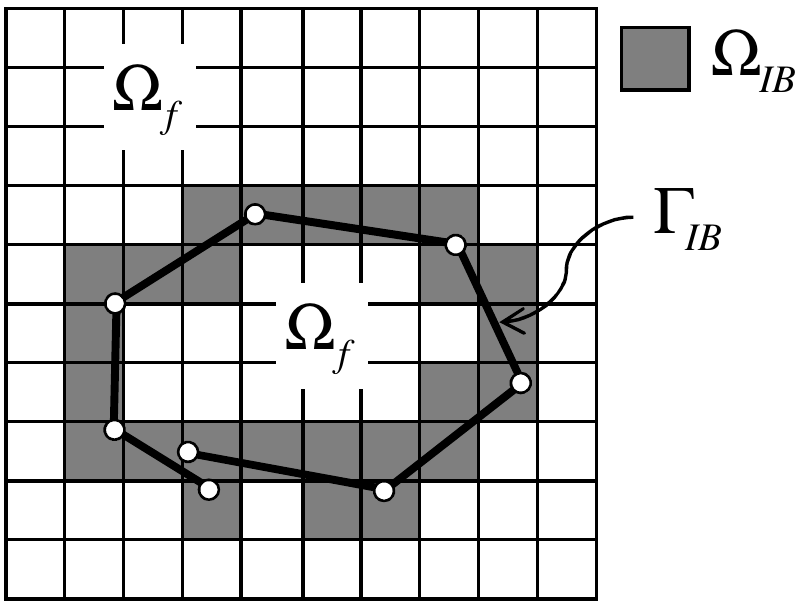}
      \vspace{0.2cm}
    \subcaption{}
    \label{fig:schematic_grid}
  \end{minipage}
  \caption{(a) Schematic showing a generic body, past which flow is to be simulated. The dirty body that has a gap, overlap, or defect is immersed into the volume $\Omega$ with boundary $\Gamma$. The body has a characteristic length scale of $L$ and a geometry thickness of $h$ over the body. $\bm{X}(s,t)$ is a vector function that gives the location of points of boundary. (b) Cartesian grid (cells) discretization of a two-dimensional computational domain and an immersed boundary (black bold line). Each white circle represents a node of the immersed geometry. The cells shaded in dark gray are wall-including cells that have a discretized wall boundary $\Gamma_{IB}$. The $\Omega_{IB}$ is the region that includes the wall-including cell, and $\Omega_{f}$ is the region in which fluid cells are filled.}
  \label{fig:grid_system}
\end{figure}

The physical domain $\Omega$ is decomposed into pairwise disjoint subdomains, the IB region $\Omega_{IB}$, which is intersecting IB and its neighbor, and the fluid region $\Omega_{f}$, which is a complement from the entire space. This satisfies the disjoint union without duplication, as follows:
\begin{align}
  \label{eq:Omega}
  & \Omega = \Omega_{f} \cup \Omega_{IB}, &
\end{align}
As shown in Fig. \ref{fig:grid_system}\subref{fig:schematic_grid}. The immersed boundary $\Gamma$ is transferred into a discretized piecewise set of IB, tessellating the original CAD face into the polygonal element as:
\begin{align}
  \label{eq:transform}
  & \Gamma_{IB}=\mathcal{C}(\Gamma) + O(e), &
\end{align}
where $\mathcal{C}$ is the transforming operator having an error of $e$. The $e$ is defined in each CAD system as a deviation from the spline to polygons. The $O(e)$ is, however, not discussed in this paper.

The discretization in the cells near the IB is adjusted to account for its presence, which is the IBM forcing term. It is then written simply as follows:
\begin{align}
  \label{eq:forcing_term}
  & \bm{f}^{n+1/2} =
  \begin{cases}
    \displaystyle -RHS^{n+1/2} + \frac {\bm{u}_{IB}^{n+1/2} - \bm{u}^n} {\Delta t}, & on \quad \Gamma_{IB}, \\
    \displaystyle 0, & elsewhere,
  \end{cases}, & \\
  & \bm{u}=\bm{u}_{IB}^{n+1/2} \quad at \quad \Gamma_{IB}, &
\end{align}
where $\bm{u}_{IB}^{n+1/2}$ is the known velocity associated with the boundary conditions on the immersed surface $\Gamma_{IB}$, and $RHS$ is the right-hand side of Eq. \eqref{eq:crank-nicolson}, which contains advection and diffusion terms in the cell centers near the IB. This is the discrete forcing approach of Mohd-Yusof \cite{Mohd-Yusof1997,Fadlun2000}, which determines the forcing directly from the numerical solution, for which there is an a priori estimate. The $\bm{f}^{n+1/2}=0$ is preserved without the presence of the IB. The definition of Eq. \eqref{eq:forcing_term} complies with the definition of \cite{Fadlun2000}. The implementation of $RHS$ depends on the time integration scheme, for example, an Adams-Bashforth explicit scheme varies the definition of constants on the head of the advection and diffusion terms of Eq. \eqref{eq:crank-nicolson}. However, it can represent the same equation systems.

Mohd-Yusof \cite{Mohd-Yusof1997} applied this discrete force when imposing the boundary condition indirectly in the domain of $\Omega_{IB}$. A method imposes boundary conditions directly in the dummy cells, which is a type of ghost cell proposed by Mittal \cite{Mittal2007}. These two differ only in the method of discretization and solve essentially the same boundary conditions. Furthermore, they neither require specific constants nor an iterative solution. Thus, these two methods can be combined.

For the indirect imposition of boundary condition, the discrete force of $\mathcal{F}$ on cells in $\Omega_{IB}$ is defined as follows:
\begin{align}
  \label{eq:discrete_forcing_term}
  \mathcal{F}(\bm{x},t) &= \sum_{k} \bm{f}_{k}^{n+1/2} \delta (\mid \bm{x} - \bm{X}(s,t) \mid) & \nonumber \\
  & \approx \sum_{k} \bm{f}_{k}^{n+1/2} \mathcal{D} (\mid \bm{x} - \bm{X}(s,t) \mid), \quad \forall \: \bm{X}(s,t) \in \Gamma, &
\end{align}
where $\bm{X}(s,t)$ is a vector function giving the location of points of IB as a function of position $s$, and time $t$. The $\bm{x}$ is the position of each cell center, as shown in Fig. \ref{fig:grid_system}. The $\delta$ is the Dirac delta function, which distributes force imposing on the momentum equations of the surrounding cells. This is the discretized form of continuous forcing function of the immersed boundary approach of Peskin \cite{Peskin1972}. Moreover, this is equivalent to the general form of the discrete forcing IBM of Mohd-Yusof \cite{Mohd-Yusof1997}, or the fictitious domain method of Glowinski \cite{Glowinski1994}. The $\delta$ is then replaced by a discrete distribution function $\mathcal{D}$. This function is now not smooth, linear, or continuous in space; however, it is advantageous to represent it as a sharp interface on the immersed boundary. The definition of $\mathcal{D}$ is described later in section \ref{axis_projection}. 

To solve Eq. \eqref{eq:correction}, as the pressure gradient is defined at the center of the control volume, the solution of the pressure is separated between even and odd points. The pressure difference of the neighbor cannot be sensed, and this sometimes causes a checker-board-like oscillation. The present study, therefore, adopts the Rhie--Chow interpolation \cite{RHIE1983} to eliminate unphysical oscillation on the collocated grid system.

The outer boundary conditions of the fluid domain are omitted here for the sake of simplifying the discussion.

\subsection{Discretization of grid system}

\begin{figure}[htb]
  \centering
  \begin{minipage}{0.45\hsize}
      \includegraphics[keepaspectratio,width=\textwidth]{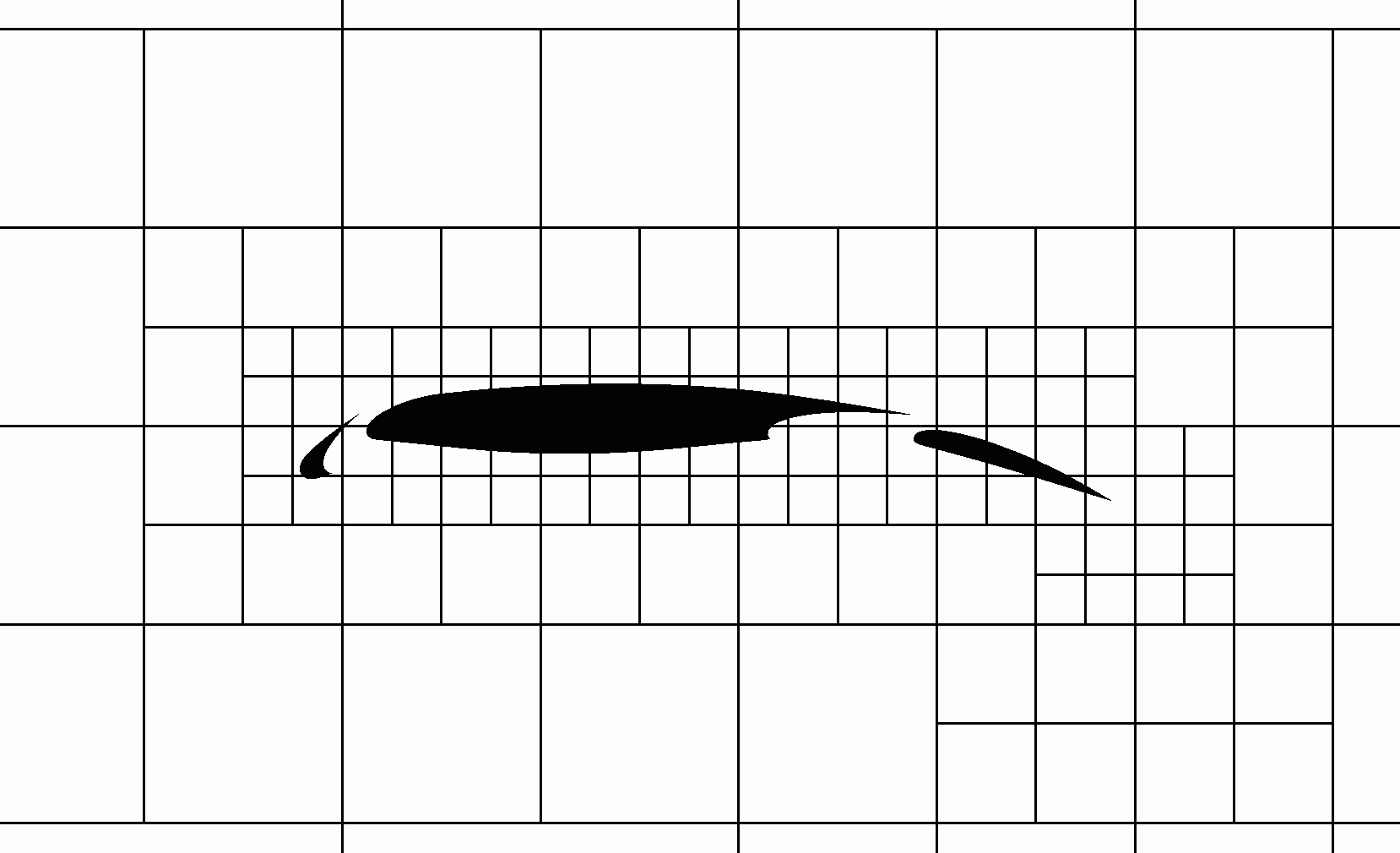}
    \subcaption{Cubes.}
    \label{fig:BCM_cube}
  \end{minipage}
  \begin{minipage}{0.45\hsize}
      \includegraphics[keepaspectratio,width=\textwidth]{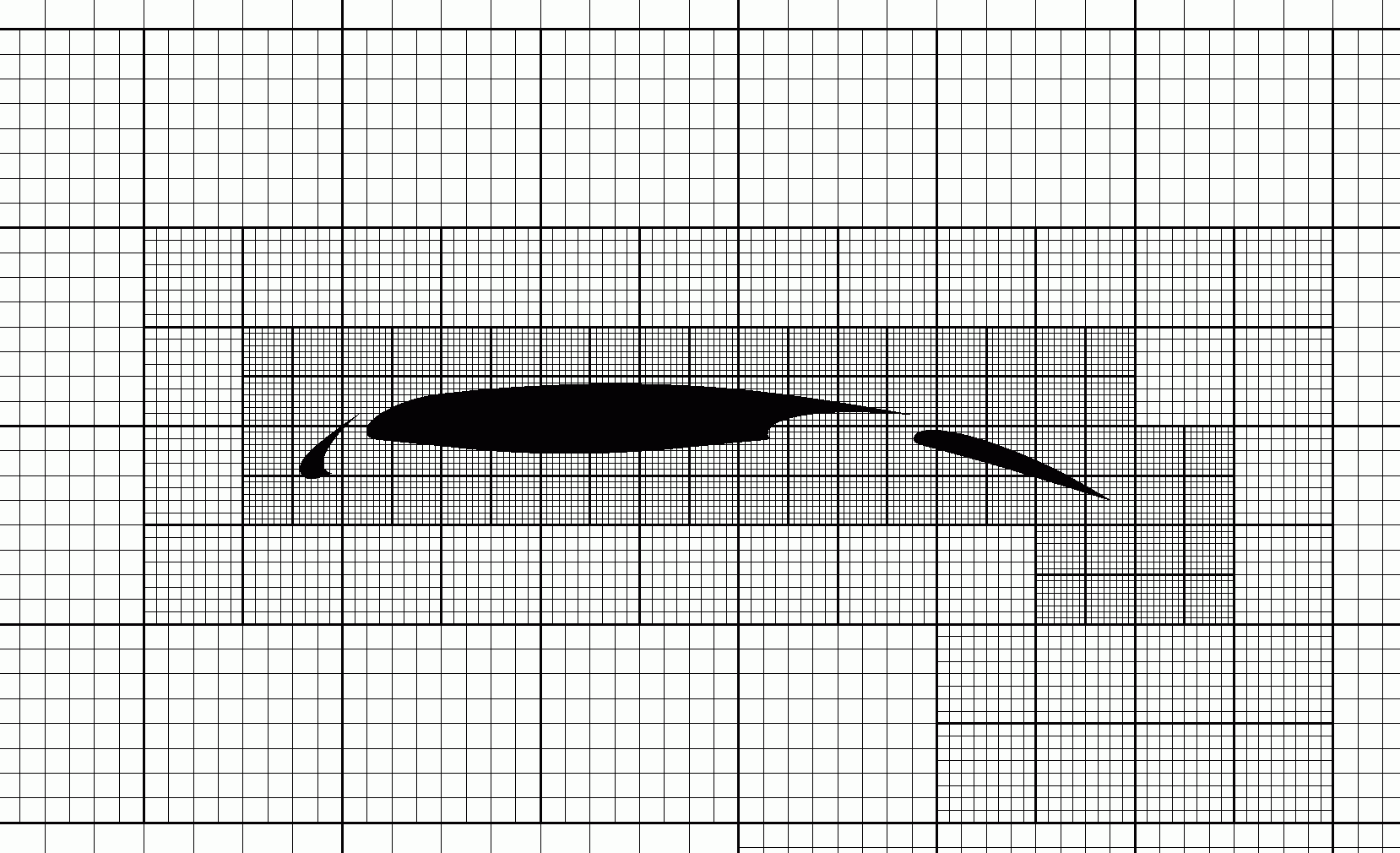}
    \subcaption{Cells.}
    \label{fig:BCM_cell}
  \end{minipage}
  \caption{BCM grid discretization example: (a) Cubes. (b) Cells. There are $8 \times 8$ cells, indicated by gray lines, in each cube, indicated by bold lines.}
  \label{fig:BCM}
\end{figure}

The building cube method (BCM) \cite{Nakahashi2003,Nakahashi2005} is employed as the basis of the grid system. The BCM is a concept that was proposed by Nakahashi in the early 2000s \cite{Nakahashi2003} to solve a flow field with a hierarchical equidistant Cartesian grid. The basic data structure is similar to that used in adaptive mesh refinement \cite{Aftosmis1998}. Unlike these approaches, however, the BCM has two fixed levels of subdivided spatial domains, which stand in the hierarchical structure. The computational space is first divided into ''cubes'' that become finer by a factor of two as they approach the geometry wall, see Fig. \ref{fig:BCM}. Each cube is then subdivided into ''cells,'' typically arranged $16 \times 16 \times 16$ in three dimensions. As each cube is composed of the same number of cells, it is easy to allocate the calculation load to each CPU equally, which is advantageous to achieving high parallelization of computational load balancing. The finest calculation unit is the cell.

Data must be exchanged between cubes, as the interface of each cube is discontinuous between adjacent cubes. Each cube has buffer areas (halo cells) at the interface, and data are exchanged between cells in this buffer. Two layers of halo cells were used in the present study. To exchange the data, two methods were applied. One was to conduct the first-order interpolation from small cubes to large cubes (i.e., an averaging operation). The other was to conduct the zero-order interpolation from large cubes to small cubes, or between cubes of the same size (i.e., a data copy operation). From the viewpoint of calculation efficiency, data were exchanged only in the orthogonal direction of each axis, not in the diagonal direction. This is because the differentiation and interpolation of the IBM, as described in section \ref{axis_projection}, are executed independently in the axial direction.

\subsection{Turbulence model}\label{turbulence_model}

For modeling eddy viscosity, the coherent structure model (CSM), derived by Kobayashi \cite{Kobayashi2005}, was used. This model is a method for dynamically determining the local constant of the Smagorinsky SGS model based on the relationship between the second invariant of the velocity gradient tensor and the energy dissipation rate of vortices. Although the dynamic Smagorinsky model \cite{Germano1991} is widely used to solve the filtering of model coefficients in a uniform direction to stabilize the solution, this is computationally intensive, and has disadvantages regarding its application to complex shapes. The CSM is a model proposed to solve it simply with a low computational cost.

The SGS model by Smagorinsky is represented as:
\begin{align}
  \label{eq:Smagorinsky}
  & \nu_{t} = C \bar{\Delta}^{2} |\bar{\bm{S}}|, \quad |\bar{\bm{S}}| = \sqrt{S_{ij} S_{ij}}, &
\end{align}
where $\bar{\Delta} = \left( \bar{\Delta_{1}} \bar{\Delta_{2}} \bar{\Delta_{3}} \right)^{1/3}$ is the filter width, given by the grid width $\bar{\Delta_{i}}$ in the $i$-th direction.

In the CSM, the model coefficient $C$ is determined as follows. The second invariant of the velocity gradient tensor $Q$ is represented as:
\begin{align}
  \label{eq:Q}
  & Q = \frac{1}{2} \left( W_{ij} W_{ij} - S_{ij} S_{ij} \right), & \\
  & \bar{\bm{W}}=\frac{1}{2} \left( \nabla \bar{\bm{u}} - \left( \nabla \bar{\bm{u}}\right)^{T} \right), \quad |\bar{\bm{W}}| = \sqrt{W_{ij} W_{ij}}, &
\end{align}
where $\bar{\bm{W}}$ is the vorticity tensor. This definition is termed the Q-criterion, and it is often used to visualize vortex structures \cite{Tanaka1993}. Kobayashi determined the model coefficients (in non-rotating flow) independently of the flow properties and Reynolds number, as follows:
\begin{align}
  \label{eq:FCS}
  & F_{CS} = Q / E, & \\
  & E = \frac{1}{2} \left( W_{ij} W_{ij} + S_{ij} S_{ij} \right), & \\
  & C = C_{1} |F_{CS}|^{3/2} , \quad C_{1} = 1 / 20. &
\end{align}
As a result, an eddy viscosity model is expressed, in which the coefficient is automatically zero in laminar flow, and is naturally damped in the wall direction. For details, refer to \cite{Kobayashi2005}.

\begin{figure}[htb]
  \centering
  \begin{minipage}{0.4\hsize}
      \includegraphics[keepaspectratio,width=\textwidth]{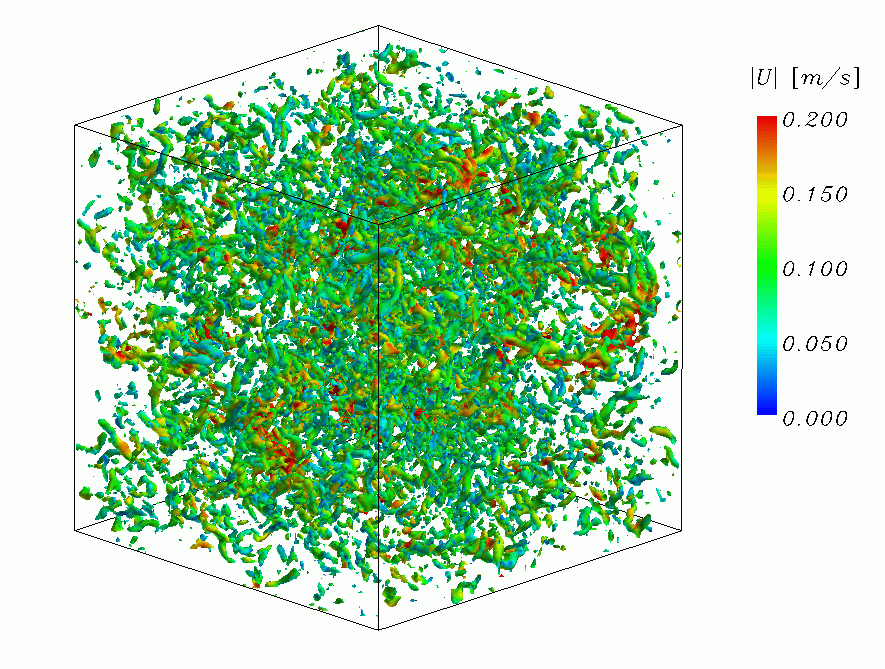}
      \subcaption{}
    \label{fig:iso_turb_Q}
  \end{minipage}
  \begin{minipage}{0.4\hsize}
      \vspace{0.6cm}
      \includegraphics[keepaspectratio,width=\textwidth]{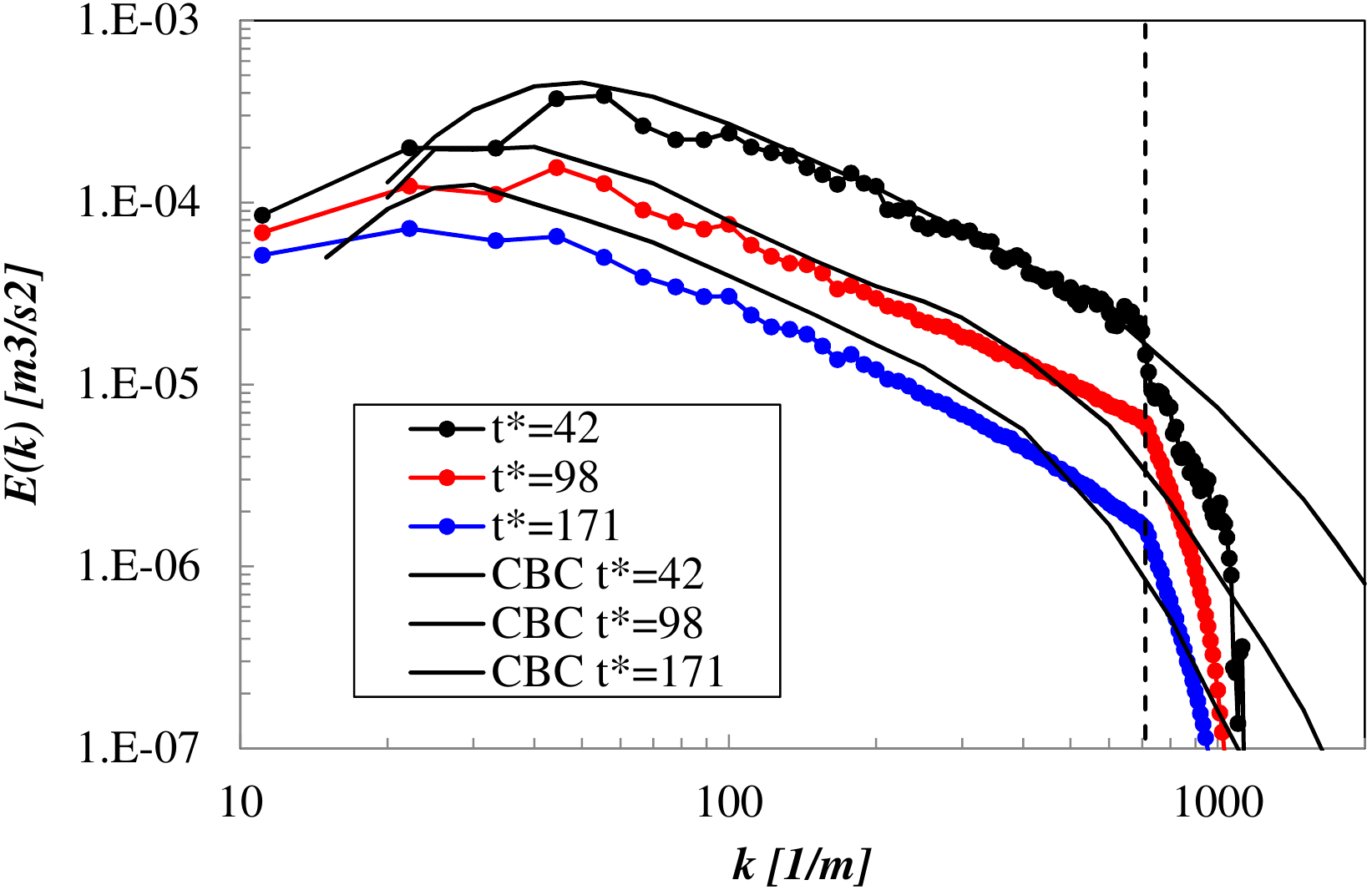}
      \subcaption{}
    \label{fig:iso_turb_E}
  \end{minipage}
  \caption{Isotropic turbulence decay energy spectrum for the CSM turbulence model. (a) The turbulence structures visualized by the Q-criterion. (b) The black circle represents the initial reference energy $E(k)$ for each wavenumber $k$; the red or blue colored circle shows simulation result. The solid line shows CBC data. The Nyquist limit is indicated via the vertical dashed line.}
  \label{fig:iso_turb}
\end{figure}

Figure \ref{fig:iso_turb} shows the basic verification results for this turbulence model on the isotropic turbulence decay energy spectrum. The homogeneous incompressible isotropic turbulence was generated in a three-dimensional cubic box using the random Fourier mode. The side length of the box was $L=0.09 \times 2 \pi \: m$, with a number of grid (cell) of $128 \times 128 \times 128$ in three dimensions. The number of Fourier modes was $5000$ and a generic dynamic viscosity $\nu = 1.0 \times 10^{-5} \: m^{2}/s$ was chosen. The initial turbulence velocity field was generated from the energy spectrum table of Comte--Bellot et al. (CBC) \cite{comte-bellot_corrsin_1971}, based on the method of Saad et al. \cite{Saad2017}. A Fourier series of an arbitrary spatial velocity field $\bm{u}$ at point $\bm{x}$ is given by:
\begin{align}
  \label{eq:isotropic_turbulence}
  \bm{u}(\bm{x}) = 2 \sum_{m=1}^{M} q_m cos(\kappa_m \hat{\bm{k}}_m \cdot \bm{x} + \psi_m) \hat{\bm{\sigma}}_m, 
\end{align}
where $M$ is the number of modes, $q_m$ is the amplitude, $\kappa_m$ is the $m$-th wavenumber, and $\hat{\bm{k}}_m \equiv (k_{x,m},k_{y,m},k_{z,m})$ is a unit direction vector. The procedure consisted of choosing random $\hat{\bm{k}}_m$ and $\psi_m$ with $q_m$ obtained from the energy spectrum such that $q_m = \sqrt{E(\kappa_m) \Delta \kappa}$. The modal direction vector $\hat{\bm{\sigma}}_m$ was evaluated enforcing the divergence-free constraint; it was selected to be orthogonal to $\hat{\bm{k}}_m$,
\begin{align}
  \label{eq:unit_direction_vector}
  \hat{\bm{k}}_m \cdot \hat{\bm{\sigma}}_m = 0, \quad \forall \: m \in \{0,1,...,M\}.
\end{align}

In Fig. \ref{fig:iso_turb}, the input spectrum corresponds to the CBC data (solid lines). The simulated data are shown in filled circles and correspond to non-dimensional times of $t^{*} = 42, 98, 171$, respectively. These results indicate a reasonably accurate solution, given the initial and decaying spectra.
  
\subsection{Topology-free immersed boundary}\label{dirty_CAD_definition}

When applying the IBM to dirty geometry data, it is necessary to consider the following issues.
\begin{enumerate}[(1)]
\item Nonconformity such as gaps, overlaps, and defects of faces
\item Penetration of faces, and the intersection of multiple faces or volumes
\item Narrow flow paths with sizes below the grid resolution
\item Zero-thickness faces or thin structures
\item Large numbers of faces on the issue (1) to (4), which are difficult to correct using manual operations
\end{enumerate}

Issue (1) inhibits the model from forming a closed volume. This often occurs when the CAD data are transferred through the intermediate format file, e.g. IGES, STEP, or Parasolid. Duplicated faces are in the same category. Issue (2) makes it difficult to identify the inside or outside of a fluid or solid region. The nested faces or volumes exist inside of other groups, and the duplicated connected edges of faces are in the same category. Issue (3) occurs when two opposing faces are placed within a short distance, or intersect at a narrow angle, forming a wedge shape region. This impedes the grid generation process itself and makes the calculation unstable. Issue (4), which is the same as issue (2), makes it difficult to identify the fluid region, further inhibiting the use of solid cells placed in thickness. Even if the data suffer from issues (1) to (4), if the number of faces in these issues is small enough, correction is easy. Therefore, such data were not categorized as dirty data in the presented method. That is, issue (5) is an essential factor.

\begin{figure}[htb]
  \centering
  \begin{minipage}{0.325\hsize}
    \includegraphics[keepaspectratio,width=\textwidth]{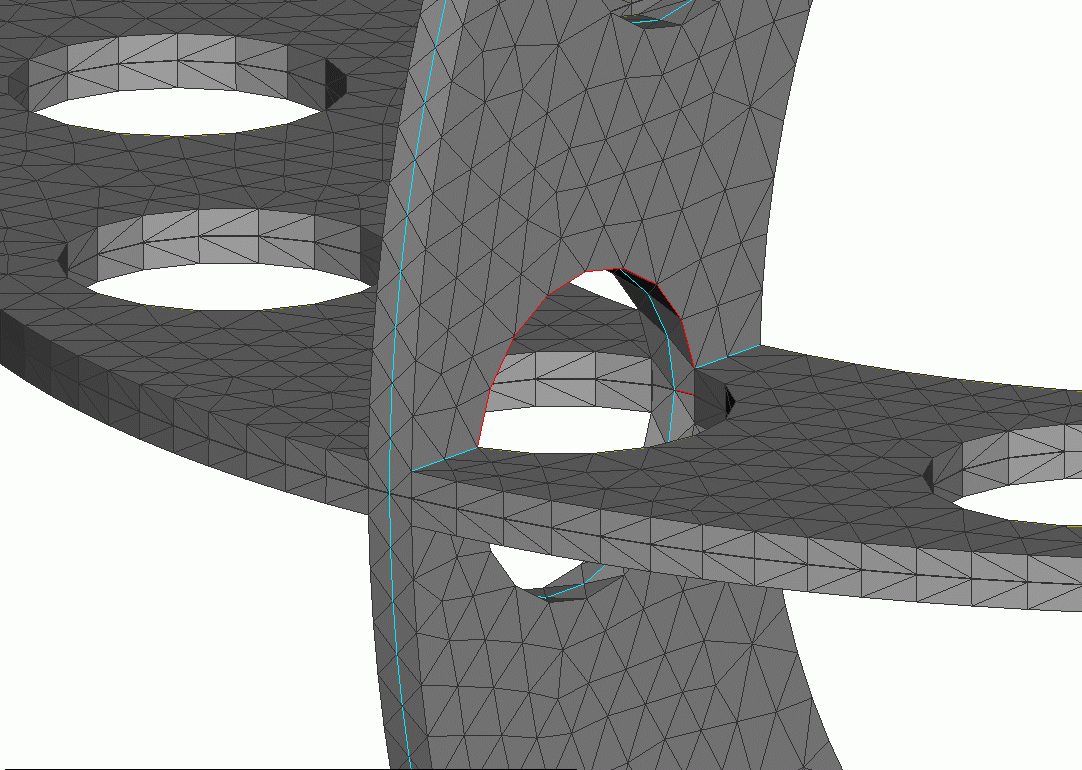}
    \subcaption{Surfaces are not joined perfectly. There are gaps between frames. Red lines indicate the edges that are not connected. Light blue lines indicate the edges that are over-connected.}
    \label{fig:frame_error}
  \end{minipage}
  \begin{minipage}{0.3\hsize}
    \includegraphics[keepaspectratio,width=\textwidth]{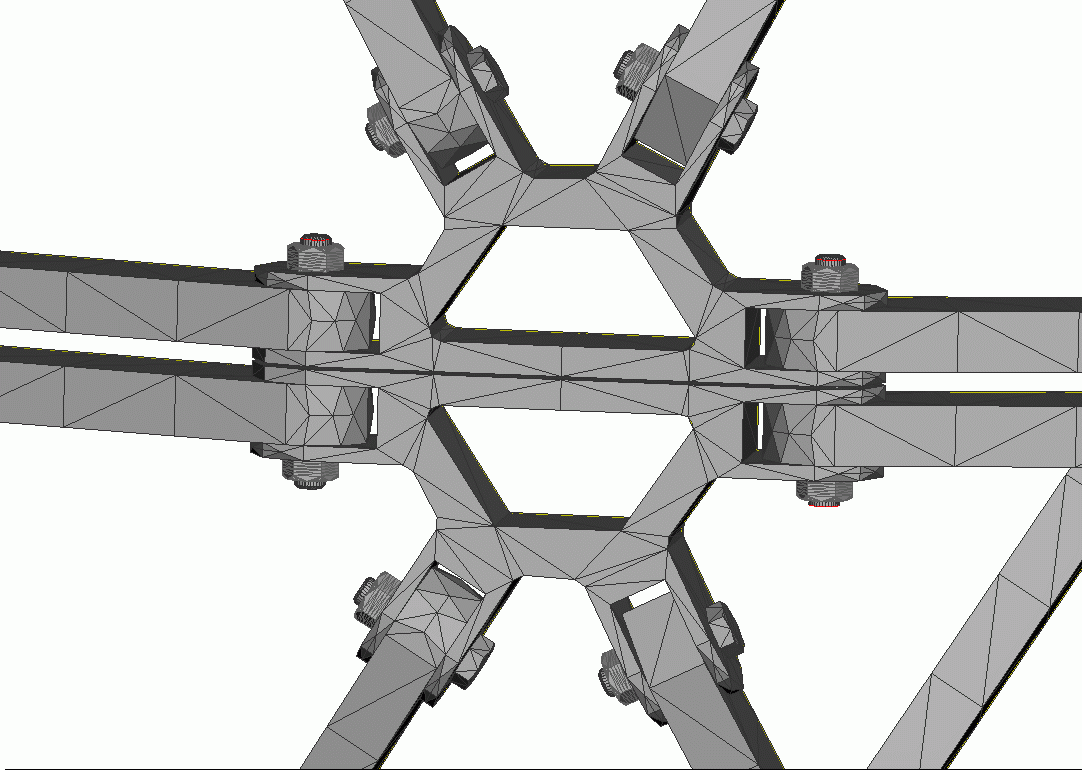}
    \subcaption{Surfaces are intersecting. The bolts are penetrating the bracket. There are narrow flow paths between the truss frames and the brackets.}
    \label{fig:geodom_error}
  \end{minipage}
  \begin{minipage}{0.3\hsize}
    \includegraphics[keepaspectratio,width=\textwidth]{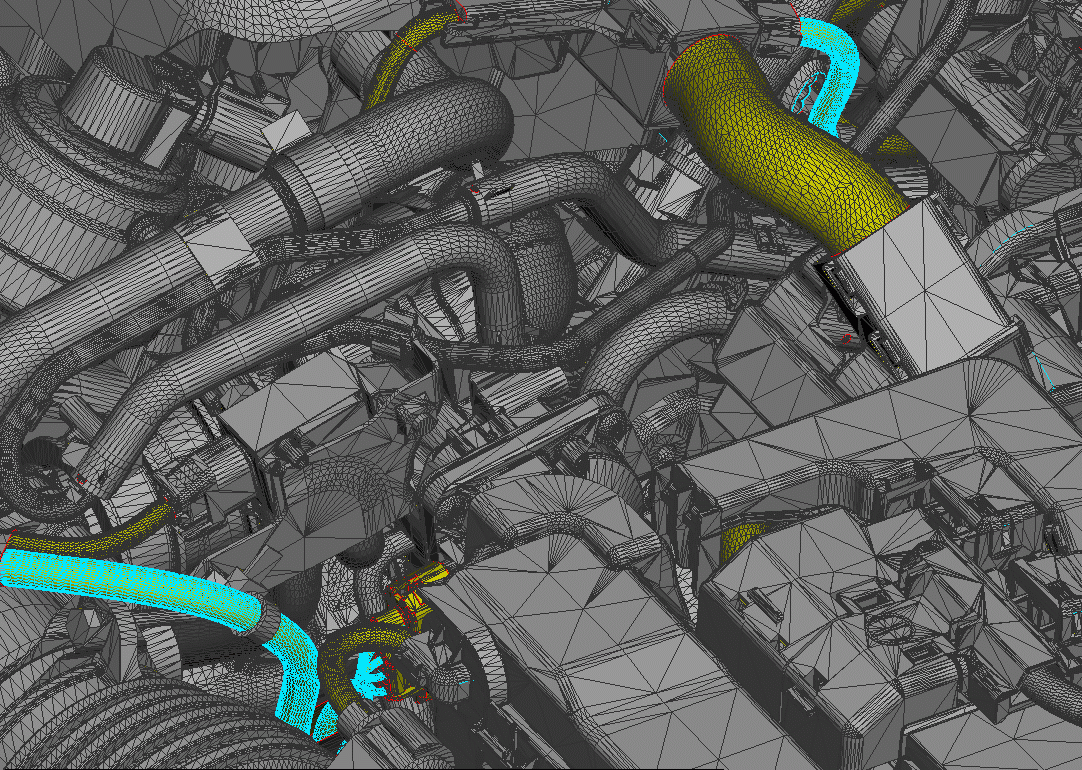}
    \subcaption{Numerous error faces. Gray indicates the front face, yellow indicates the back face. There are many narrow flow paths, gaps, overlaps, and zero-thickness faces.}
    \label{fig:engine_error}
  \end{minipage}
  \caption{Magnified view of the problematic geometry area: (a) an intersection of frames, (b) a truss frame bracket tightened with bolts and nuts, and (c) an example of the automobile engine bay.}
  \label{fig:geometry_error}
\end{figure}

Figure \ref{fig:geometry_error} shows examples of geometrical errors. In Fig. \ref{fig:geometry_error}\subref{fig:frame_error}, there is  issue (1) between the frames. The red colored edges indicate gaps or overlaps. The light blue colored edges indicate duplicated connected edges. There are also small gaps between the frames; therefore, to divide the space volume into either fluid or solid, we need to fill in all the gaps, i.e. this is issue (3). In Fig. \ref{fig:geometry_error}\subref{fig:geodom_error}, there is issue (1) on the bolt-top faces and issue (2) between the bolts and frames. issue (3) is seen between the frames and brackets. The gap the flow should pass through must be reserved, and the gap the flow should not pass must be filled. In Fig. \ref{fig:geometry_error}\subref{fig:engine_error}, there are a large number of faces exhibiting issues (1), (2), (3), and (4). Some parts consist of thin plates, and some parts are facing in opposing directions. There are many narrow flow paths. It is difficult to check everything by hand. However, if these checks are not completed, it will not even be possible to determine the fluid region for grid generation.

Regarding the unstructured grid or mesh-less method, it is difficult to solve (1) to (4), though surface wrapping can avoid these issues. In contrast, the Cartesian grid would be generated with no regard to these issues. The immersed boundary would still be represented through some means of the surface grid covering the boundary $\Gamma_{IB}$. The immediate task therefore is to incorporate appropriate modifications on the equations of the IBM in the vicinity of the dirty boundaries.

In the continuous forcing IBM (Peskin's type \cite{Peskin1972}) and the discrete forcing IBM (Mohd-Yusof's type \cite{Mohd-Yusof1997}), the forcing term in Eq. \eqref{eq:discrete_forcing_term} can be formulated, including the non-conformal body of issue (1) and the zero-thickness face of issue (4). However, issue (2) cannot be solved, as the solution inside of the solid has to be canceled out manually. Generally, velocity and pressure are set to zero, or the unphysical flow is left in the solid region. Issue (3) adversely affects the weighting of the force distribution. The use of a distribution function makes it difficult to obtain a sharp solution, and calculations for narrow (below the grid resolution; i.e., insignificant) flow channels are considered to be errors. In the sharp discrete forcing representation (Mittal's type \cite{Mittal2007}), the cells including the immersed boundary and its adjacent (i.e., the so-called ghost cells) are defined on issues (1) and (3). However, continuously filled solid cells are required that satisfy the well-posed definition for ghost cells. These must be determined by defining a closed volume beforehand, otherwise, the search fails and the calculation of interpolation becomes either ill-conditioned or unfeasible. Therefore, issues (2) and (4) remain. Further, if issue (1) occurs, the calculation of the intersection can fail due to the gap on the boundary.

Solving issues (1) and (2) mathematically requires searching for all combinations of face borders to form closed volumes. Note that arbitrary shape data have an enormous number of faces, and the searching should be finished within a finite time. Even if methods that could solve using a relatively simple problem are formulated, and it would be anticipated that they could apply to a fluid dynamics problem, there is no guarantee that they would be able to solve complex shape data in a finite time; this is known as the P versus NP problem. For instance, full-vehicle geometry data have hundreds of thousands of CAD surfaces, or tens of millions of face elements. A better approach is to avoid forming closing volumes, allowing the flow to split into the inner and outer side locally in space. If local stencils of flow solution are separated near the face, this can be easily realized.

In this study, a method to solve issues (1) and (3) is proposed based on the ghost-cell boundary condition method proposed by Mittal \cite{Mittal2007}, coupled with the continuous forcing method of Peskin's type IBM \cite{Peskin1972}. At the same time, issue (4) is solved by using the dummy cell definition to have an arbitrary placed solid region that splits the regions into inner and outer locally. The formulation of the IBM was modified by projecting the interpolation onto the axial direction, to avoid failures when searching for interpolation points. The method was able to solve issue (2), and displayed a high affinity with the solution for issue (3). Issue (3) was not completely solved, but the method made this issue easier to address. Finally, the issue that (1) causes in the intersection calculation was avoided by the adjustment of the parameter on the ray-tracing algorithm.

\subsubsection{Arbitrary dummy cell construction}\label{dummy_cell}

\begin{figure}[htb]
  \centering
  \begin{minipage}{0.3\hsize}
    \vspace{0.4cm}
    \includegraphics[keepaspectratio,width=\textwidth]{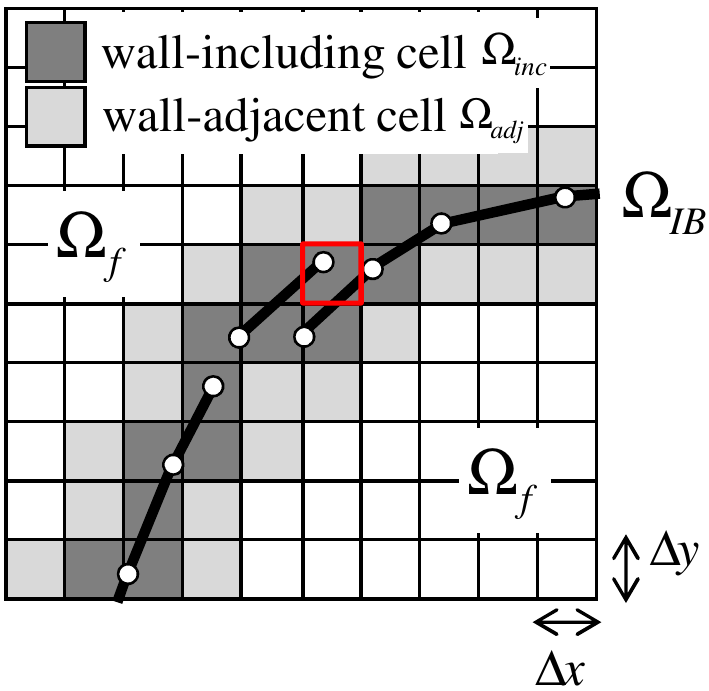}
    \subcaption{Background cells.}
    \label{fig:dummy_cell_1}
  \end{minipage}
  \begin{minipage}{0.3\hsize}
      \includegraphics[keepaspectratio,width=\textwidth]{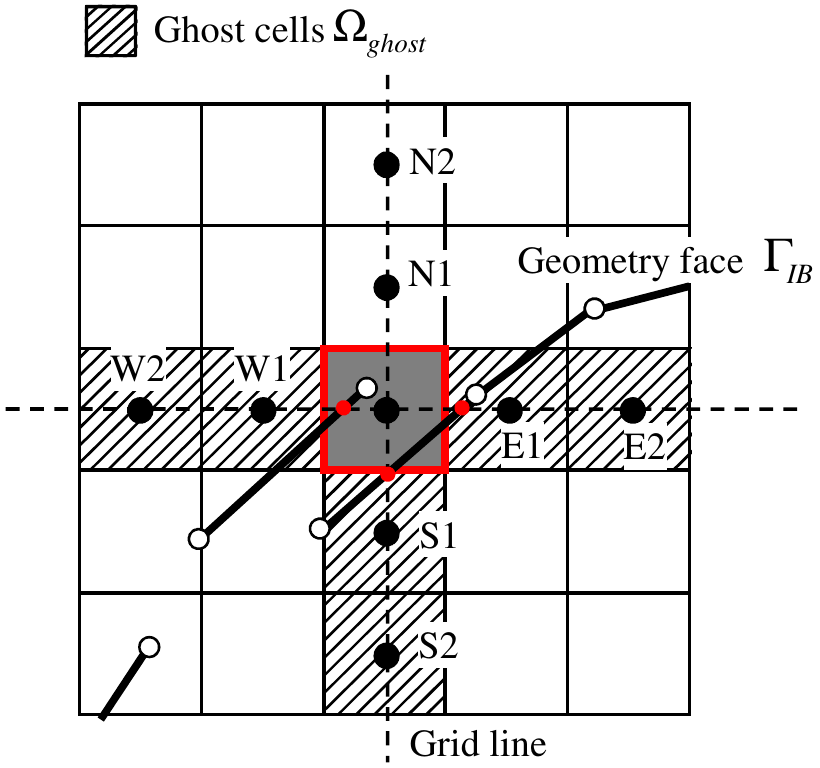}
    \vspace{0.2cm}
    \subcaption{Arbitrary dummy cell in $\Omega_{inc}$.}
    \label{fig:dummy_cell_2}
  \end{minipage}
  \caption{Arbitrary dummy cell definition: (a) The cells shaded dark gray are wall-including cells $\Omega_{inc}$ that have discretized wall boundaries $\Gamma_{IB}$. The cells shaded light gray are cells that neighbor the wall-including cells, and are referred to as wall-adjacent cells $\Omega_{adj}$. (b) The dummy cell for a wall-including cell shown by the red box in (a). The black line is the immersed wall boundary. Black circles labeled N1, N2, S1, S2, E1, E2, W1, and W2 are cell centers. Red circles indicate intersecting points. The white circle is a node of the geometry face. Shaded cells are ghost cells $\Omega_{ghost}$.}
  \label{fig:arb_dummy_cell}
\end{figure}

Region $\Omega_{IB}$ is enlarged by one or two layers of cell width and defined again as region $\Omega_{dum}$. A virtual cell data structure, called a dummy cell, is employed in $\Omega_{dum}$ to impose the boundary condition directly on Mittal's. Figure \ref{fig:arb_dummy_cell} shows a conceptual diagram of the dummy cells. The cells shaded dark gray are fluid cells $\Omega_{inc}$ that include wall boundary information $\Gamma_{IB}$. The cells shaded light gray are adjacent fluid cells $\Omega_{adj}$ that neighbor the wall-including cells. Consequently, they are defined as:
\begin{align}
  \label{eq:dummy_cell_region}
  & \Omega_{dum} = \Omega_{inc} \cup \Omega_{adj}. &
\end{align}
Both types of cells have their dummy cells. Each dummy cell has $5 \times 5 \times 5$ cell information in three-dimensional axial directions. Consequently, the stencil calculation satisfies second-order difference at the maximal in one side direction. Figure \ref{fig:arb_dummy_cell}\subref{fig:dummy_cell_2} shows a dummy cell of a wall-including cell, which is a part-fluid cell. The fluid variables of dummy cells were copied from the background grid. There was an intersection point between the wall boundary and the grid line, shown as red circles in Fig. \ref{fig:arb_dummy_cell}\subref{fig:dummy_cell_2}. The grid line is shown as a dashed line connecting each cell center. In the directions of S1, E1, and W1, intersecting points are present; therefore, cells S1, S2, E1, E2, W1, and W2 are treated as solid cells (i.e., the ghost cells). The ghost cells are defined as $\Omega_{ghost} \subseteq \Omega_{dum}$. This feature allows the splitting of regions locally to inner/outer side, viewed from the cell center, due to the existence of ghost cells. This does not require searching for entire fluid/solid regions in surrounding cells; the multiple face intersection or the zero-thickness walls are thus not an issue. In the same way, the wall-adjacent cells have their dummy cells.

The intersection was calculated using the ray-tracing algorithm. There are many options to choose a fast and accurate algorithm. In this study, the Tomas Moller \cite{Moller1997} method was chosen. This algorithm solves a $3 \times 3$ matrix consisting of a ray direction viewed from the cell center and a normal vector of triangle and is designed to be stable by appropriately setting the error of the determinant.

This virtual data, in addition to the background, were designed to be accessed by the 'if' branch of the application code when the stencil operation of differentiation reaches $\Omega_{inc}$ or $\Omega_{adj}$. The 'if' branch was applied only once in the loop of advection, diffusion, and pressure equations. The penalty for computational performance was thus small. However, for datasets that have ''watertight'' geometries, all cells within the solid region were also calculated as being fluid. This wastes computational resources. Additionally, it is not surprising that this technique requires more computing memory for duplicated data.

\subsubsection{Axis projection of interpolation}\label{axis_projection}

\begin{figure}[htb]
  \centering
  \begin{minipage}{0.3\hsize}
    \vspace{0.3cm}
    \includegraphics[keepaspectratio,width=\textwidth]{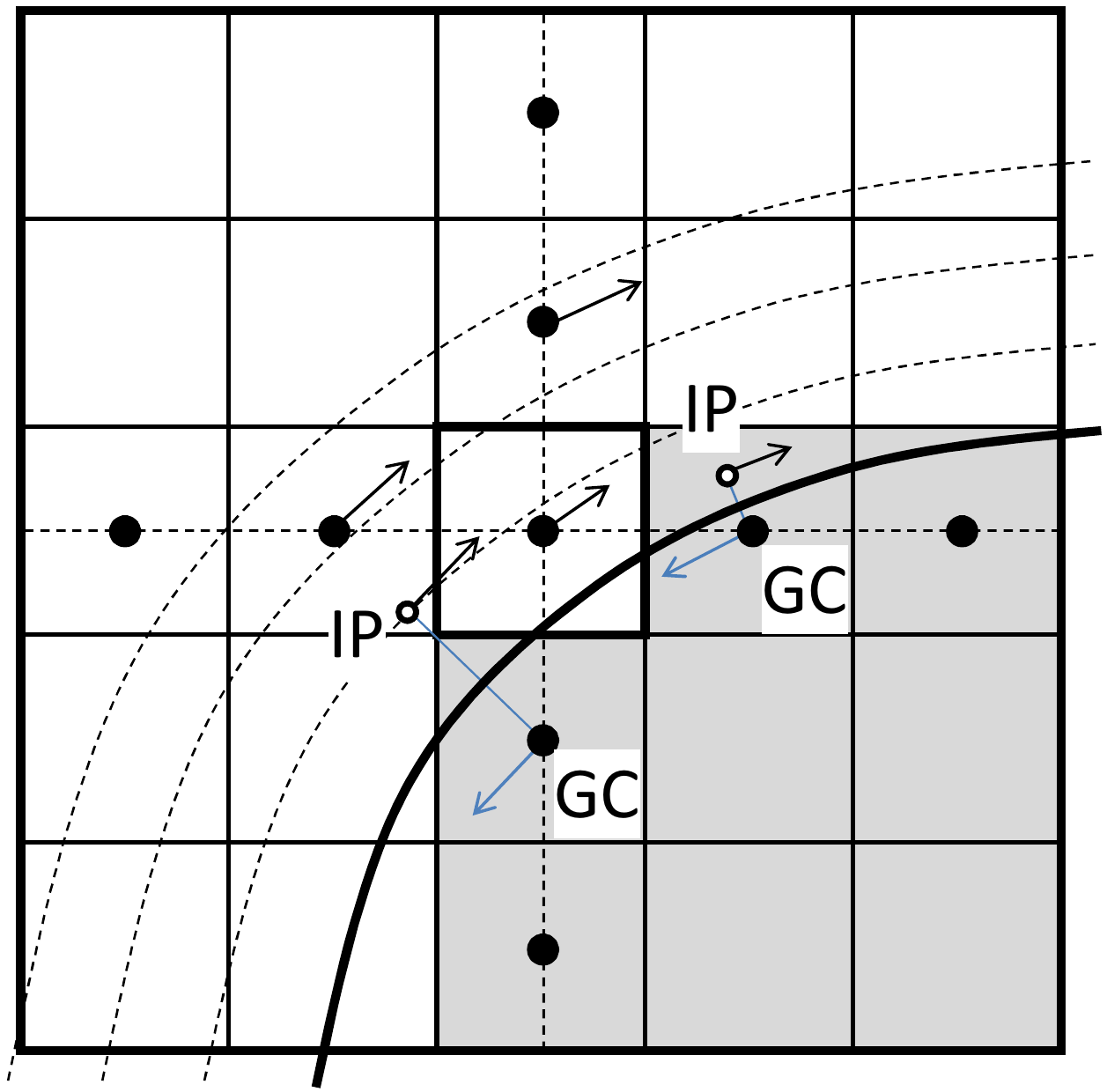}
    \subcaption{Ghost cell and imaginary point in Mittal's IBM.}
    \label{fig:wall_boundary_1}
  \end{minipage}
  \begin{minipage}{0.305\hsize}
    \includegraphics[keepaspectratio,width=\textwidth]{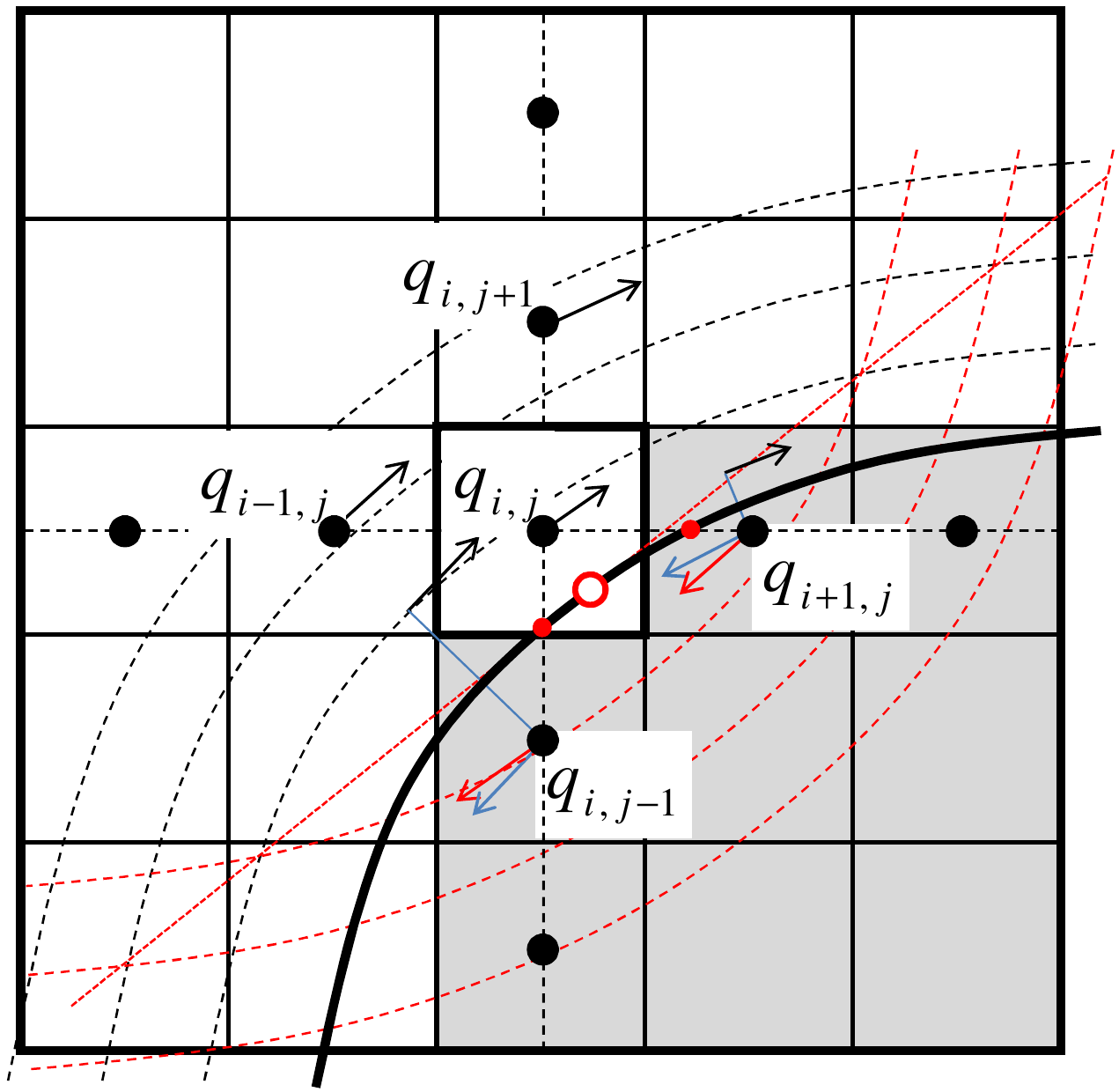}
    \subcaption{Axis-projected IBM.}
    \label{fig:wall_boundary_2}
  \end{minipage}
  \caption{Two-dimensional schematic view of the definition of the IBM wall boundary condition on the dummy cell. GC indicates a ghost-cell center, while IP indicates an imaginary point. A black dashed line indicates a flow field function, while a red dashed line indicates the sign-inverted function of the flow field. The red circle marks an intersection point of the perpendicular line from the concerned cell center to the surface.}
  \label{fig:wall_boundary}
\end{figure}

The schematic of the direct boundary condition IBM proposed by Mittal \cite{Ye1999,Mittal2005,Mittal2007,Mittal2008,Bergmann2013}  in the two-dimensional case is shown in Fig. \ref{fig:wall_boundary}\subref{fig:wall_boundary_1}. The ghost cell is indicated by a black circle labeled GC in the figure. The imaginary point is indicated by a white circle labeled IP; it is defined by the distance from the ghost-cell center and the wall. The IP is on the opposite side of the intruding fluid region and has the same distance from the wall in the direction orthogonal to the wall. The fluid velocity at the imaginary point was calculated by the two-dimensional interpolation of the surrounding fluid cells, such as bilinear interpolation (trilinear interpolation for three-dimensional case) or polynomial interpolation. The sign-inverted velocity was then imposed directly onto the ghost cell on the solid side. This scheme is shown by the blue arrows in Fig. \ref{fig:wall_boundary}\subref{fig:wall_boundary_1}.

The fluid quantity, e.g., velocity, at the imaginary point, expressed as $q_{IP}$, was calculated using the Cartesian coordinates $(x, y)$ of an imaginary point. The ghost-cell quantity $q_{GC}$ was then imposed directly as the boundary condition of $\Gamma_{IB}$. The interpolation can be described as follows:
\begin{align}
  \label{eq:Mittal}
  & q_{IP} = c_1 xy + c_2 x + c_3 y + c_4, & \\
  & [ C ] = \left[ c_1 \, c_2 \, c_3 \, c_4 \right]^{\mathrm{T}} = \left[ V \right]^{-1} [ q ], \nonumber & \\
  & V = \left[
    \begin{array}{cccc}
      xy_1 & x_1 & y_1 & 1 \\
      xy_2 & x_2 & y_2 & 1 \\
      xy_3 & x_3 & y_3 & 1 \\
      xy_4 & x_4 & y_4 & 1
    \end{array}
    \right], \nonumber \\
  & [ q ] = \left[ q_1 \, q_2 \, q_3 \, q_4 \right]^{\mathrm{T}}, & \\
  & q_{GC} = \zeta q_{IP} + \eta = \zeta \sum_{i=1}^{4} \alpha_{i} \, q_{i} + \eta, &
\end{align}
where $[ C ]$ is the matrix containing the four unknown constants, using the corresponding Vandermonde matrix $V$. It can be rewritten as $\alpha_i$, which is a coefficient of interpolation of fluid quantity $q_i$ weighted by the distance from surrounding four points to the imaginary point. For the Dirichlet boundary condition, $\zeta = -1$ and $\eta = 2 q_{IB}$, where $q_{IB}$ is a quantity on the boundary. For the Neumann boundary condition, $\zeta = 1$ and $\eta = (\nabla q)_{IB} \cdot \Delta \bm{n}$, where $\Delta \bm{n}$ is the normal vector on the boundary. The effectiveness of this method has been actively discussed in the field of biological CFD applications in recent years. The literature has been summarized by Mittal \cite{Mittal2005}, so we will not go into detail here.

The ghost cells in this study were located in the arbitrary dummy cells; The method is therefore already superior for complex geometries. However, the feasibility of the surrounding points also needs to be guaranteed. For instance, when tiny geometries with sizes below the grid resolution or multiple intersections of volumes occurred, the searching of fluid cells to be used for interpolation fails, and interpolation, therefore, becomes unfeasible. In practice, this problem has been avoided by establishing a case selection procedure or a special exception treatment. Mittal \cite{Mittal2007} applied the replacement of a body-normal operator and $q_i$ by imposing the boundary condition directly when the ghost cell is included in the surrounding points for the interpolation. This is not always a valid countermeasure, however, and is problematic depending on the complexity of the geometry.

In the present study, an axis projection method of interpolation is proposed. The imaginary field is defined as a function of the fluid quantity $f(x,y)$ on the position of $(x,y)$, which is indicated as a red dashed line in the two-dimensional schematic view in Fig. \ref{fig:wall_boundary}\subref{fig:wall_boundary_2}, this is equivalent to a sign-inverted field of the flow quantity, shown as a black dashed line. Now, the imaginary point is selected on the grid line in a one-dimensional form. This is equivalent to select first-order terms only in the Taylor expansion of function $f(x,y)$, that is, high-order terms are discarded. It is possible to maintain the necessary accuracy by selecting the appropriate number of orders, however, the first order was chosen here intentionally. Hence, the two-dimensional interpolation was projected onto the one-dimensional calculation. In this case, the searching of the imaginary point has become easy whenever the ray-tracing algorithm has been done feasible.

Consequently, the immersed boundary interpolation can be expressed in first-order form. Mittal's definition (Eq. \eqref{eq:Mittal}) was rewritten for the Dirichlet boundary condition for a no-slip wall following the two-dimensional schematic in Fig. \ref{fig:wall_boundary}\subref{fig:wall_boundary_2} as:
\begin{align}
  \label{eq:q_Mittal}
 & q_{k} = - \sum_{k} \alpha_{k} \, q_{k}, \quad (Dirichlet); &
\end{align}
where $q_{k}$ is the fluid quantity at position $(k)$, and $k$ is the number of surrounding points. The $\alpha_{k}$ is a coefficient of the interpolation weighting with the distance from the surrounding points to the imaginary point.

The present study started with Gibou's idea \cite{Gibou2002} that the extrapolation scheme can be used for the calculation of the ghost-cell velocity. Gibou et al. applied the adjacent fluid node value to the ghost cell, as well as the boundary value, in the axial direction to satisfy a Dirichlet boundary condition. That is a one-dimensional extrapolation scheme for the ghost cell, is described by:
\begin{align}
  \label{eq:q_Gibou}
 & q_{i+1,j} = \frac{q_{IB}+(\theta - 1) q_{i,j}}{\theta}, \quad (Dirichlet); &
\end{align}
where $q_{i,j}$ is the fluid quantity at position $(i,j)$, and $\theta$ is a weighting factor defined by $\theta = \mid x - X \mid / \Delta x$. Gibou pointed out that the above interpolation scheme behaves poorly for small $\theta$ in terms of conservation. Therefore, they took the drastic countermeasure of imposing $q_{i,j} = q_{IB}$, where the perturbation of the interface location does not degrade the overall second-order accuracy of the two-dimensional and three-dimensional solutions. 

Following this idea, the distribution function $\mathcal{D}$ of direct forcing immersed boundary in Eq. \eqref{eq:discrete_forcing_term} was written as a linear interpolation:
\begin{align}
  \label{eq:distribution_function}
  & \mathcal{D} (\mid x - X \mid) =
  \begin{cases}
  \displaystyle \frac{\mid x - X \mid}{\Delta x + \mid x - X \mid}, & \quad for \; x \in \; \Omega_{dum} - \Omega_{ghost}, \\
  \displaystyle \frac{\Delta x - \mid x - X \mid}{\Delta x + \mid x - X \mid}, & \quad for \; x \in \; \Omega_{ghost}, \\
  \displaystyle 0, & \quad for \; x \in \; \Omega_{f},
  \end{cases} &
\end{align}
where $\Omega_{ghost}$ indicates the solid region of ghost cells. The interpolation for the force distribution on the fluid side of $\Omega_{dum}$ was designed as a linear interpolation to match the force distribution on the solid side. In other words, the distribution function has a width of $\Delta x + d$ for Peskin's continuous forcing method \cite{Peskin1972}, where $d = \mid x - X \mid$ is the distance of each axis from the center of the concerned cell to the body of the intersecting point. As long as the same distribution function is applied to both the fluid and solid sides, the conservation law will be satisfied.

In light of Gibou's warning, the non-slip boundary condition in Eq. \eqref{eq:q_Gibou} was rewritten to avoid extrapolation, as:
\begin{align}
  \label{eq:q_axis}
  & q_{i+1,j} =
  \begin{cases}
    \vspace{3mm}
    \displaystyle \frac{\Delta x}{d} q_{IB} + \frac{d - \Delta x}{d} q_{i,j}, & on \quad 0.5 \leq d / \Delta x < 1.0, \\
    \displaystyle 2 q_{IB} + \frac{2d - \Delta x}{\Delta x} q_{i-1,j} - \frac{2d}{\Delta x} q_{i,j}, & on \quad 0.0 \leq d / \Delta x < 0.5.
  \end{cases}
\end{align}

\begin{figure}[htb]
  \centering
    \begin{minipage}{0.5\hsize}
      \includegraphics[keepaspectratio,width=\textwidth]{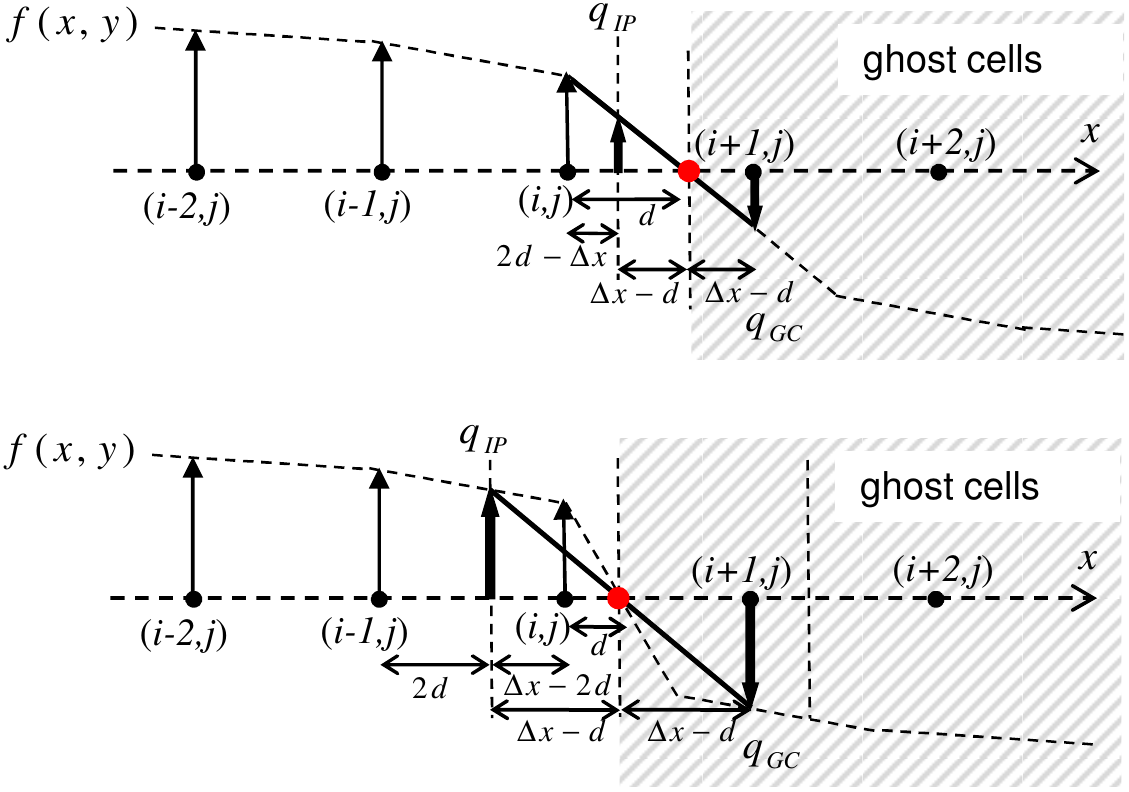}
    \end{minipage}
  \caption{One-dimensional grid stencil lines and definition of the forcing for the ghost cell. Upper: on $0.5 \leq d / \Delta x < 1.0$. Lower: on $0.0 \leq d / \Delta x < 0.5$. The cell size on this axis is given by $\Delta x$, and $d$ indicates the distance from the cell center to the intersection point of the characteristic plane, which is marked by a red circle. The $q_{i,j}$ is a flow quantity of the cell. The $q_{IP}$ is a quantity on the imaginary point, and $q_{GC}$ is a quantity on the ghost cell.}
  \label{fig:immersed_forcing}
\end{figure}

In Fig. \ref{fig:wall_boundary}\subref{fig:wall_boundary_2}, a red dotted line indicates the characteristic plane on the geometric gravity center of surrounding cells, the white circle on the red bold line is the gravity center, and a red circle indicates an intersecting point. The fluid quantity at the imaginary point was then calculated through linear interpolation, and the sign-inverted quantity was imposed onto the ghost cell, as shown in Fig. \ref{fig:immersed_forcing}. This is also expressed as $q_{i+1,j}$ in Eq. \eqref{eq:q_axis}, and is shown as a red arrow in Fig. \ref{fig:wall_boundary}\subref{fig:wall_boundary_2}. This process can be understood as the sum of blue arrows replaced by the sum of red arrows in Fig. \ref{fig:wall_boundary}\subref{fig:wall_boundary_2}. This is not always correct, however, when the flow field function of $f(x,y)$ is smooth enough and the distance from the wall to concerning cell center is small, this assumption becomes acceptable. The reason is that the immersed ghost-cell information is always referred to as the divergence form; the interpolated values are not used directly in the solution but are used as the sum of it. Although this calculation includes a certain amount of error, the main subject of this article is to examine the degree of acceptance for this assumption.

The implementation in three dimensions is straightforward, as it was independently carried out for each axial component. However, the discretization error at the wall was of the first order and served as a restraint on the accuracy. As Tseng \cite{Tseng2003} and Nishida \cite{Nishida2009} pointed out, care should be taken when using the method for flows having a high $Re$. The accuracy of interpolation should be improved by employing a higher-order interpolation. This method would also be easy to enhance by applying, for example, a polynomial interpolation, Hermite interpolation, or a level-set function. However, for the sake of simplicity, the numerical tests in this study used a first-order interpolation. Therefore, near the wall, this method behaves as velocity damping.

The formulation was stable and satisfies mass conservation for any shape topology. All ghost cells were well-posed, irrespective of shape complexity, and there was no instability due to the small cut cell or small distance. Note that this solution did not require an iterative method or uncertain constants. This method converged rapidly even if the SOR was employed to solve the entire linear equation system. Additionally, as the BCM had the data structure of the cube unit, it must refer to adjacent cubes when the interpolation was performed at the cube interface. This would deteriorate the computation efficiency if the diagonal direction was referred in addition to the axis direction. However, when axis projection was used, a reference to the diagonal cube could be omitted.

The software framework described in the present study was fully parallelized with a message passing interface (MPI) for communication between the calculation nodes of the distributed memory platforms and open multi-processing (OpenMP) within each node in operations at the cube level. This included pre- and post-processing---for further details on performance, refer to Jansson \cite{Jansson2018} and Onishi \cite{OnishiSC17}.

\subsection{Spatial and temporal accuracy}

This section describes the results of grid convergence studies carried out with the current method. The effect of the boundary condition of the immersed boundary on the accuracy of the solution is examined according to the second-order accurate spatial discretization used for regular fluid cells.

The two-dimensional unsteady Taylor-Green decaying vortex problem is considered as:

\begin{align}
  \label{eq:Taylor-Green}
 & u_{1}(x,y,t) = -cos \; \pi x \; sin \; \pi y \; e^{-2 \pi^{2} t / Re}, & \\
 & u_{2}(x,y,t) = sin \; \pi x \; cos \; \pi y \; e^{-2 \pi^{2} t / Re}, & \\
 & p(x,y,t) = -\frac{1}{4} (cos \; 2 \pi x + cos \; 2 \pi y ) e^{-4 \pi^{2} t / Re}. &
\end{align}

\begin{figure}[htb]
 \centering
 \begin{minipage}{0.3\hsize}
   \includegraphics[keepaspectratio,width=\textwidth]{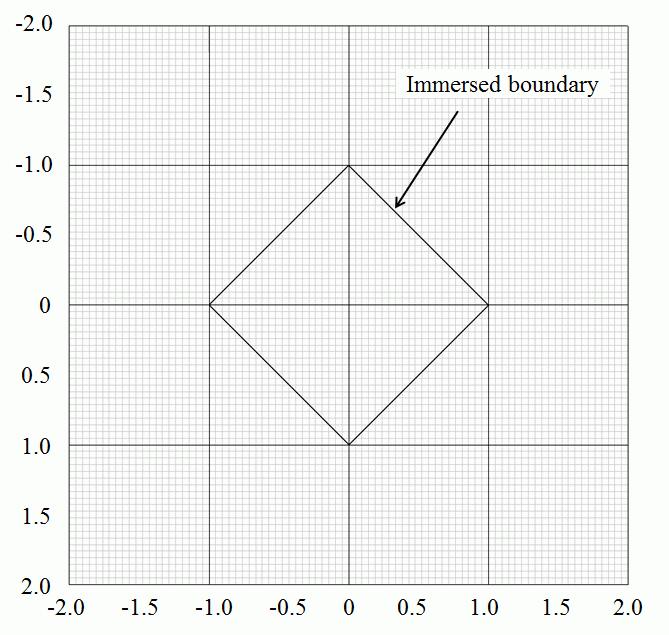}
  \subcaption{Square immersed boundary.}
  \label{fig:Taylor-Green_square}
 \end{minipage}
 \begin{minipage}{0.3\hsize}
   \includegraphics[keepaspectratio,width=\textwidth]{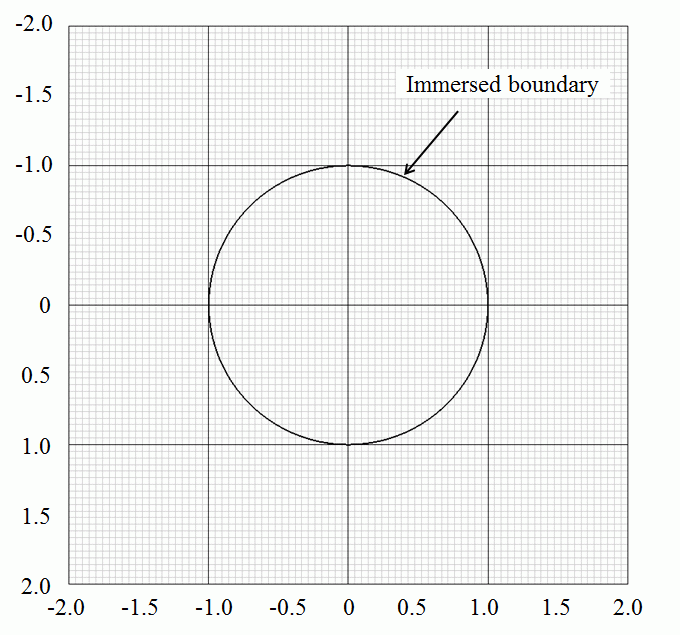}
  \subcaption{Circular immersed boundary.}
  \label{fig:Taylor-Green_circular}
 \end{minipage}
 \caption{Grid system, computational domain, and immersed boundary for the vortex decay problem. Thin black lines: cubes, gray lines: cells, thick black lines: immersed boundary.}
 \label{fig:Taylor-Green}
\end{figure}

\begin{figure}[htb]
 \centering
 \begin{minipage}{0.3\hsize}
   \includegraphics[keepaspectratio,width=\textwidth]{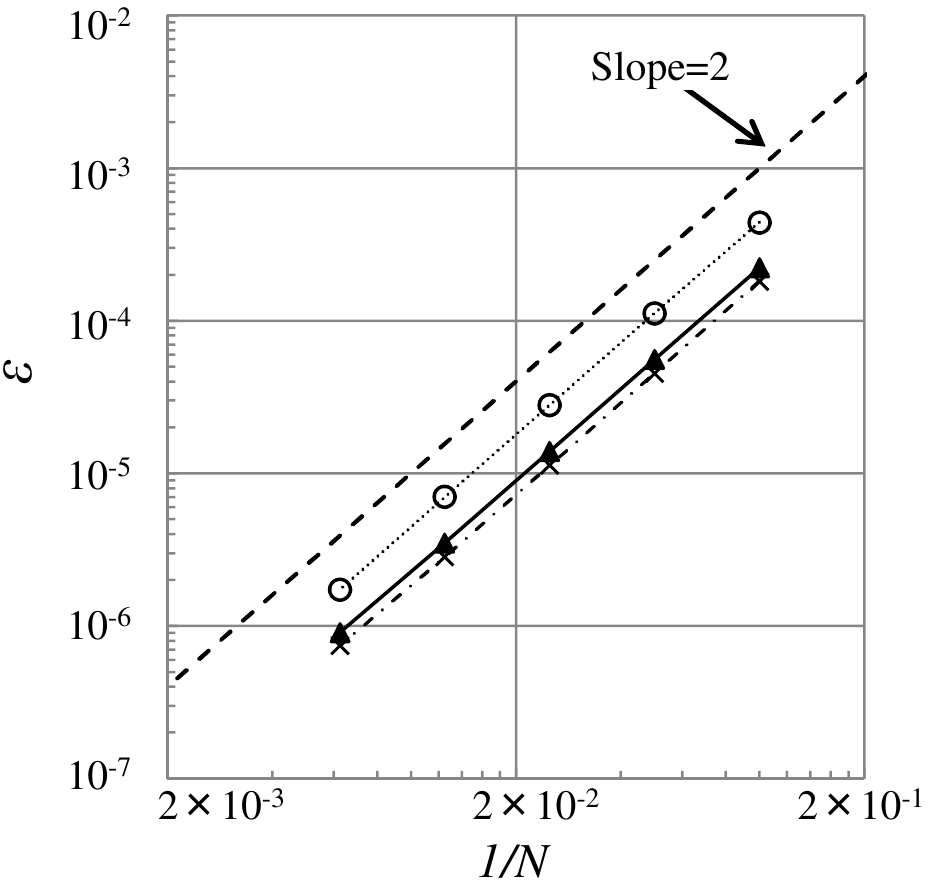}
  \subcaption{Without IBM having regular fluid cells only.}
  \label{fig:Taylor-Green_error_BCM}
 \end{minipage}
 \begin{minipage}{0.3\hsize}
   \includegraphics[keepaspectratio,width=\textwidth]{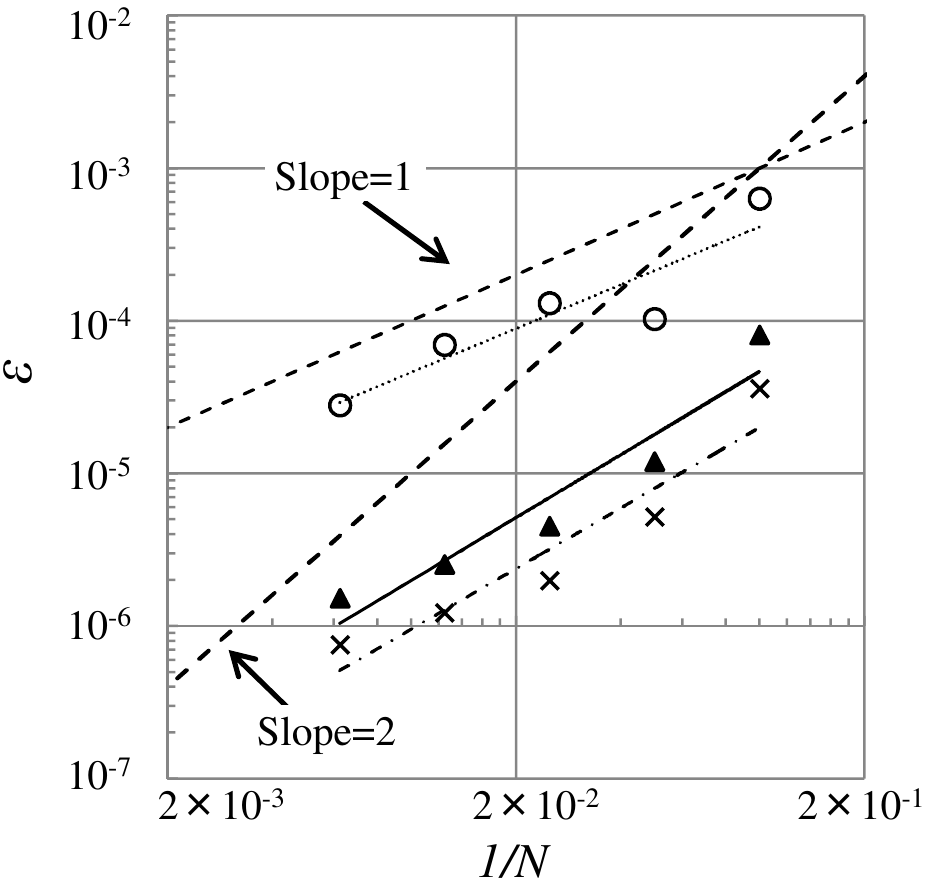}
  \subcaption{Topology-free IBM having a square immersed boundary.}
  \label{fig:Taylor-Green_error_square}
 \end{minipage}
 \begin{minipage}{0.3\hsize}
   \includegraphics[keepaspectratio,width=\textwidth]{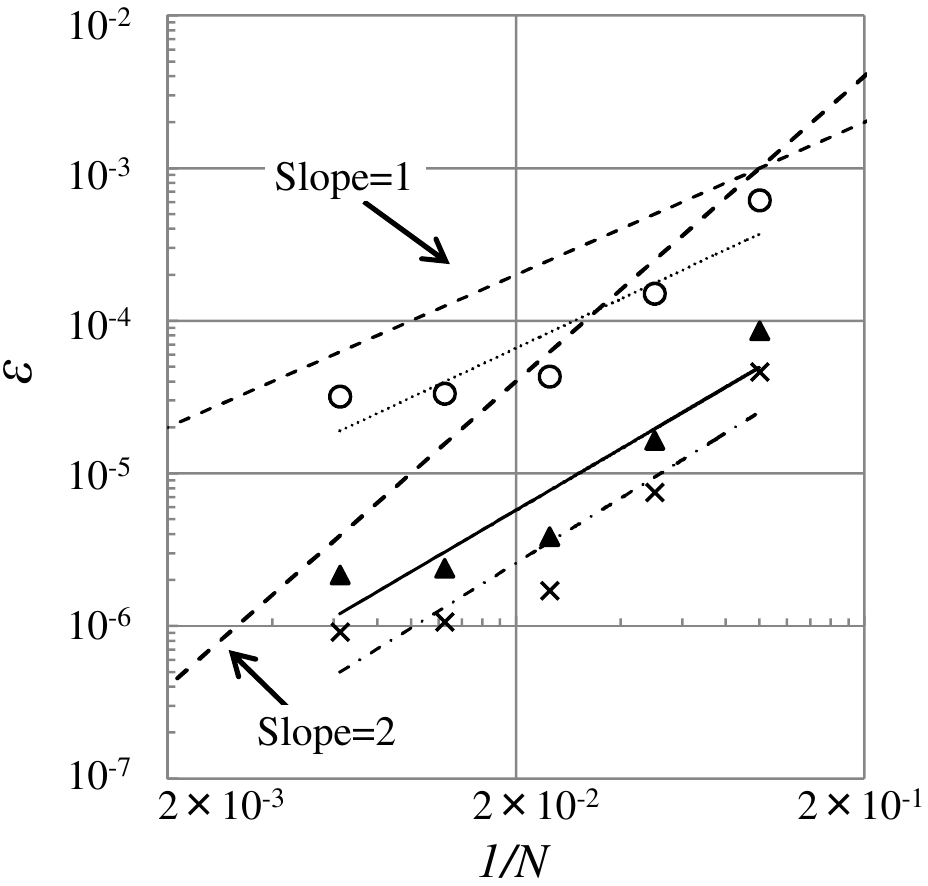}
  \subcaption{Topology-free IBM having a circular immersed boundary.}
  \label{fig:Taylor-Green_error_circular}
 \end{minipage}
 \caption{$L_{1}$, $L_{2}$, and $L_{\infty}$ norms of the error for the streamwise velocity ($u_{1}$) component versus the computational cell size. $\times : L_{1}, \blacktriangle : L_{2}, \bigcirc : L_{\infty}$.}
 \label{fig:Taylor-Green_error}
\end{figure}

Figure \ref{fig:Taylor-Green} shows the grid system, computational domain $(-2.0 \leq x, y \leq 2.0)$, and immersed boundary (thick black line) for a square cylinder having a length of $\sqrt{2}$, and a circular cylinder having a radius of $1.0$. A uniform BCM grid that has two levels of discretization was used; i.e., cubes (black lines) and cells (gray lines).

Five sizes of Cartesian grids were used by fixing the number of cubes to 16 and the number of cells in each direction to 40, 80, 160, 320, and 640. Computations were performed by varying the mesh size, but by keeping the non-dimensional uniform timestep of $1.0 \times 10^{-4}$. The solution was integrated using the Adams-Bashforth explicit time integration scheme for $3.0 \times 10^{3}$ timesteps. The second-order central spatial difference scheme was used, while the BiCGStab method was used to solve the Poisson equation. The initial velocity condition was set at $t = 0$, and the velocities at the immersed boundary in time were obtained from the analytical solution of Eq. \eqref{eq:Taylor-Green}. The periodic boundary condition was used on the boundary of the computational domain. The Reynolds number was set to $Re = U_{0} L / \nu = 100$, where $U_{0}$ is the initial maximum velocity, $L$ is the size of a vortex, and $\nu$ is the kinematic viscosity.

Figure \ref{fig:Taylor-Green_error} shows variations in the $L_{1}$ and $L_{2}$ error norms, and the maximum error norm $L_{\infty}$ in the streamwise velocity component $u_1$. The error was calculated from the analytical solution. It is plotted against $1/N$, where $N$ is the number of cells in each direction. The second-order and first-order lines are indicated by the dashed lines. Figure \ref{fig:Taylor-Green_error}\subref{fig:Taylor-Green_error_BCM} shows that the original BCM solver without IBM had a second-order accuracy in space. Meanwhile, Fig. \ref{fig:Taylor-Green_error}\subref{fig:Taylor-Green_error_square} shows that the slope of $L_{2}$ on the square boundary was approximately 1.37. The slope of the circular boundary was approximately 1.34, which nearly matches the first order. This is due to the influence of the first-order interpolation by the axis projection. However, as far as the three points in the right half of the coarse grid are concerned, we consider that convergence of $L_{2}$ close to second-order accuracy was obtained. These values were 2.08 and 2.23 respectively. As the cell size decreased, the shape around each cell approached a simpler plane. Hence, we consider that the accuracy reached its limit, as the reference point for interpolation became less than one. This can be solved by increasing the number of reference points; however, in this study, no particular correction was made. Therefore, in the following calculations, the accuracy of the turbulence near the wall may have been sacrificed. However, as the Cartesian grid tended to be relatively coarse at high-$Re$ flows, this property is also considered to be a useful feature of the artificial damping provided around the wall.

\section{Results and discussion}

Here, we calculated the flow around the Ahmed body to verify the accuracy of the method. Furthermore, we selected three subjects that could not be analyzed by the existing IBM and demonstrated the usefulness of our method. Three-dimensional flow was simulated past a flat plate, for a wide range of angles of attack. This case does not apply to Mittal's type of IBM. The flow past a sphere for various geometries that have dirty topology was also simulated. Some cases are not applicable to Mittal's type of IBM and Peskin's type of IBM. Our method was then applied to full-vehicle aerodynamic analysis and city area wind environment analysis, to demonstrate its ability for simulating flows with highly complex geometries that have dirty topology. These have not been successfully reported using the existing IBM.

\subsection{Ahmed body}
The Ahmed body \cite{Ahmed1984} is a geometry that reproduces a flow field by simply simulating the aerodynamic phenomenon around an automobile. It is commonly used for basic flow validations on the bluff body. In this study, this model was used to simulate the basic velocity field validation using established measurement data \cite{Lienhart2003}.

The chosen model had 25 degrees of slant angle, the inclined back face angle of the rear body. The model had a size of $1044 \times 389 \times 288 mm$ with a floating distance of $50 mm$ from the floor, supported by four legs. The calculation conditions were adapted to the measurement conditions in the low-speed wind tunnel, using the laser Doppler anemometer of Lienhart et al. \cite{Lienhart2003}. The inlet uniform velocity was set to $40 m/s$, and the Reynolds number, based on the height of the body, was set at $Re = 768,000$. The hybrid scheme was used for spatial discretization in the second-order central-difference scheme, with 10\% blending of the quadratic upstream interpolation for the convective kinematics (QUICK) scheme. The Crank-Nicolson implicit scheme was used for time integration.

\begin{figure}[htbp]
 \centering
 \begin{minipage}{0.8\hsize}
   \includegraphics[keepaspectratio,width=\textwidth]{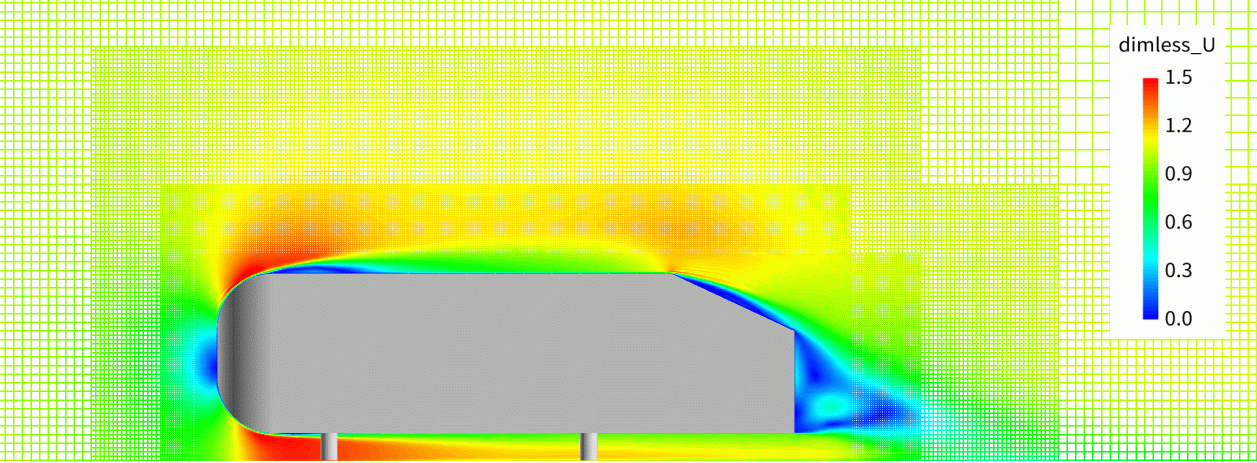}
   \subcaption{Time-averaged velocity magnitude on the center section, mapped onto the calculation cells of the fine grid case.}
  \label{fig:Ahmed_velmag}
  \vspace{2mm}
 \end{minipage}
 \begin{minipage}{0.8\hsize}
   \includegraphics[keepaspectratio,width=\textwidth]{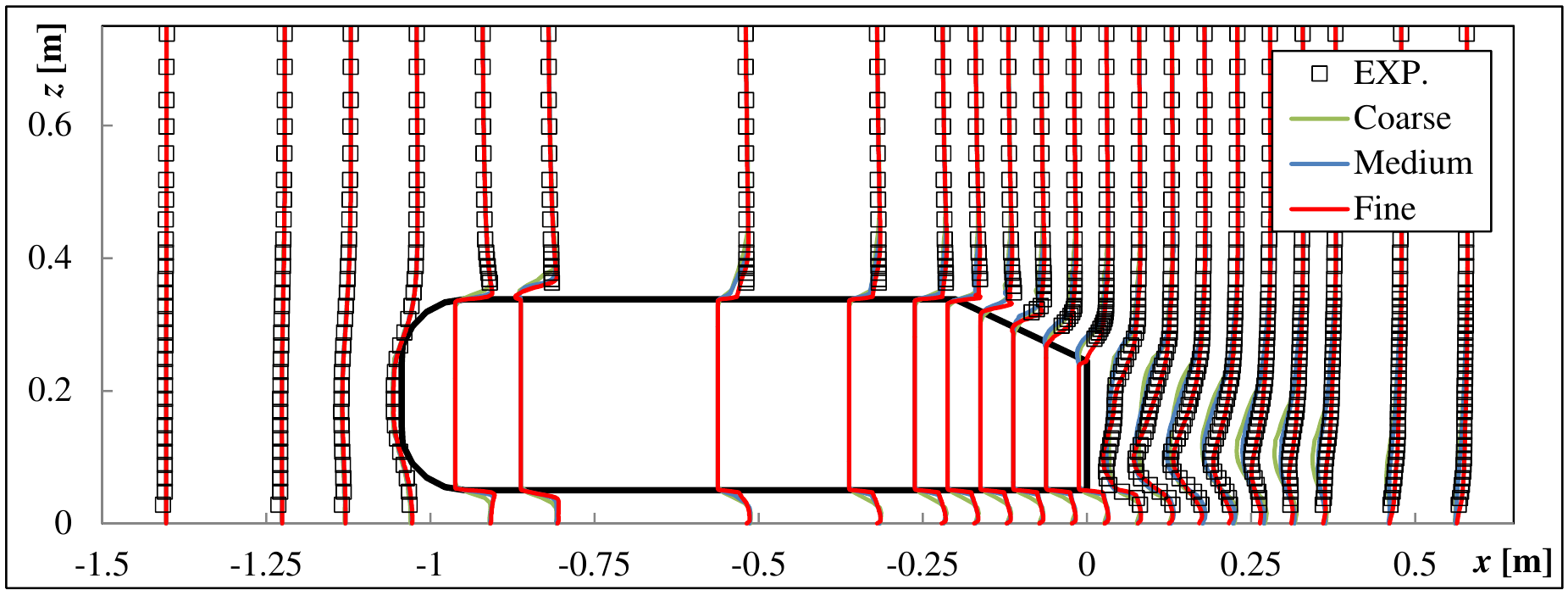}
   \subcaption{Profile of the streamwise component of the time-averaged velocity at three different resolutions.}
  \label{fig:Ahmed_whole_ave}
  \vspace{2mm}
 \end{minipage}
 \begin{minipage}{0.45\hsize}
   \includegraphics[keepaspectratio,width=\textwidth]{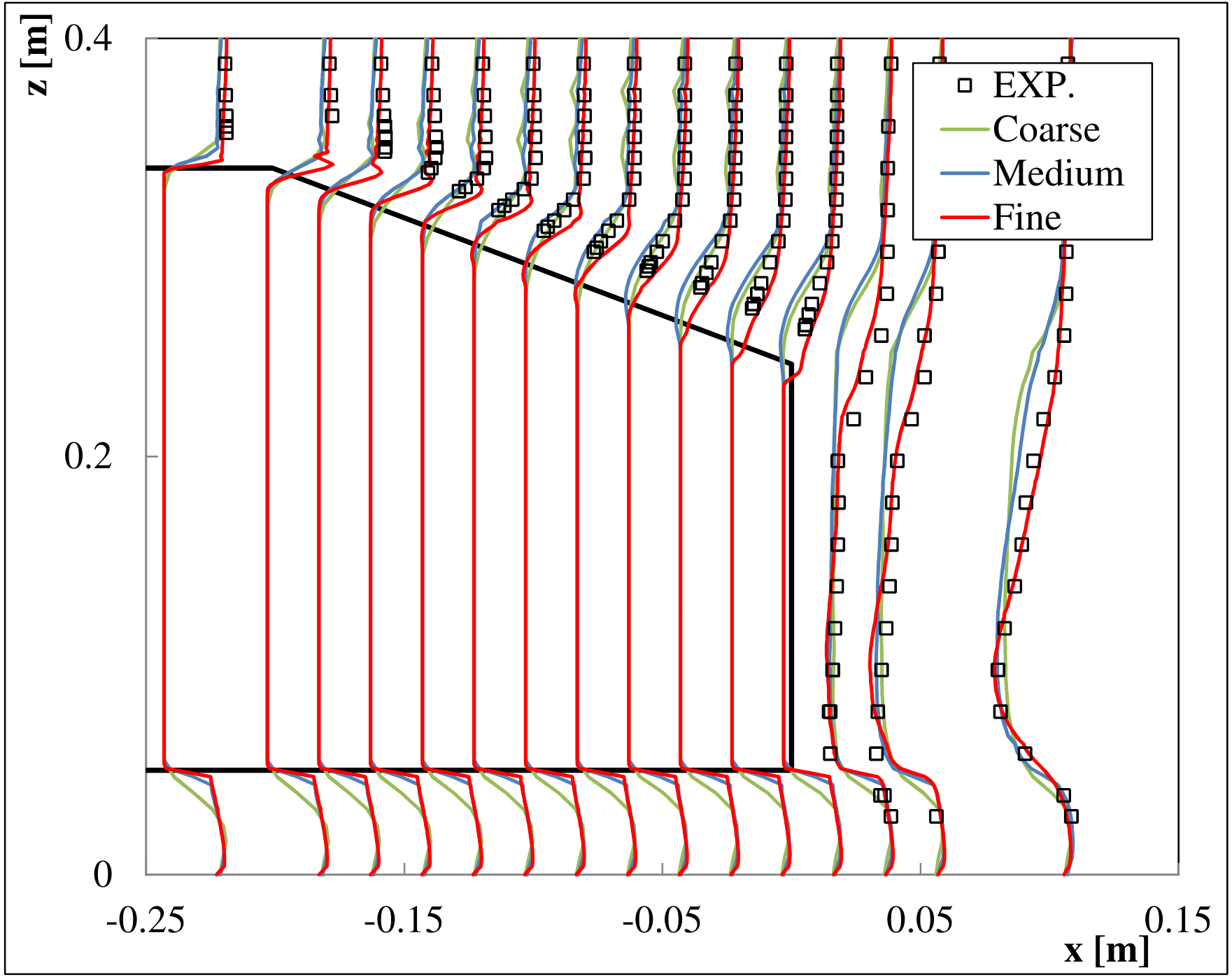}
   \subcaption{Time-averaged velocity profile of the streamwise component around the rear slant.}
  \label{fig:Ahmed_xz_ave}
 \end{minipage}
 \begin{minipage}{0.45\hsize}
   \includegraphics[keepaspectratio,width=\textwidth]{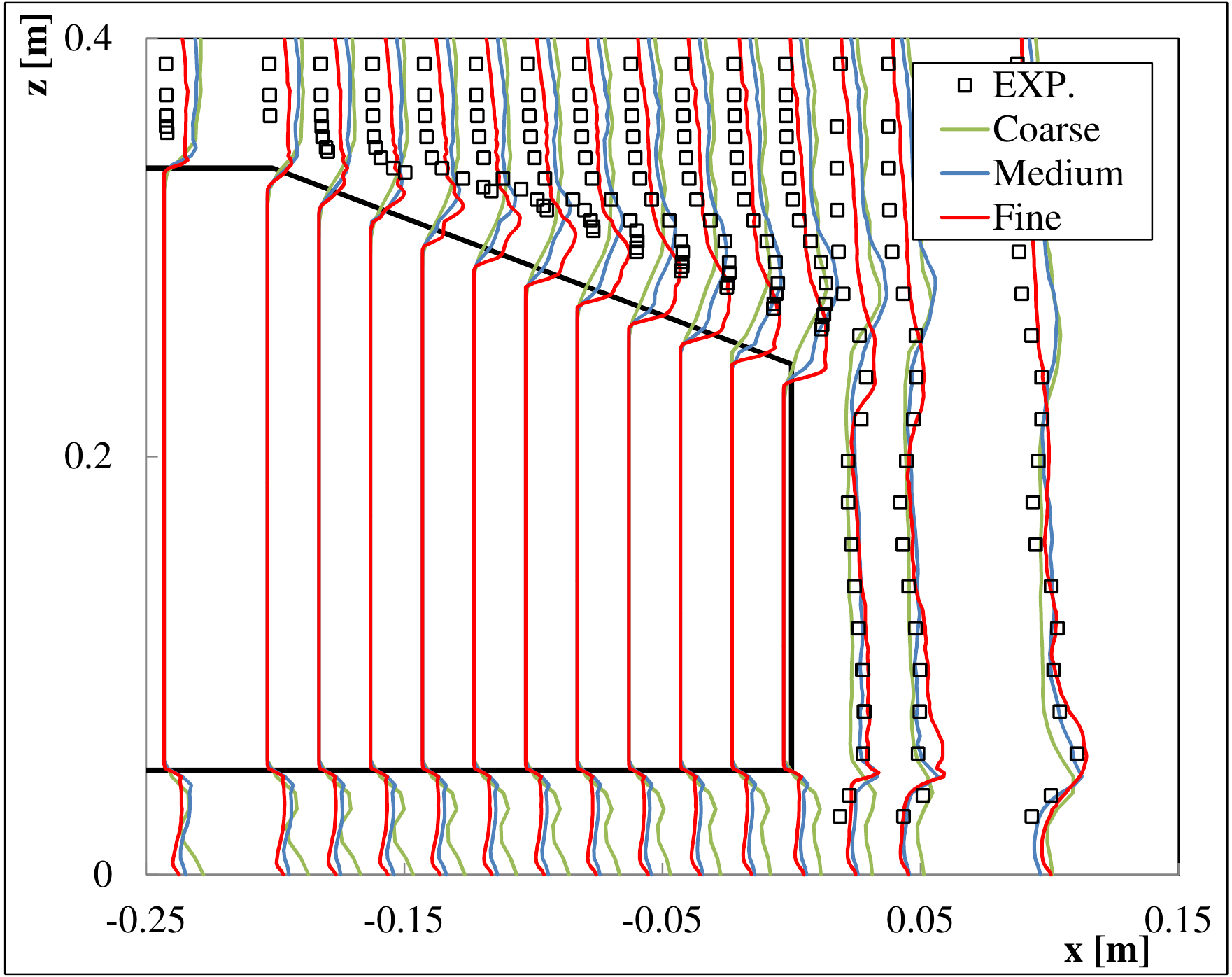}
   \subcaption{Root-mean-square velocity profile of the streamwise component around the rear slant.}
  \label{fig:Ahmed_xz_rms}
 \end{minipage}
 \caption{Streamwise velocity field prediction results of the Ahmed model. Black square symbols indicate experimental results. Green, blue, and red lines indicate simulation results with coarse, medium, and fine grids, respectively.}
 \label{fig:Ahmed_body}
\end{figure}

The solutions of three different grid resolutions were compared. Figure \ref{fig:Ahmed_body}\subref{fig:Ahmed_velmag} shows a contour plot of the time-averaged velocity magnitude on the center section, mapped onto the calculation cells. The smallest grid resolution was $1.95 mm$ for the finest grid, it was twice this value for the middle grid, and four times for the coarsest grid. The number of cubes were 676, 1,936 and 8,292, respectively; each cube was divided into $16 \times 16 \times 16$ cells. Figure \ref{fig:Ahmed_body}\subref{fig:Ahmed_whole_ave} shows the velocity profiles of the streamwise component of the time-averaged velocity at three resolutions. The results of all three cases were in reasonable agreement with the experimental results. A detailed profile around the rear slant is shown in Fig. \ref{fig:Ahmed_body}\subref{fig:Ahmed_xz_ave} and \subref{fig:Ahmed_xz_rms}. The time-averaged velocity component in the coarse and medium cases showed deviation from the experimental value in the upper region of the body wake. In contrast, the fine case showed good agreement up to the downstream of the body wake. The root-mean-square velocity component showed reasonable improvement in the fine case, though a deviation from the experimental value was seen in the upper region of the body. Consequently, we concluded that the basic characteristics of turbulent flow around a bluff body, including fluctuating components, were captured by appropriately selecting the resolution of the computational grid.

\subsection{Flow past a flat plate}

\begin{figure}[htbp]
 \centering
 \begin{minipage}{0.4\hsize}
   \includegraphics[keepaspectratio,width=\textwidth]{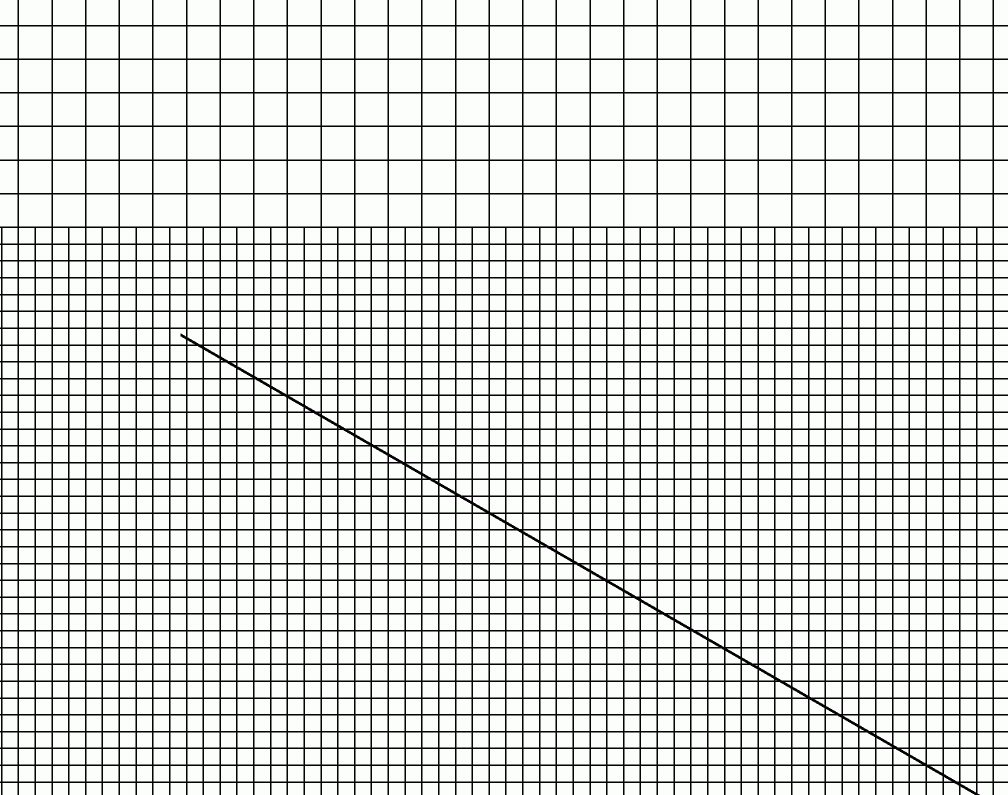}
  \subcaption{Magnified view of the computational grid in the region of the top of the plate.}
  \label{fig:flat_plate_grid}
 \end{minipage}
 \hspace{0.4cm}
 \begin{minipage}{0.4\hsize}
   \includegraphics[keepaspectratio,width=\textwidth]{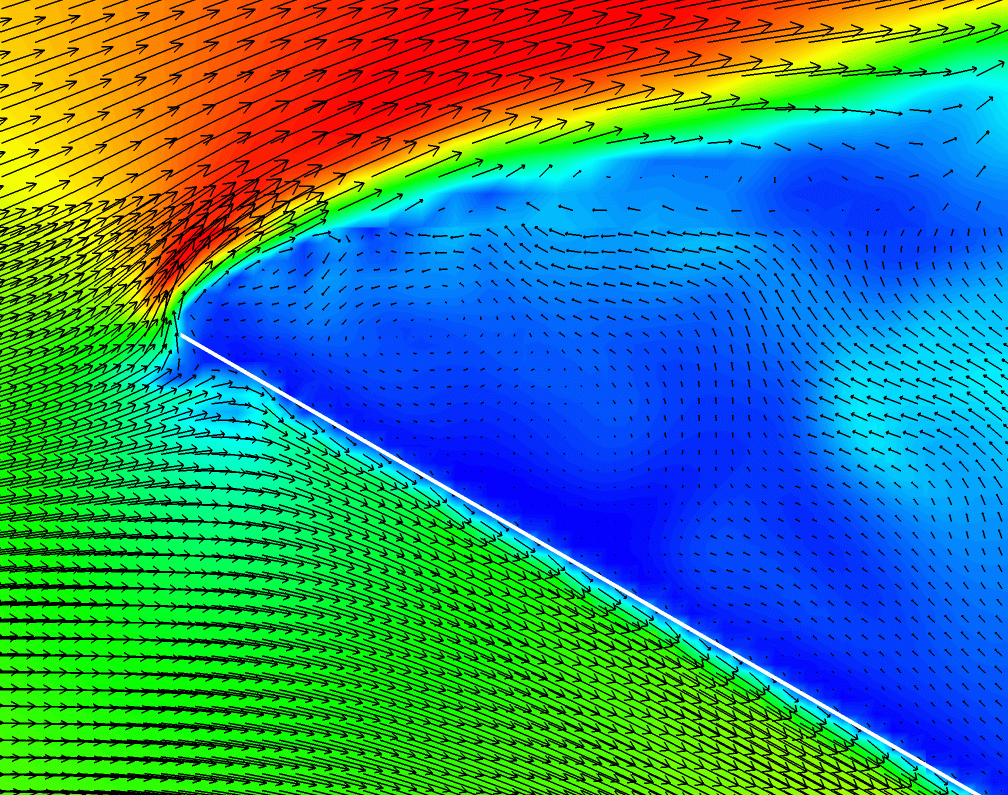}
  \subcaption{Magnified view of the velocity field in the region of the top of the plate.}
  \label{fig:flat_plate_velvec}
 \end{minipage}
 \begin{minipage}{0.6\hsize}
   \includegraphics[keepaspectratio,width=\textwidth]{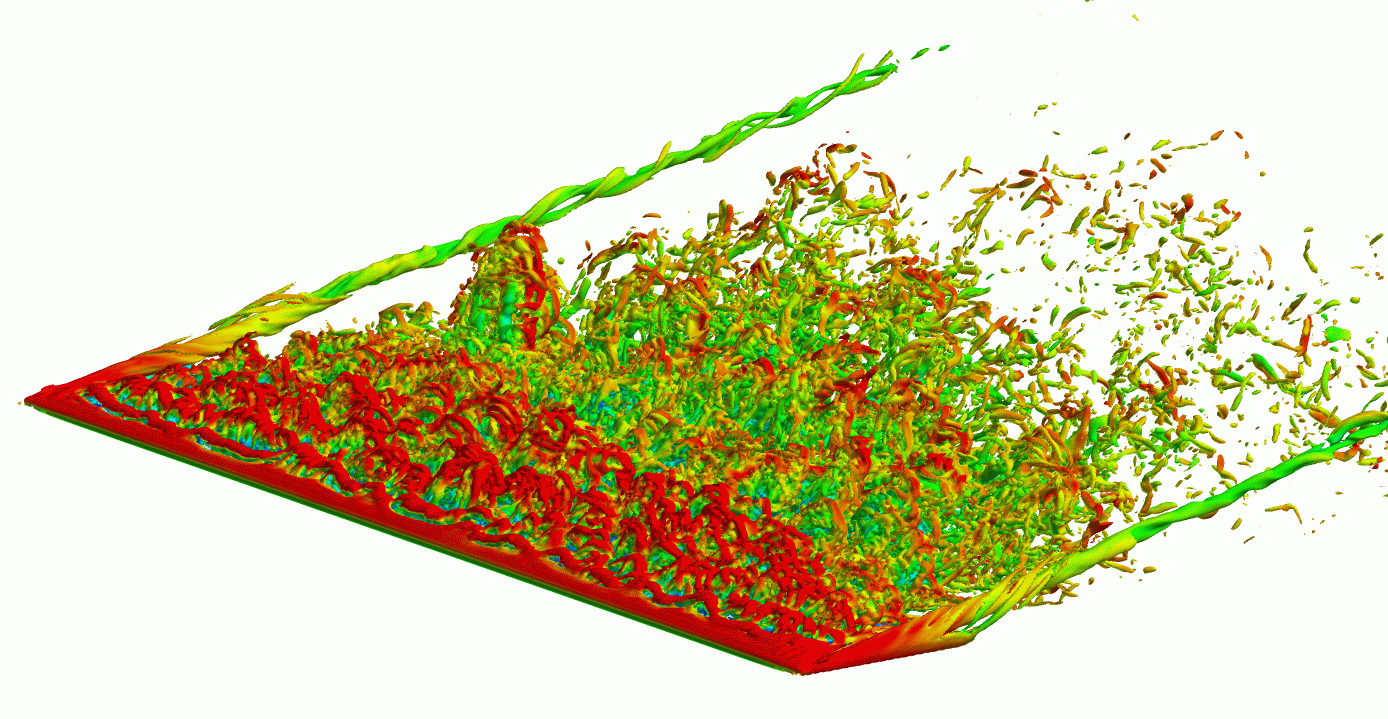}
   \subcaption{Vortex structure on the case with $\alpha = 12$, showing an iso-surface of the Q-criterion colored by the non-dimensional velocity magnitude.}
   \label{fig:flat_plate_Q}
 \end{minipage}
 \caption{Overview of the computational grid and the velocity field results of flow past a thin flat plate.}
 \label{fig:flat_plate_overview}
\end{figure}

The flow past a flat plate having zero thickness was simulated. This case does not apply to Mittal's type of IBM. The calculation was performed in three dimensions for the flat plate, with an aspect ratio of 6. The flat plate had a smooth surface with a thickness of zero. The lift coefficient and drag coefficient for a given angle of attack were compared with the experimental results obtained by Okamoto et al. \cite{Okamoto1996,Okamoto2011}. Okamoto used a rectangular specimen having a thickness-chord length ratio of less than 1.7\%. The minimum cell width based on the chord length $C$ was set to $\Delta x = 9.77 \times 10^{-3} \times C$ as shown in Fig. \ref{fig:flat_plate_overview}\subref{fig:flat_plate_grid}, and the vicinity of the wall surface was subdivided according to the angle of attack. The number of cubes was 12,856 to 43,600, while the number of cells was 52,658,176 to 178,585,600. For the calculation conditions, $Re$ was set to $1.14 \times 10^{4}$ according to the experiment of Okamoto. The time increment was set to $\Delta t = 2.0 \times 10^{-3}$ for non-dimensional time, and the analysis termination time was set to 100.0 in non-dimensional time. The computational domain had dimensions of $40C \times 20C \times 40C$, and the slip boundary conditions were applied to the sidewall of the computational domain. The spatial discretization used the second-order central difference, and the time integration used the Crank-Nicolson implicit scheme.

An example of the flow field result is shown in Fig. \ref{fig:flat_plate_overview}\subref{fig:flat_plate_velvec}. The color contours indicate the non-dimensional velocity magnitude and are overlaid by velocity vectors. This shows that the flows on the pressure side and suction side were successfully separated. The turn of flow around the corner was well predicted, where the computation became unstable in the conventional CFD method with an unstructured grid or body-fitted structured grid. The geometric data consisted of shell elements without thickness information. The geometry appears to have a thickness in Fig. \ref{fig:flat_plate_overview}\subref{fig:flat_plate_velvec}, however, this is for the purpose of visualization only. Figure \ref{fig:flat_plate_overview}\subref{fig:flat_plate_Q} is shown for reference, to show that the flow was three-dimensional. The visualization of the vortex structure shows that a strong trailing vortex appeared at the plate tip.

\begin{figure}[htbp]
 \centering
 \begin{minipage}{0.4\hsize}
   \includegraphics[keepaspectratio,width=\textwidth]{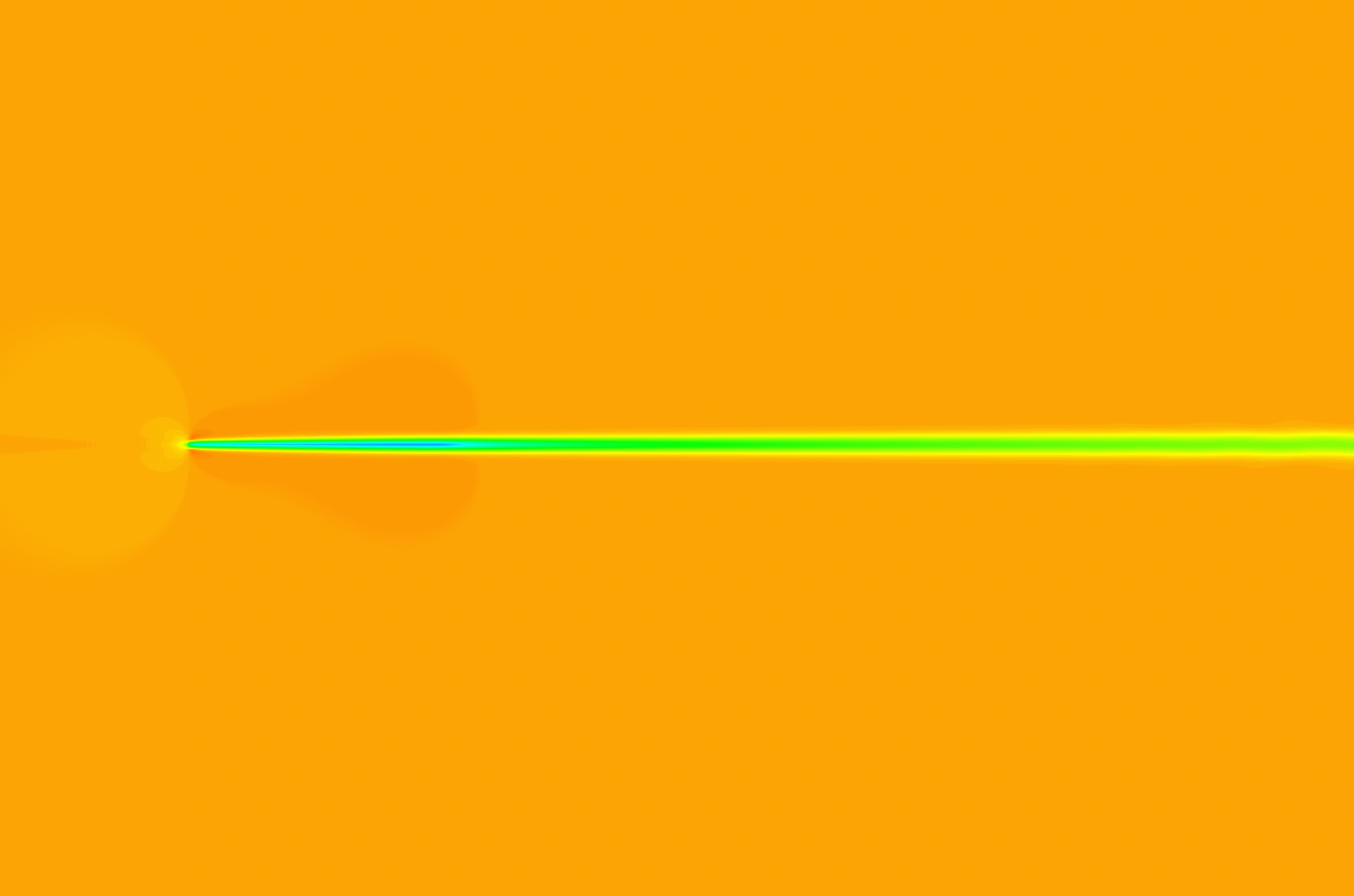}
   \subcaption{Result of $\alpha = 0 [deg.]$.}
  \label{fig:flat_plate_a0}
 \end{minipage}
 \begin{minipage}{0.4\hsize}
   \includegraphics[keepaspectratio,width=\textwidth]{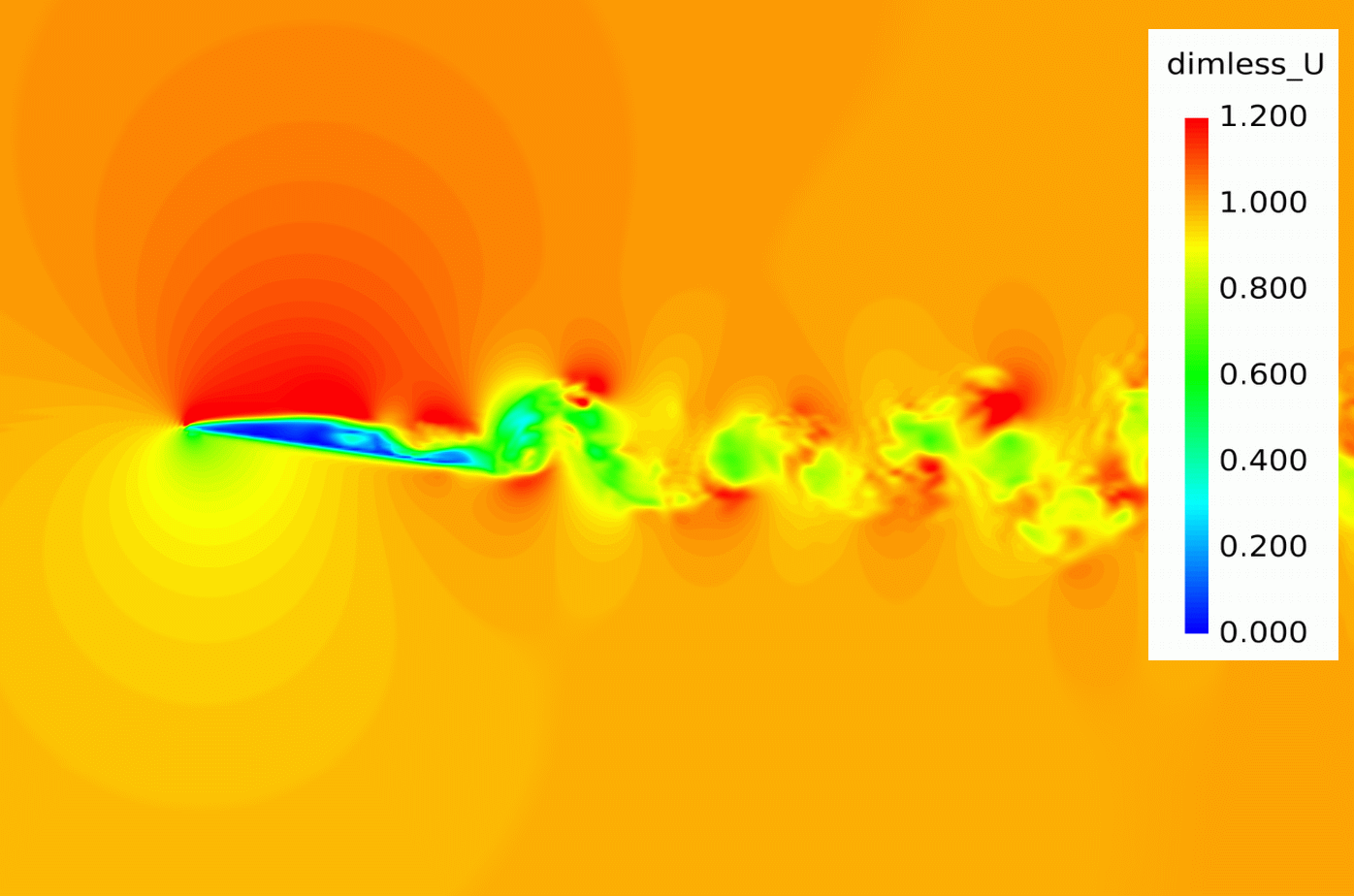}
   \subcaption{Result for $\alpha = 8 [deg.]$.}
  \label{fig:flat_plate_a8}
 \end{minipage}
 \begin{minipage}{0.4\hsize}
   \includegraphics[keepaspectratio,width=\textwidth]{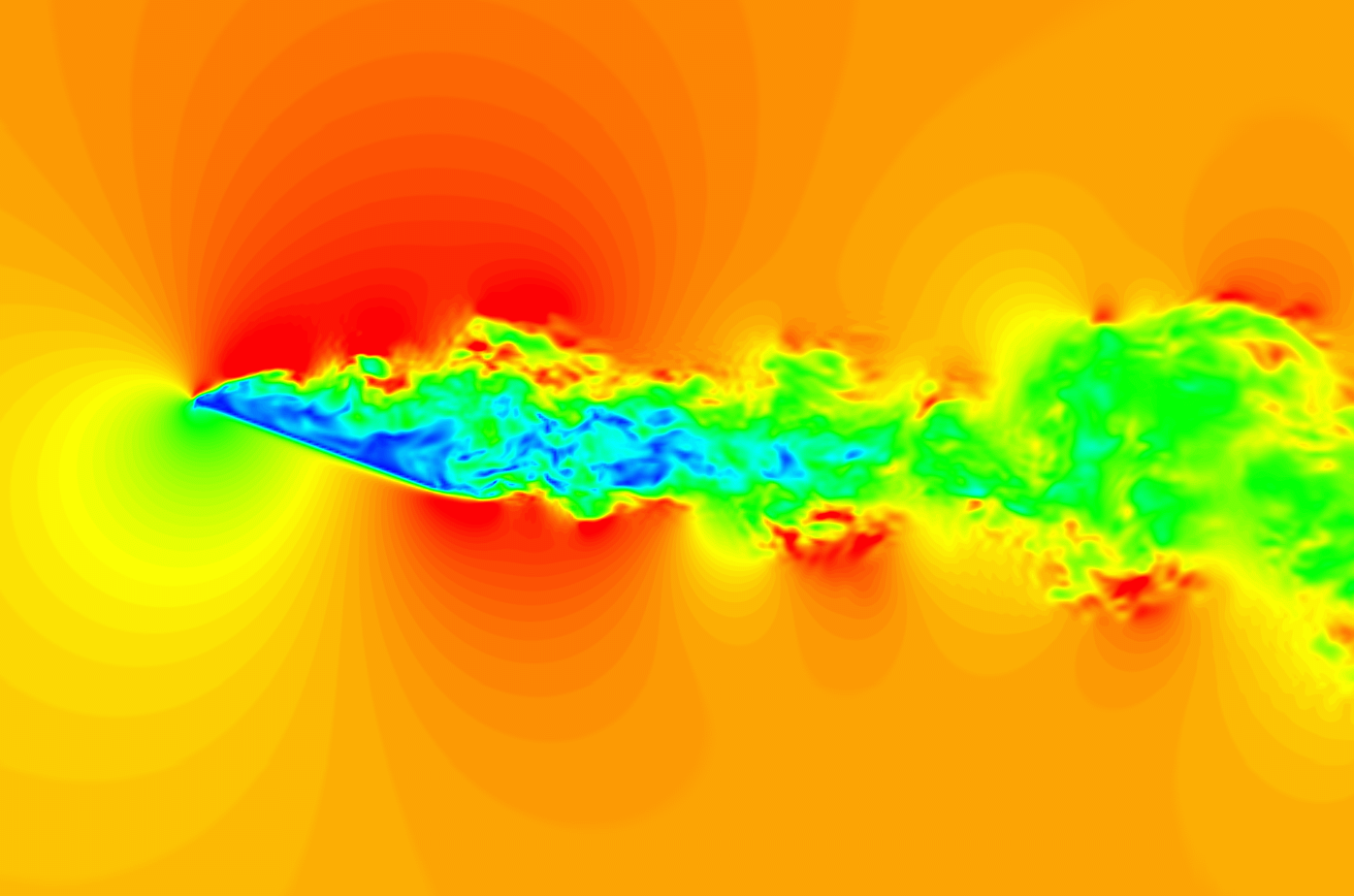}
   \subcaption{Result for $\alpha = 20 [deg.]$.}
  \label{fig:flat_plate_a20}
 \end{minipage}
 \begin{minipage}{0.4\hsize}
   \includegraphics[keepaspectratio,width=\textwidth]{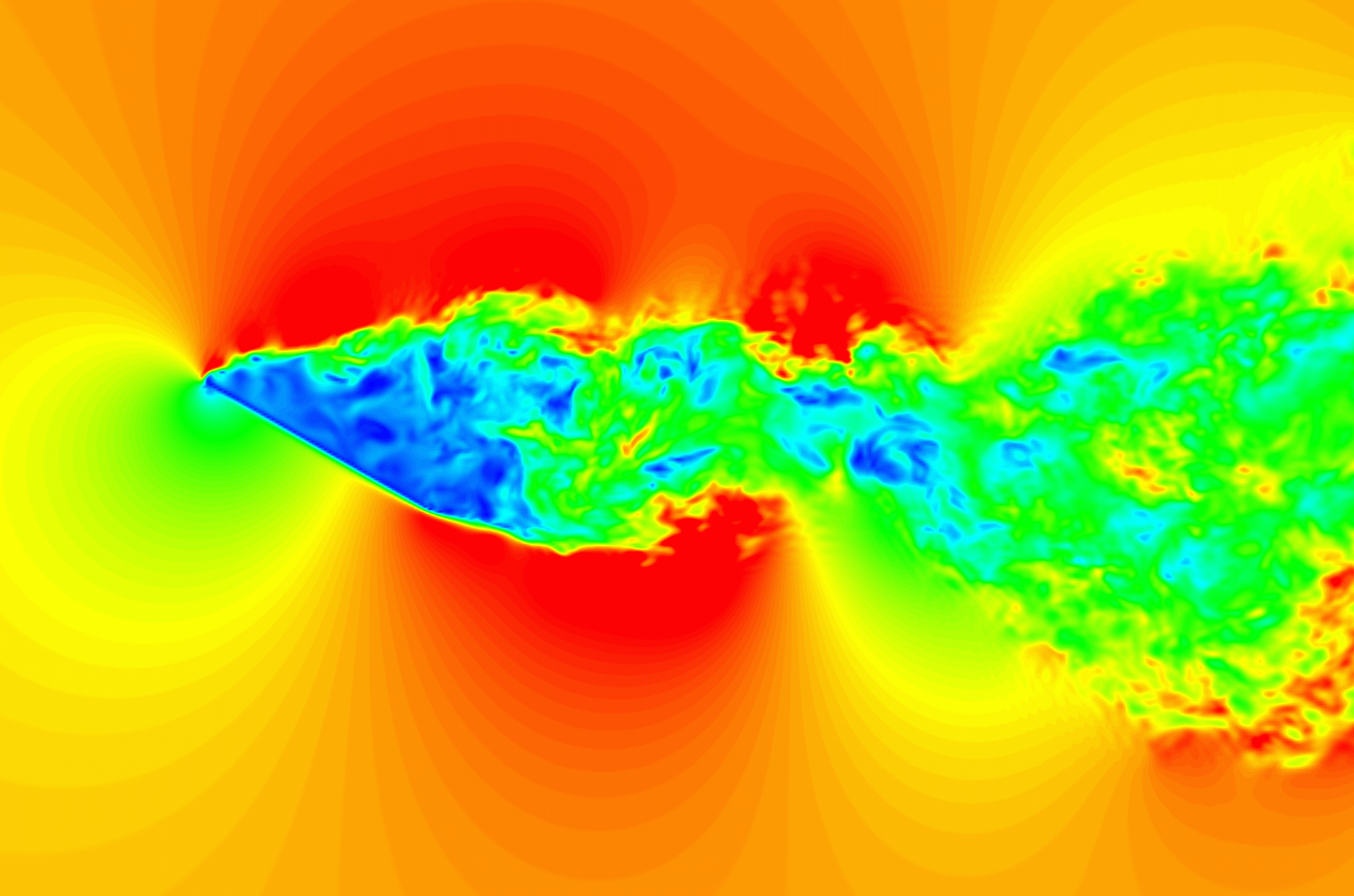}
   \subcaption{Result for $\alpha = 30 [deg.]$.}
  \label{fig:flat_plate_a30}
 \end{minipage}
 \begin{minipage}{0.4\hsize}
   \includegraphics[keepaspectratio,width=\textwidth]{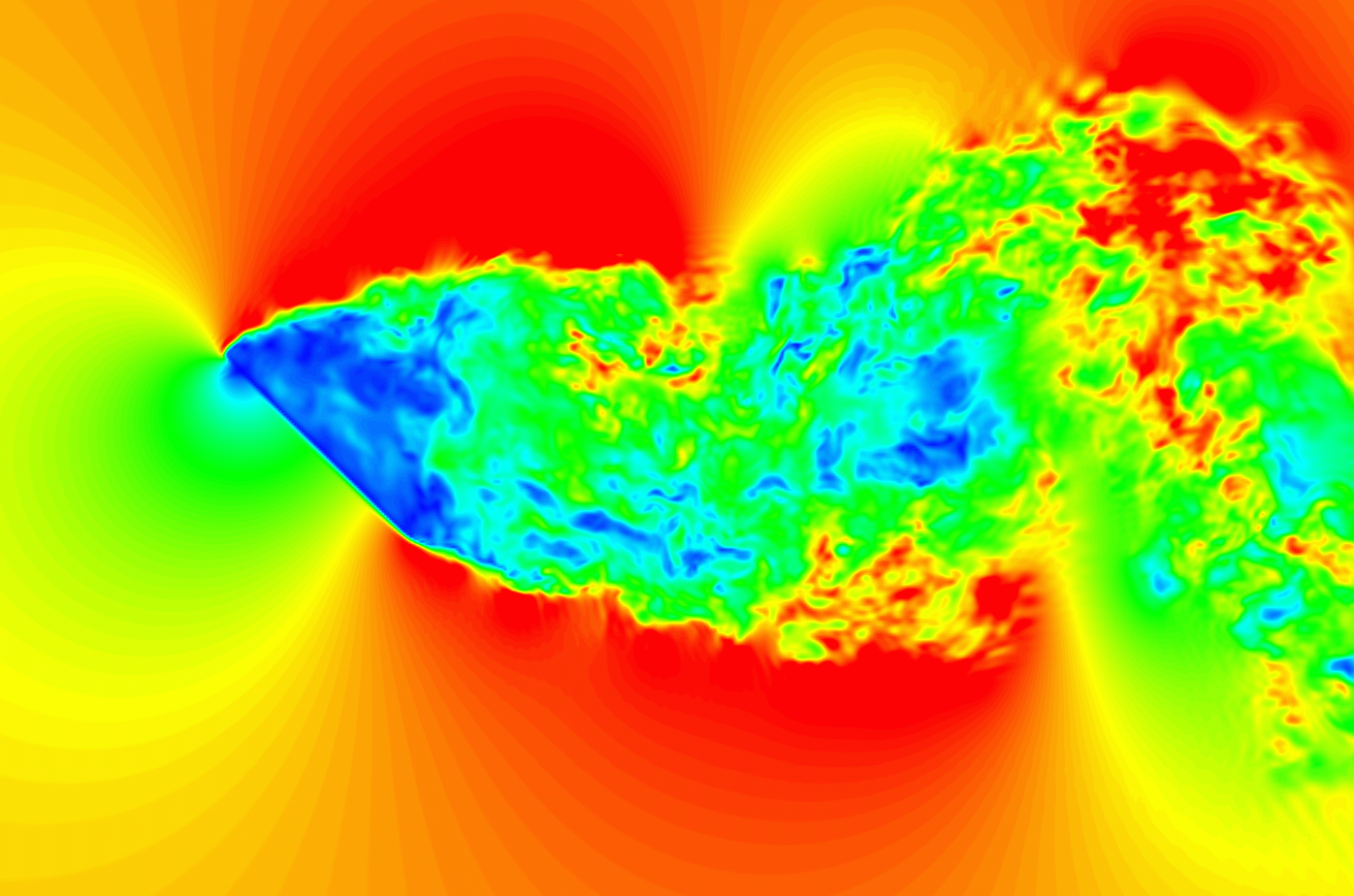}
   \subcaption{Result for $\alpha = 45 [deg.]$.}
  \label{fig:flat_plate_a45}
 \end{minipage}
 \begin{minipage}{0.4\hsize}
   \includegraphics[keepaspectratio,width=\textwidth]{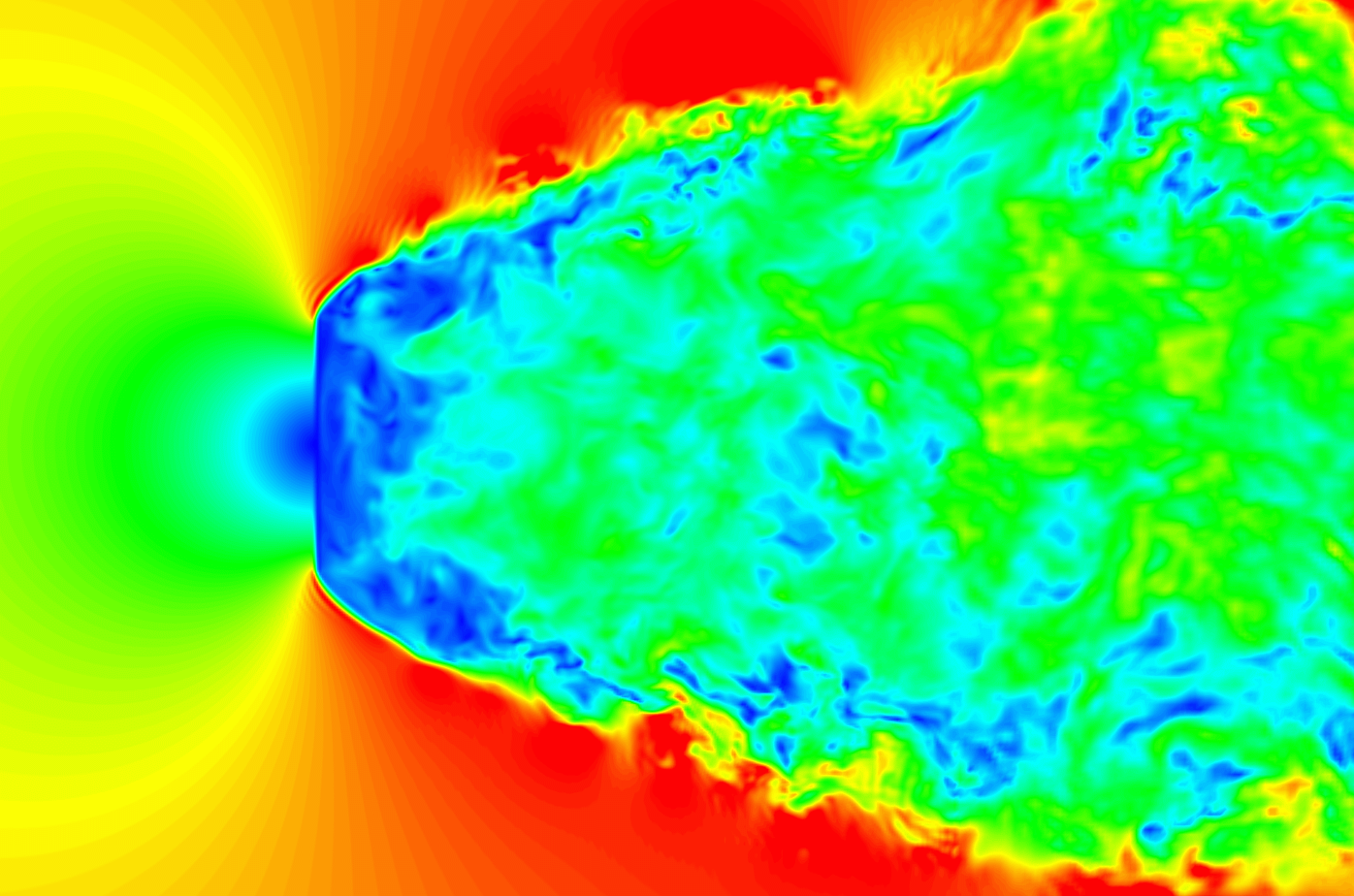}
   \subcaption{Result for $\alpha = 90 [deg.]$.}
  \label{fig:flat_plate_a90}
 \end{minipage}
 \caption{Non-dimensional velocity magnitude of the center section of flow passing a flat plate, with $\alpha$ representing the angle of attack.}
 \label{fig:flat_plate_velmag}
\end{figure}

Figure \ref{fig:flat_plate_velmag} shows the contour plots of the instantaneous non-dimensional velocity magnitude at the termination time for each angle of attack $\alpha$. Flow separation at the leading-edge markedly appeared when $\alpha$ exceeded approximately 8 degrees, and it was observed that the separated shear layer flowed downward, while forming a vortex sheet. Especially near $\alpha = 30$ degrees, it was observed that the separation from the trailing edge released a large vortex. This is similar to the flow structure observed experimentally by Okamoto et al. \cite{Okamoto2011}.

\begin{figure}[htbp]
 \centering
 \begin{minipage}{0.4\hsize}
   \includegraphics[keepaspectratio,width=\textwidth]{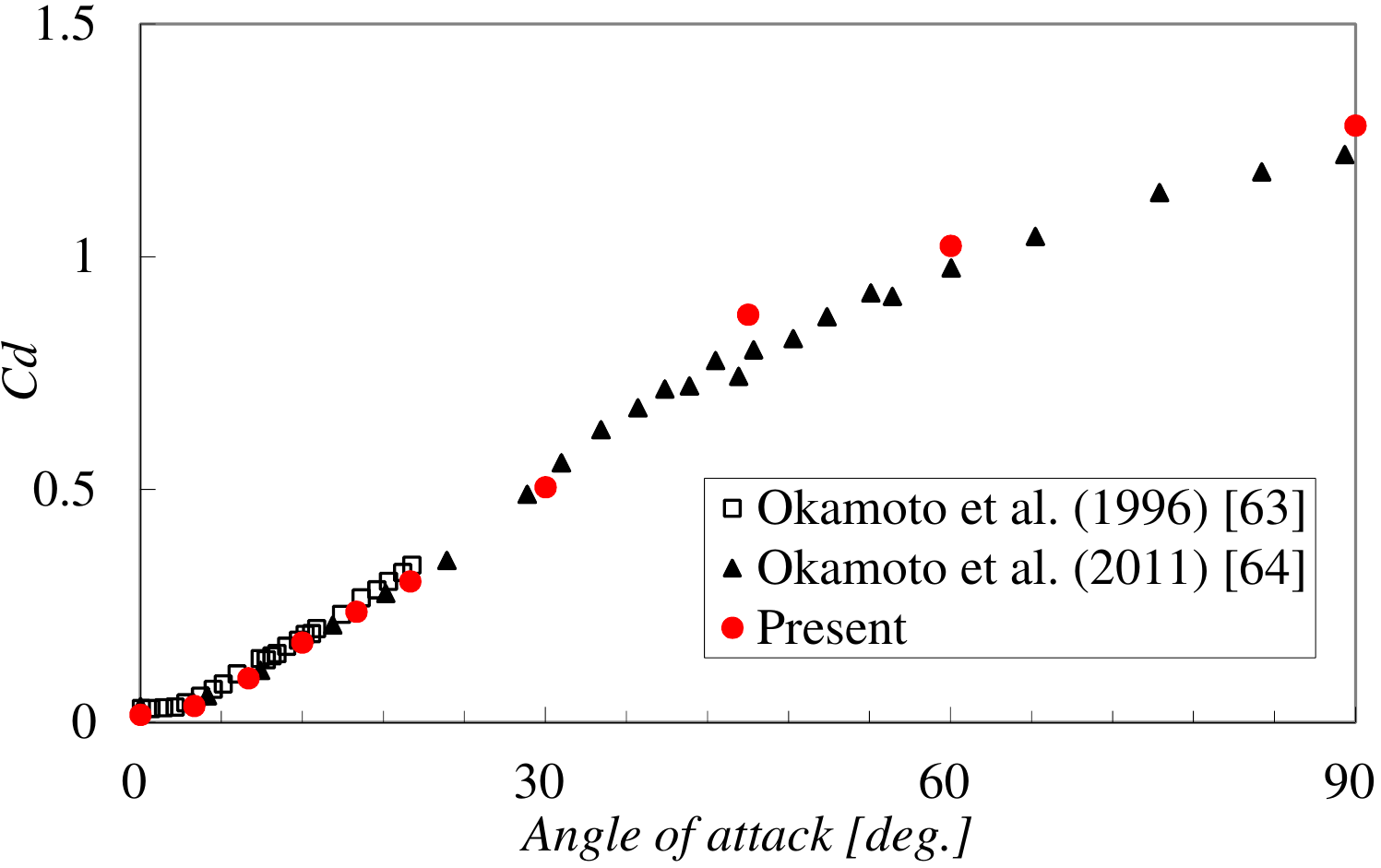}
  \subcaption{Drag coefficient.}
  \label{fig:flat_plate_cd}
 \end{minipage}
 \begin{minipage}{0.4\hsize}
   \includegraphics[keepaspectratio,width=\textwidth]{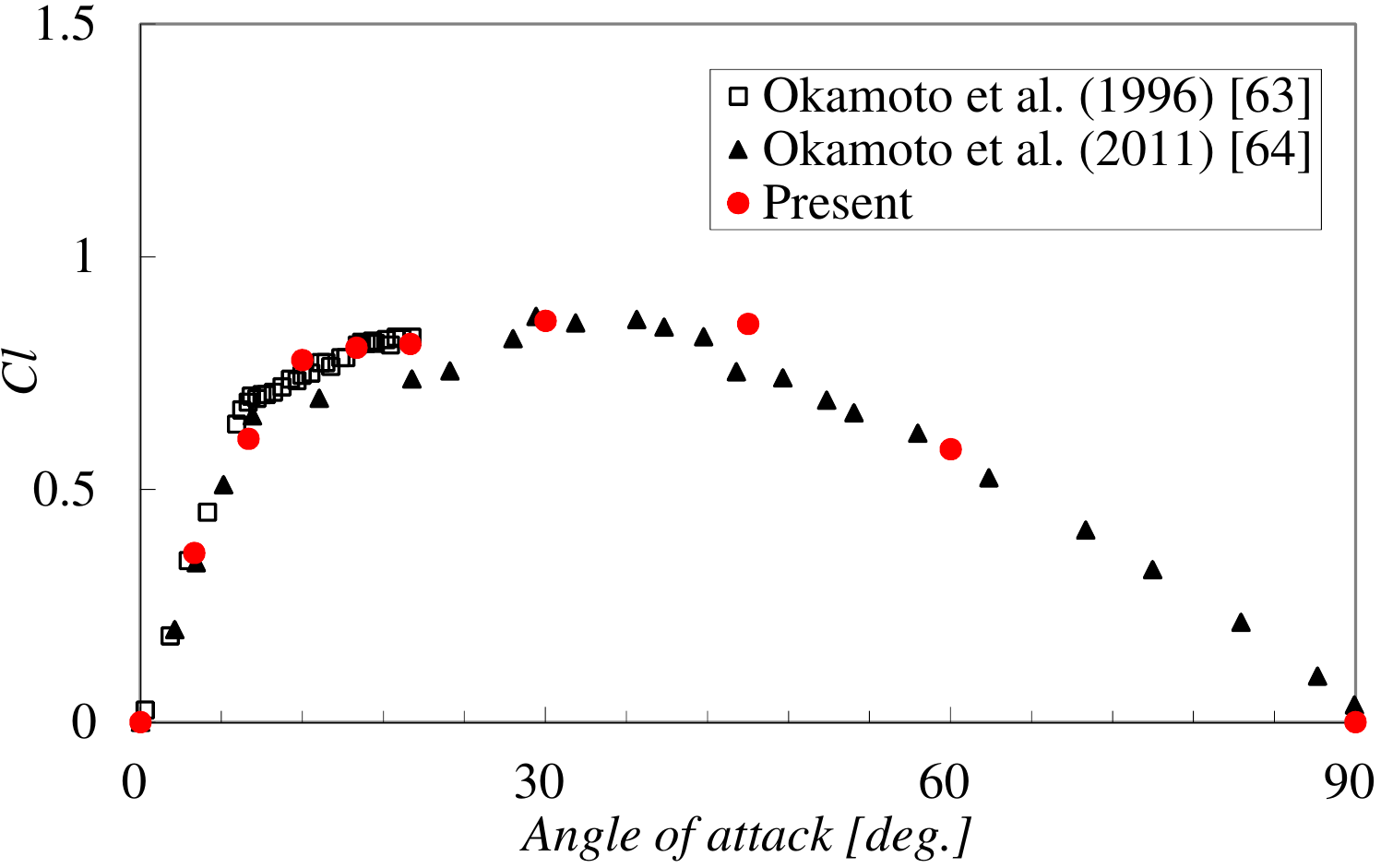}
  \subcaption{Lift coefficient.}
  \label{fig:flat_plate_cl}
 \end{minipage}
 \caption{Drag and lift coefficient profiles for various angles of attack of flow passing a flat plate.}
 \label{fig:flat_plate_coeff}
\end{figure}

Figure \ref{fig:flat_plate_coeff} summarizes the predicted lift coefficient ($Cl$) and drag coefficient ($Cd$) for each $\alpha$. $Cl$ and $Cd$ were defined using the main flow direction velocity $U_{0}$, the streamwise force integral $f_x$, the normal direction force integral $f_y$, and the density $\rho$, as:

\begin{align}
  \label{eq:cd_flat_plate}
 & Cl = f_y / (\frac{1}{2} \rho U_{0}^{2} 6 C^{2}), \qquad Cd = f_x / (\frac{1}{2} \rho U_{0}^{2} 6 C^{2}). &
\end{align}

The force acting on the surface was integrated (near-field method) by employing Shepard's method, which takes an inverse-distance-weighted average for vertices on a surface:

\begin{align}
  \label{eq:force}
 & f(\bm{X}) = \frac{\sum_{k=1}^{m} \mathcal{S} (\bm{X}) f_{k}}{\sum_{k=1}^{m} \mathcal{S} (\bm{X})}, 
 \qquad {\rm where} \; \mathcal{S}(\bm{X}) = \frac{\delta (\bm{n} \cdot (\bm{X} - \bm{x}^{(k)}))}{\mid\bm{X}-\bm{x}^{(k)}\mid^{2}}. &
\end{align}

Here, the weighting function $\mathcal{S} (\bm{X})$ is defined according to the distance between the surface vertex $\bm{X}$ and the surrounding fluid grid point $\bm{x}^{(k)}$. The $m$ is the number of surrounding fluid points. Additionally, $\delta$ is the Dirac delta function relating to the dot product of the surface normal $\bm{n}$ and the direction vector. It is introduced to calculate $f(\bm{X})$ separately for the front and back faces of the surface.

The experimental results reference the two configurations of Okamoto et al. obtained in 1996 \cite{Okamoto1996} and 2011 \cite{Okamoto2011}. The thickness of the plate was $0.3 mm$ (i.e., the thickness-chord length ratio was 1.00\%) for the 1996 configuration, and $0.5 mm$ (i.e., the thickness-chord length ratio was 1.67\%) for the 2011 configuration. In the 1996 experiment, sharpening the front edge by tapering it affected the behavior of the leading-edge separation, starting at an $\alpha$ of approximately 8 degrees. Hence, a migration of the profiles of $Cl$ and $Cd$ also occurred. This shows the strong effect of the plate thickness. In the 2011 experiment, the leading-edge was not sharpened. At a low $\alpha$, the leading-edge separation was dominant. The experimental results differed near 12 and 20 degrees, as it was presumed that the thickness of the plate was affecting the trailing edge separation as well as the leading-edge separation. Figure \ref{fig:flat_plate_coeff} shows that the results for $\alpha$ up to 12 degrees matched well the experimental results of 1996. As this calculation had zero thickness, it is reasonable to approach the result for a low thickness-chord length ratio. The behavior gradually shifted from the 1996 result to the 2011 result after 20 degrees, while it followed the 2011 result afterward. The detailed mechanism is not yet clear; however, we consider that the present calculation captured the tendencies of both experiments well, and quantitatively reproduced the state of the flow field around a thin plate.

\subsection{Flow past a dirty sphere}

\begin{figure}[htbp]
  \centering
  \begin{minipage}{0.8\hsize}
    \centering
    \begin{minipage}{0.3\hsize}
      \includegraphics[keepaspectratio,width=\textwidth]{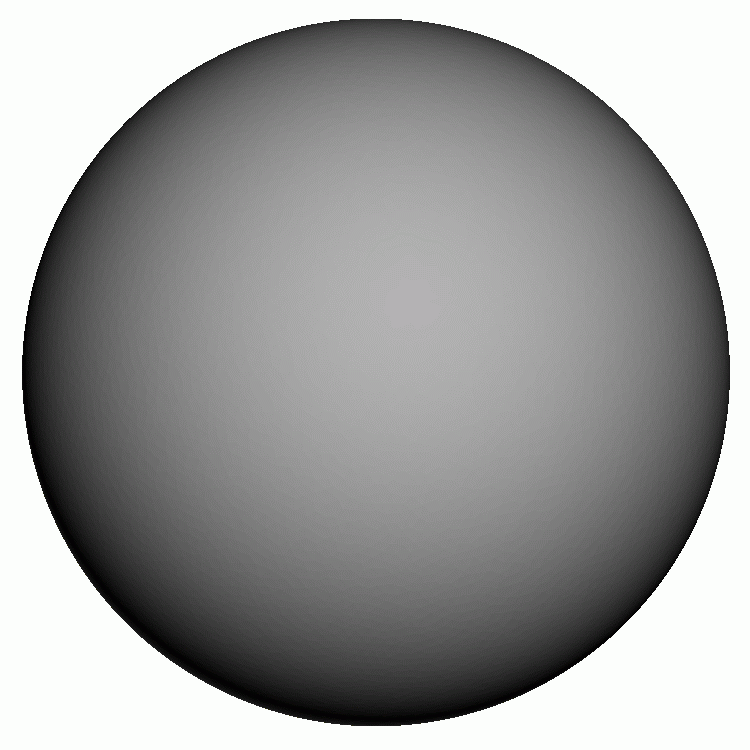}
      \subcaption{}
      \label{fig:sphere_clean}
    \end{minipage}
    \begin{minipage}{0.3\hsize}
      \includegraphics[keepaspectratio,width=\textwidth]{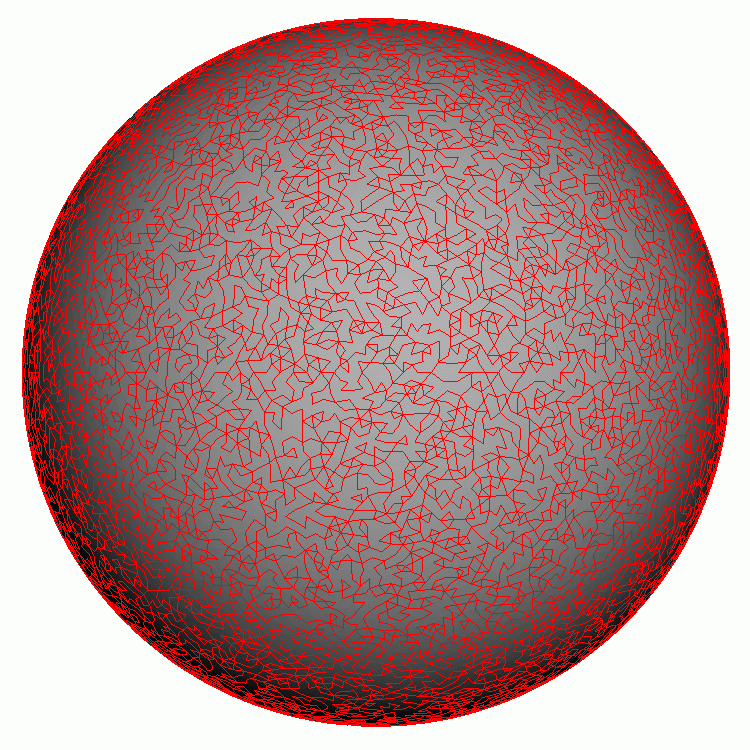}
      \subcaption{}
      \label{fig:sphere_dirty}
    \end{minipage}
  \end{minipage}
  \begin{minipage}{0.8\hsize}
    \centering
    \begin{minipage}{0.25\hsize}
      \includegraphics[keepaspectratio,width=\textwidth]{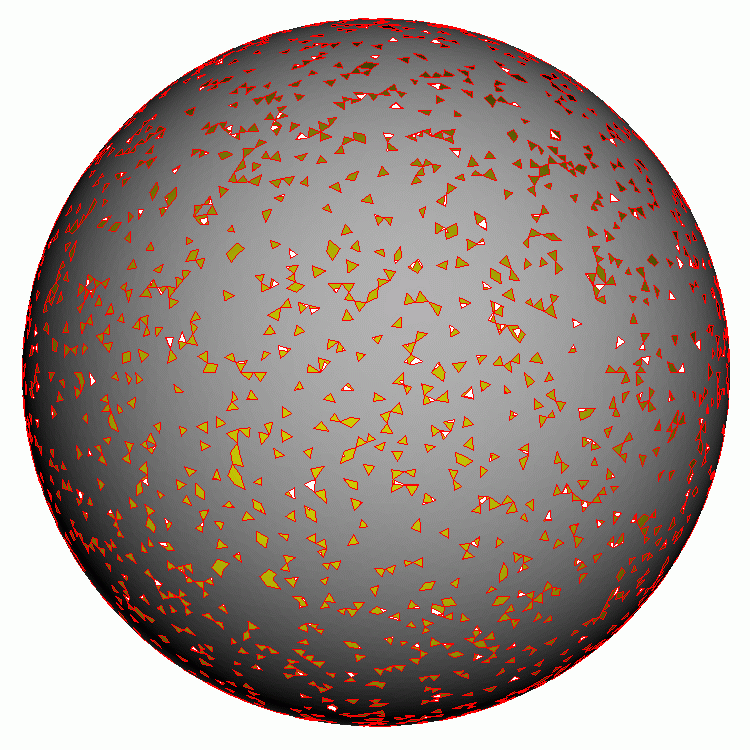}
      \subcaption{}
      \label{fig:sphere_dirty10}
    \end{minipage}
    \begin{minipage}{0.25\hsize}
      \includegraphics[keepaspectratio,width=\textwidth]{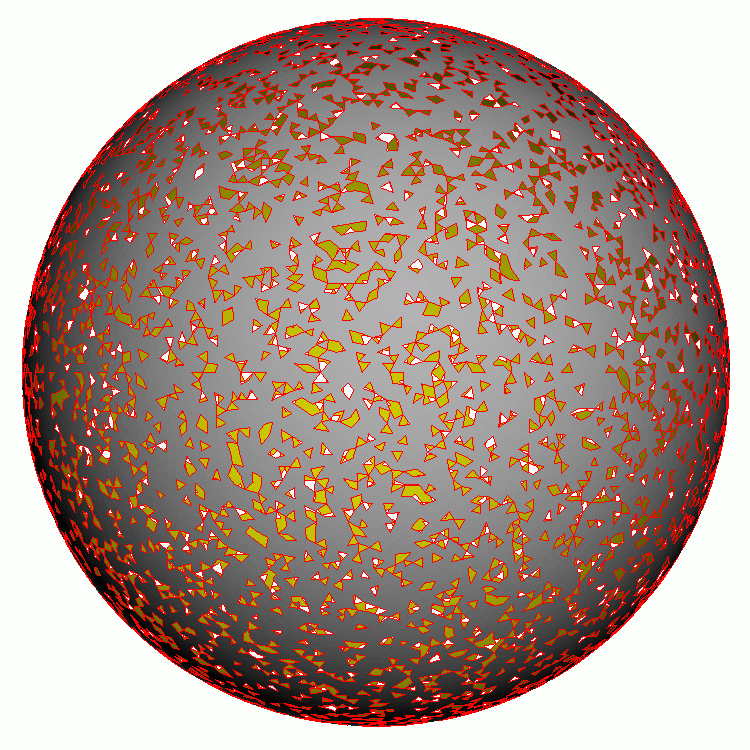}
      \subcaption{}
      \label{fig:sphere_dirty20}
    \end{minipage}
    \begin{minipage}{0.25\hsize}
      \includegraphics[keepaspectratio,width=\textwidth]{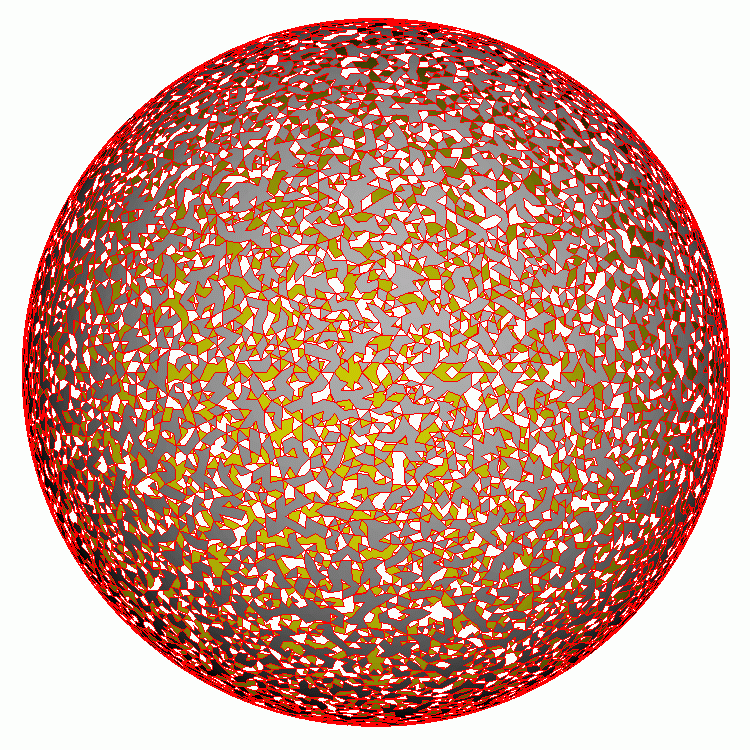}
      \subcaption{}
      \label{fig:sphere_dirty50}
    \end{minipage}
  \end{minipage}
  \begin{minipage}{0.8\hsize}
    \centering
    \begin{minipage}{0.25\hsize}
      \includegraphics[keepaspectratio,width=\textwidth]{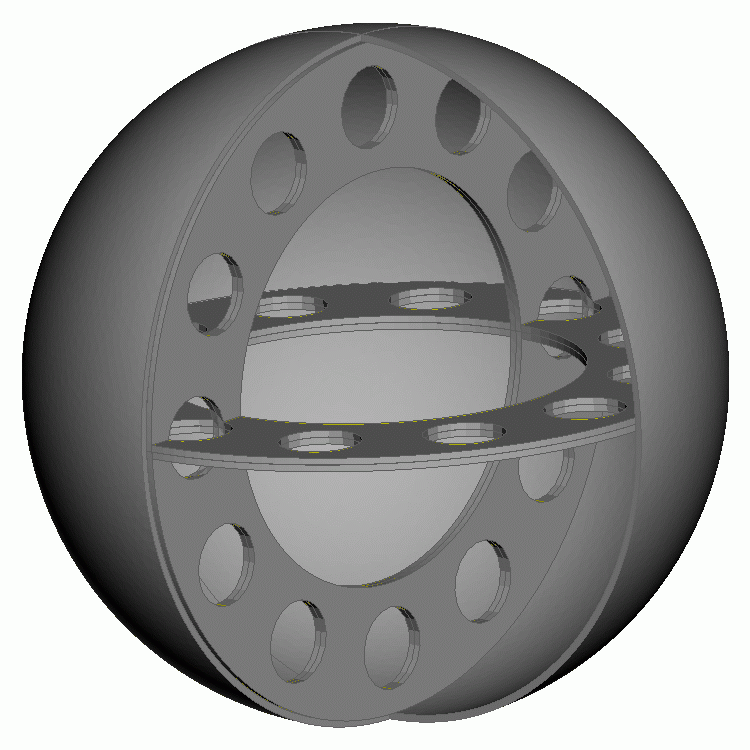}
      \subcaption{}
      \label{fig:sphere_frame}
    \end{minipage}
    \begin{minipage}{0.25\hsize}
      \includegraphics[keepaspectratio,width=\textwidth]{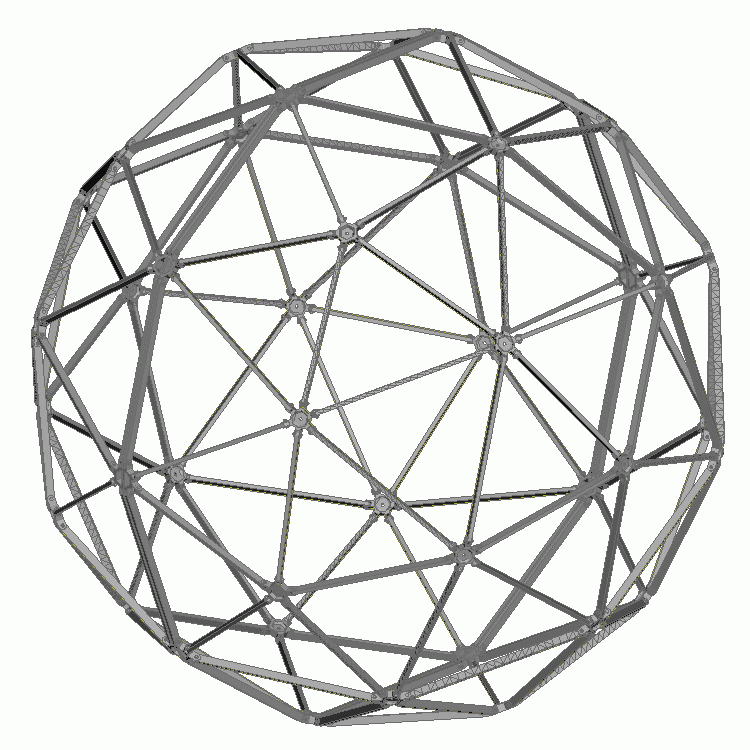}
      \subcaption{}
      \label{fig:geodom}
    \end{minipage}
    \begin{minipage}{0.25\hsize}
      \includegraphics[keepaspectratio,width=\textwidth]{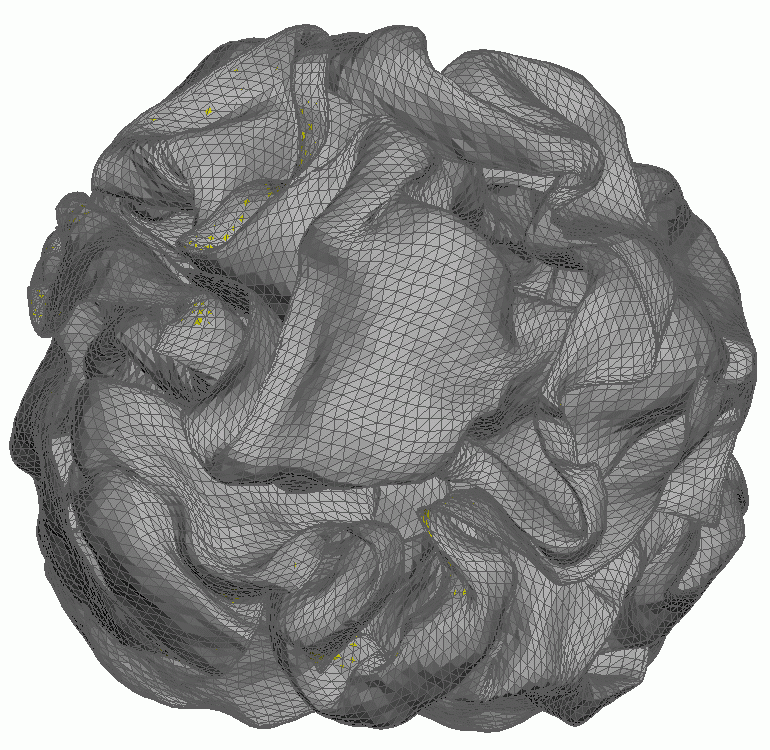}
      \subcaption{}
      \label{fig:crumpled_paper}
    \end{minipage}
  \end{minipage}
  \caption{Snapshot of the dirty sphere geometries: (a) clean surface, (b) dirty surface (red line indicates that the gap between two faces is not connected), (c) dirty surface of 10\% face reduction, (d) dirty surface of 20\% face reduction, (e) dirty surface of 50\% face reduction, (f) combination of frame and skin surfaces displaying in cut model, (g) geodesic dome shape, and (h) crumpled thin soft material ball model that has a thickness of $0.001D$ and a diameter of approximately $0.9-1.1D$.}
  \label{fig:sphere_geometry}
\end{figure}

A simulation of the flow around a sphere was conducted for various geometries that have dirty topology. Figure \ref{fig:sphere_geometry} shows eight configurations of geometries, a clean surface, a dirty surface with many gaps in which two faces are not connected, a dirty surface in which the area of each face is reduced by 10\% of the entire area, a dirty surface with 20\% of the face reduced, a dirty surface with 50\% of the face reduced, a combination of frame and skin surfaces, a ''geodesic dome'' shape comprising truss frames and joints, and a crumpled thin soft material ball. The dirty models in which the faces were reduced did not apply to Mittal's type IBM. The frame and skin model is a common topology for industrial products. This model also does not apply to Mittal's or Peskin's type of IBMs without a clean-up geometry that distinguishes the inner and outer regions. The geodesic dome shape is a product installed especially in hemispherical shape for specific industries, such as an athletic gyms for children, camping site tents, and three-dimensional scanner measurements. The crumpled ball shape shows a folded product made of soft material such as paper or cloth. It is listed here to simulate cases where the sheets fold to form many narrow channels. These geometries do not apply to Mittal's or Peskin's type IBMs. There are small overlaps and gaps between the frames and the skins. The trusses and the joints also have gaps and overlaps. The thickness of the frame and skin was $0.01D$, where $D = 1.0$ is the diameter of the sphere. The width of the truss and joint was $0.01D$. The crumpled ball was made of a material whose thickness was $0.001D$. There were places in the data where the surfaces intersected or were not joined perfectly, as shown in Fig. \ref{fig:geometry_error}\subref{fig:frame_error} and \subref{fig:geodom_error}, for example, a contact surface joined by bolts and nuts, bonded frame surfaces, duplicated surfaces on truss frame, and penetration between soft materials at high curvature areas. Obtain a clean shape for the calculation generally requires a few days of work; however, no preparatory work was needed for this research.

\begin{figure}[htbp]
  \centering
  \begin{minipage}{0.4\hsize}
    \includegraphics[keepaspectratio,width=\textwidth]{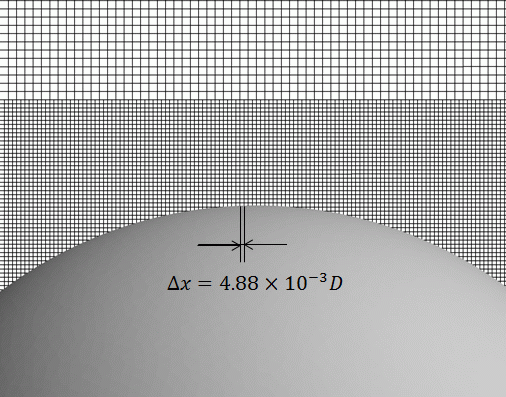}
    \caption{Magnified view of computational grids for the simulation of the flow past a dirty sphere with diameter $D$.}
    \label{fig:sphere_grid}
  \end{minipage}
  \hspace{2mm}
  \begin{minipage}{0.5\hsize}
    \includegraphics[keepaspectratio,width=\textwidth]{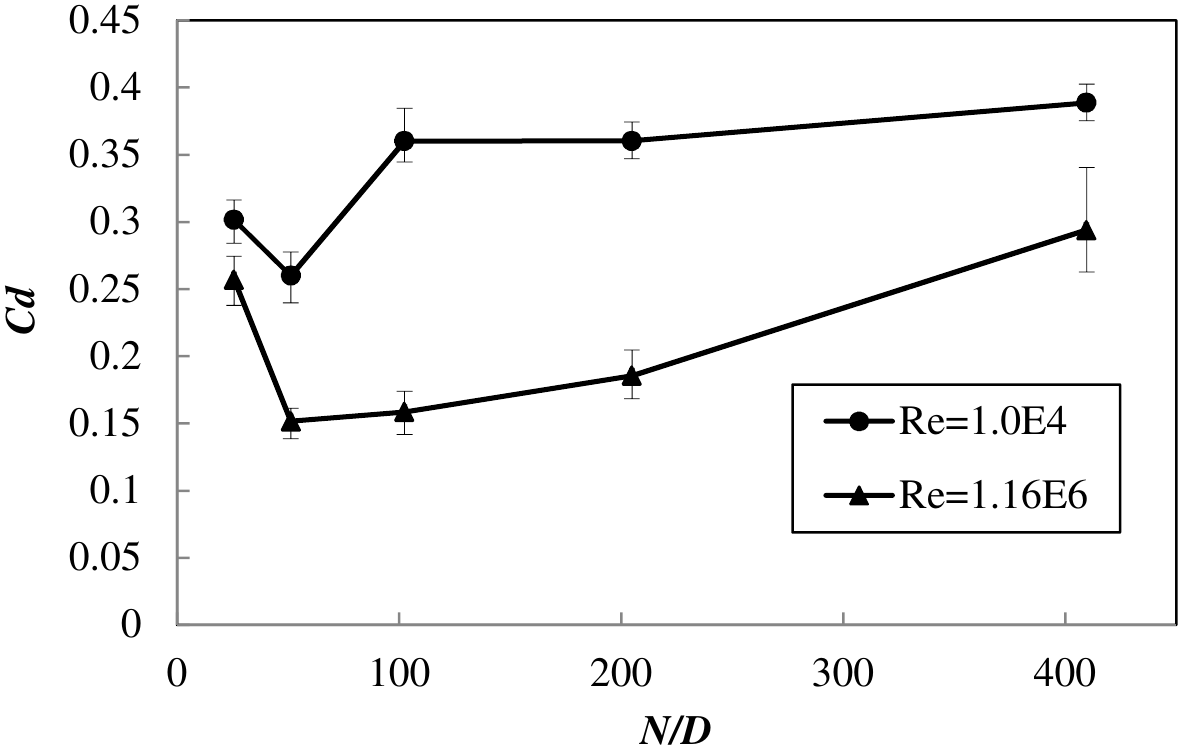}
    \caption{Grid resolution convergence by $Cd$ at $Re_{b} = 1.0 \times 10^{4}$ and $Re_{a}=1.14 \times 10^{6}$ versus the number of cells $N$ for a diameter $D$. Error bars indicate the minimum and maximum values of time history after stabilization was achieved.}
    \label{fig:sphere_grid_conv}
  \end{minipage}
\end{figure}

The $Re$ of $1.0 \times 10^{4}$ (below the critical $Re$, hereafter referred to as $Re_{b}$) and $1.14 \times 10^{6}$ (above the critical $Re$, referred to as $Re_{a}$) were adopted for the flow condition. The calculation grid had a cell size of $\Delta x = 4.88 \times 10^{-3} D$. This corresponded to a resolution that divided the diameter of the sphere into about 205. The grid resolution convergence test revealed that the solution fluctuated sensitively with grid resolution. Figure \ref{fig:sphere_grid_conv} shows the results of $Cd$ with respect to the division number for $D$, at $Re_{b}$ and $Re_{a}$. The $Cd$ values at the coarse resolution were underestimated at $Re_{b}$, and were overestimated at $Re_{a}$. This was due to the separation point shifting either upstream or downstream along the surface of the sphere. The solution at $Re_{a}$ varied significantly depending on the resolution. The separation points at $Re_{a}$ were almost similar for all cases; however, the pressure of the back face of the sphere differed, especially at the finest resolution. Considering the laminar boundary-layer thickness of Schlichting \cite{schlichting1968boundary}, the following points were confirmed. The boundary-layer thickness $\delta_{B}$ was defined as:
\begin{align}
  \label{eq:Schlichting}
 & \delta_{B} = 3 \times \sqrt{\frac{(D/2) \nu}{U_{0}}}, &
\end{align}
where $U_{0}$ is the characteristic velocity. The $\delta_{B}$ was $2.12 \times 10^{-2}$ for $Re_{b}$, and was $1.99 \times 10^{-3}$ for $Re_{a}$. The cell size used here had less than one cell at $Re_{a}$ within the laminar boundary layer, therefore the solution was highly sensitive due to the resolution around the boundary layer. Actually, in the case that had a small error (a single precision of seven decimal places), by merging the geometry face node to reduce the number of duplicated nodes, we obtained a different flow field containing an error of about 10\%. This sensitivity was enhanced by the stepwise distribution of pressure on the sphere surface, and we infer that the artificial viscosity was locally enlarged depending on the resolution. Here, we consider that the medium resolution of $N/D = 205$ was reasonable for the choice, as a stable solution was shown before this resolution.

The number of cubes was 16,256 and the number of cells was 66,584,576. The computational domain had dimensions of $40D \times 20D \times 20D$. The grid was generated on a workstation, described in Tab. \ref{table:resource_WS}, using dedicated software that had a GUI. As Cartesian grids were generated, a computation time of only about 10 minutes was required. The Crank-Nicolson implicit scheme was used for time integration. The (non-dimensional) solution time was 30.0, with a time increment of $\Delta t = 2.0 \times 10^{-4}$. The overall computation time was approximately 6.5 hours, using the 32 nodes of the supercomputer listed in Tab. \ref{table:resource_OFP}. The calculation was executed by 64 MPI parallel processes, with a 4-thread OpenMP (hyper-thread) on each node.

\begin{table}[htb]
  \caption{Computational resources used in the grid generation and visualization.}
  \label{table:resource_WS}
  \centering
  \begin{tabular}{cp{5.9cm}}
    \hline
      & Grid generation \& visualization \\
    \hline \hline
    CPU & Intel Xeon E5-2687W, 3.10GHz, 8cores \\
    Main memory & 64GB DDR3, 1600MHz \\
    Graphic board & NVIDIA Quadro5000 2.5GB \\
    \hline
  \end{tabular}
\end{table}

\begin{table}[htb]
  \caption{Computational resources used in the flow computation.}
  \label{table:resource_OFP}
  \centering
  \begin{tabular}{cp{5.9cm}}
    \hline
      & Flow computation \\
    \hline \hline
    CPU & Intel Xeon Phi 7250, 1.4GHz, 68cores \\
    Main memory & 96GB(DDR4)+16GB(MCDRAM)/node \\
    Network & Intel Omni-Path \\
    \hline
  \end{tabular}
\end{table}


The hybrid scheme was used for spatial discretization in the second-order central-difference scheme, with a 10\% blending of the QUICK scheme. Here, the turbulence model by CSM was not applied. In the calculation using the 100\% second-order central difference and the CSM turbulence model, the solution showed unphysical oscillation at the cube boundaries of different sizes. Therefore, a numerical experiment was conducted; when 5\% of the upwind difference by QUICK was blended and when 10\% was blended. The results showed that the oscillation was successfully omitted. However, the separation point shifted upstream of the sphere. We consider that it had a larger numerical viscosity. Therefore, by omitting the eddy viscosity of CSM, an adjustment was made to obtain an appropriate (case-specific) numerical viscosity as a whole. This corresponds to a technique called ''implicit LES.''

\begin{figure}[htbp]
 \centering
 \begin{minipage}{0.4\hsize}
   \includegraphics[keepaspectratio,width=\textwidth]{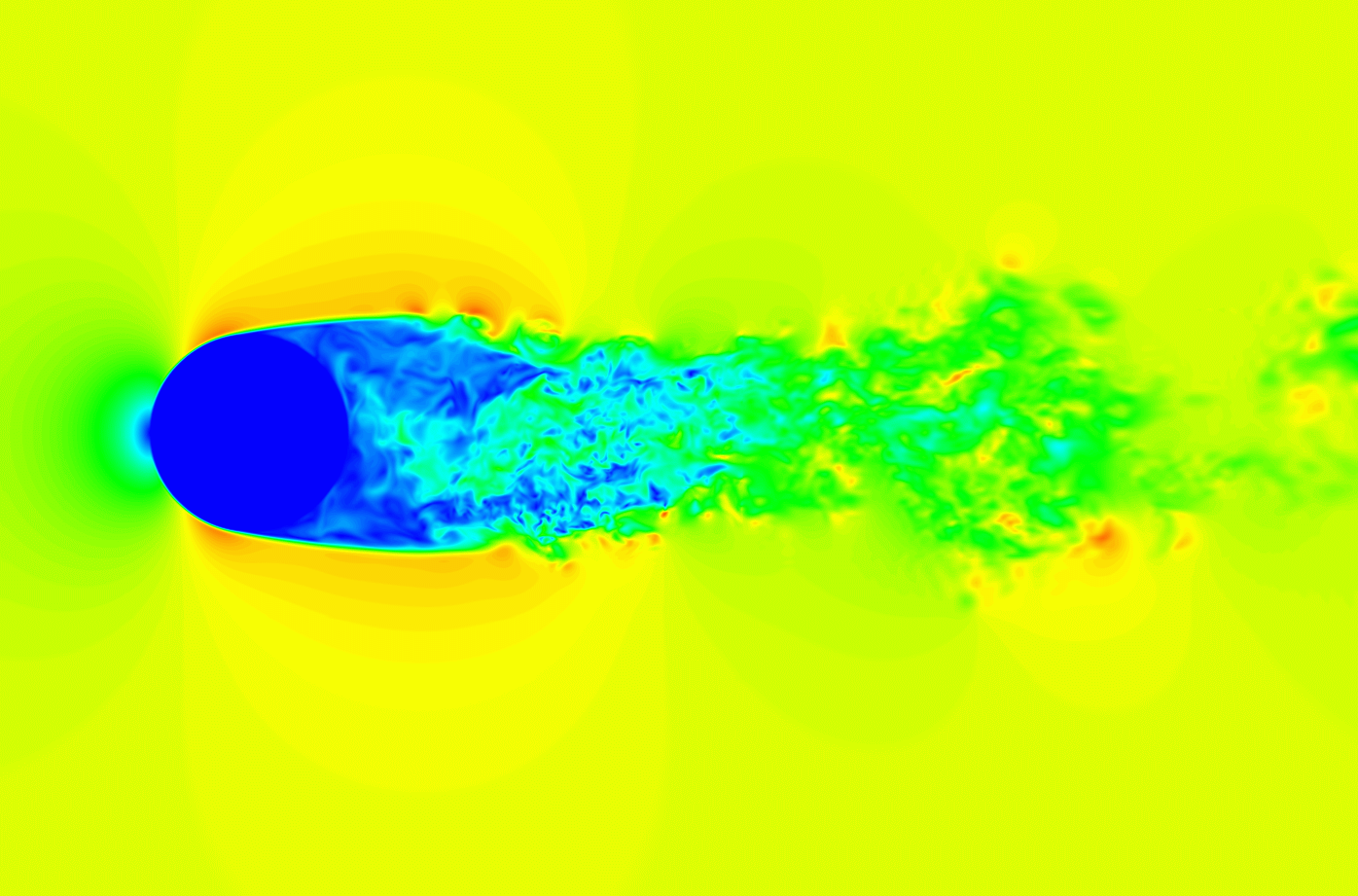}
   \subcaption{Instantaneous value for a clean surface.}
   \label{fig:sphere_moderate_ins}
 \end{minipage}
 \begin{minipage}{0.4\hsize}
   \includegraphics[keepaspectratio,width=\textwidth]{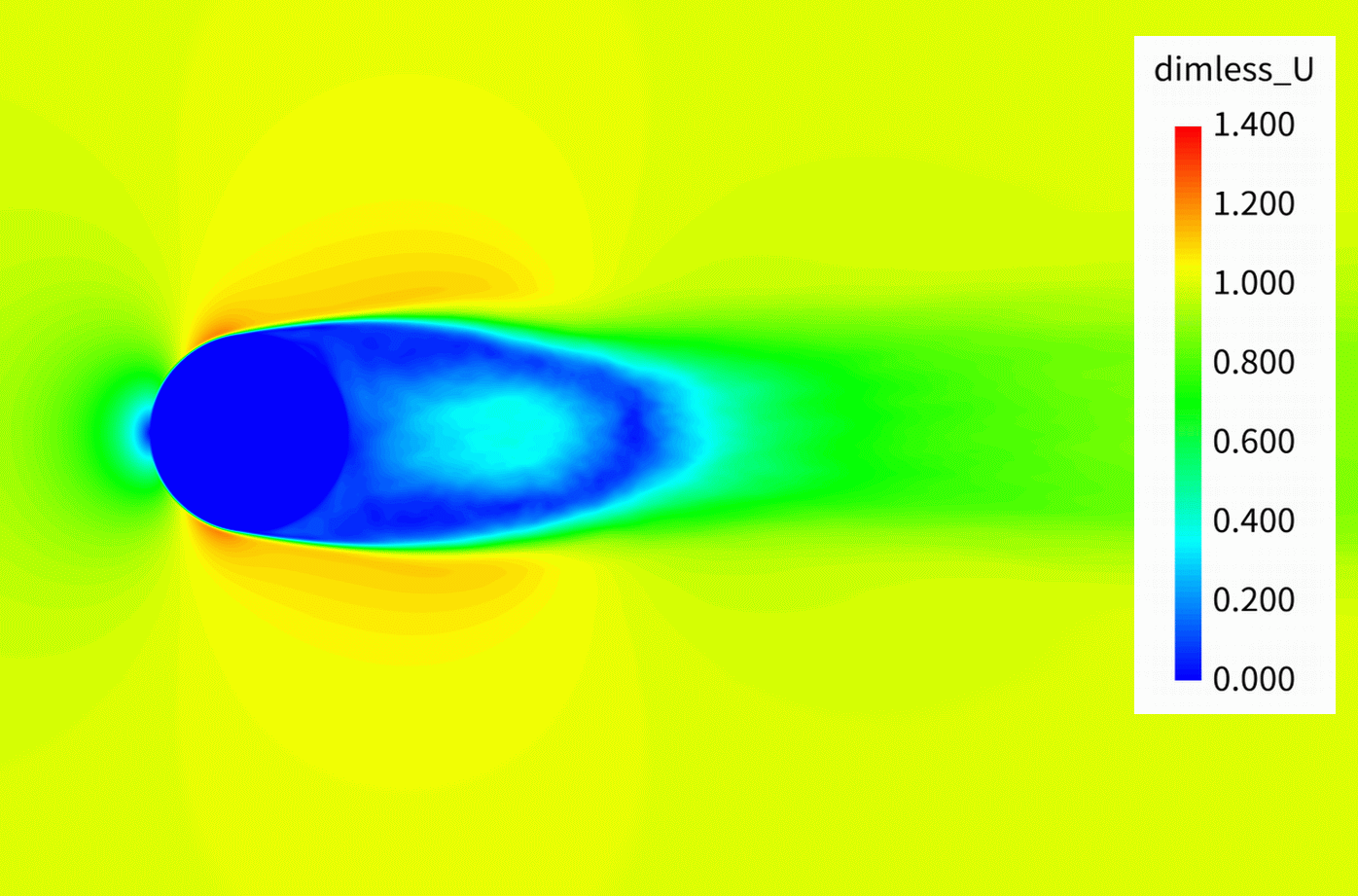}
   \subcaption{Time-averaged value for a clean surface.}
   \label{fig:sphere_moderate_ave}
 \end{minipage}
 \caption{Non-dimensional velocity magnitude in the center section for $Re_{b} = 1.0 \times 10^{4}$ flows around a sphere.}
 \label{fig:sphere_moderateRe}
\end{figure}

\begin{figure}[htbp]
 \centering
 \begin{minipage}{0.4\hsize}
   \includegraphics[keepaspectratio,width=\textwidth]{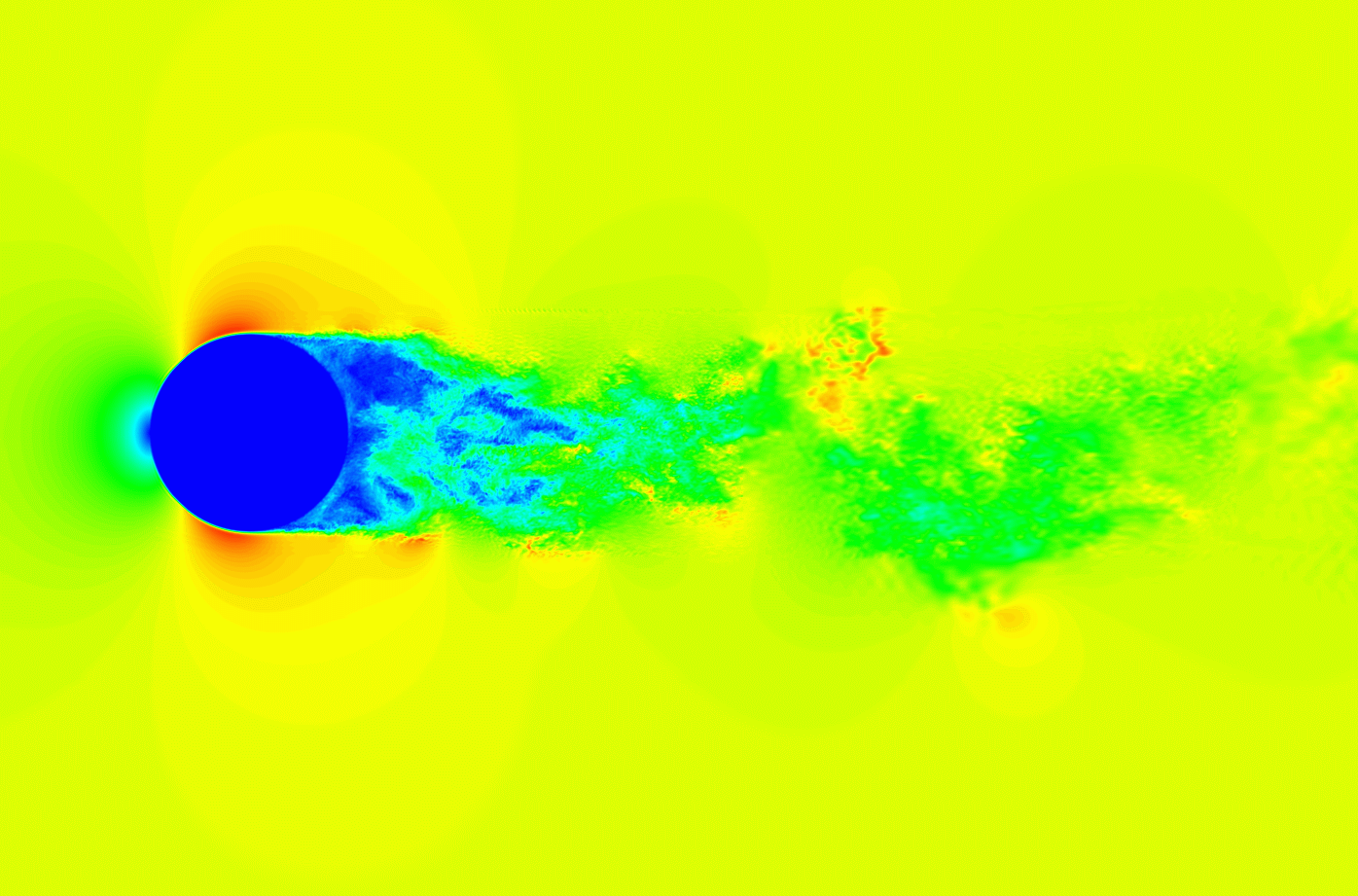}
   \subcaption{Instantaneous value for a clean surface.}
   \label{fig:sphere_clean_ins}
 \end{minipage}
 \begin{minipage}{0.4\hsize}
   \includegraphics[keepaspectratio,width=\textwidth]{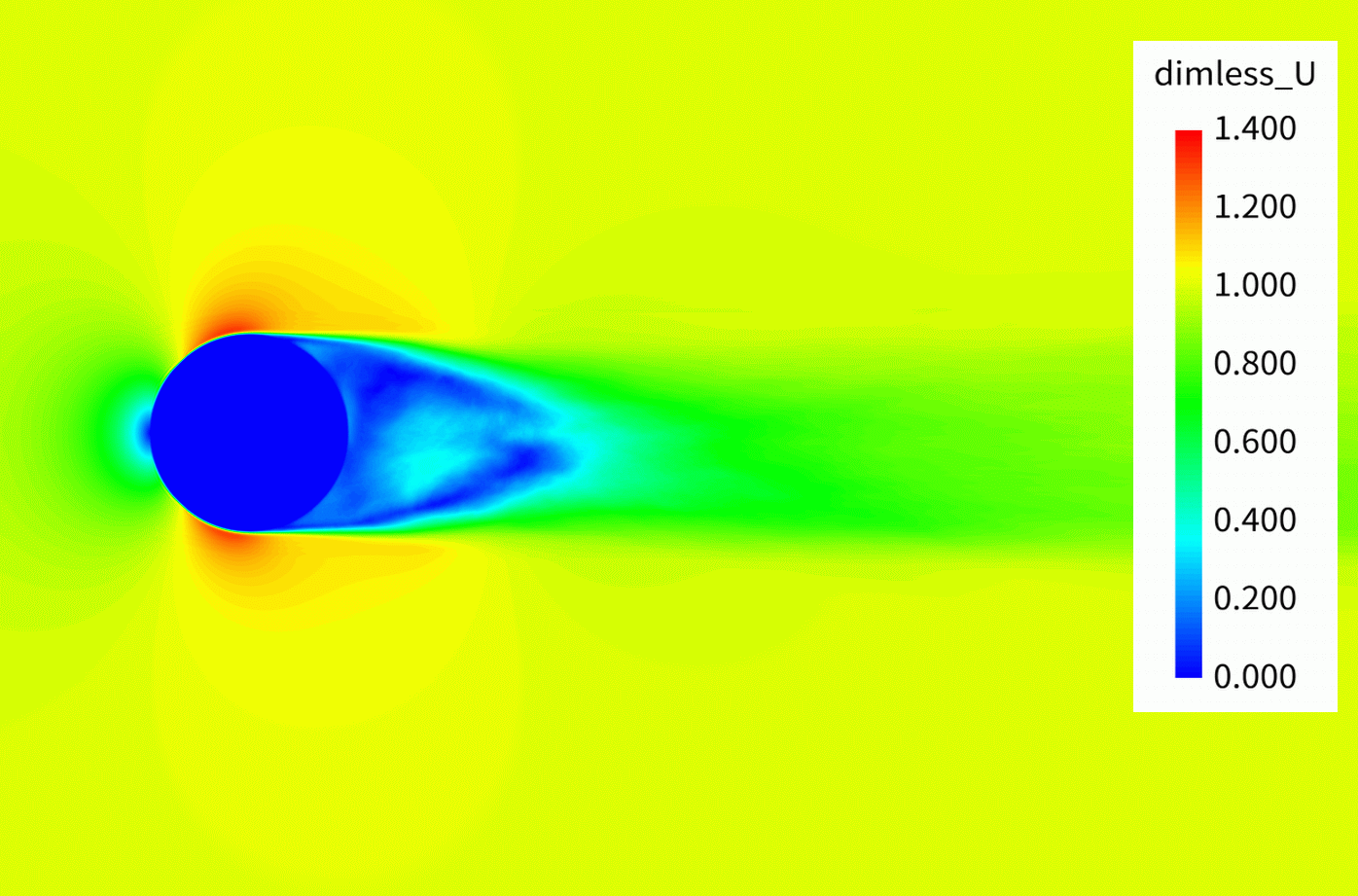}
   \subcaption{Time-averaged value for a clean surface.}
   \label{fig:sphere_clean_ave}
 \end{minipage}
 \begin{minipage}{0.4\hsize}
   \includegraphics[keepaspectratio,width=\textwidth]{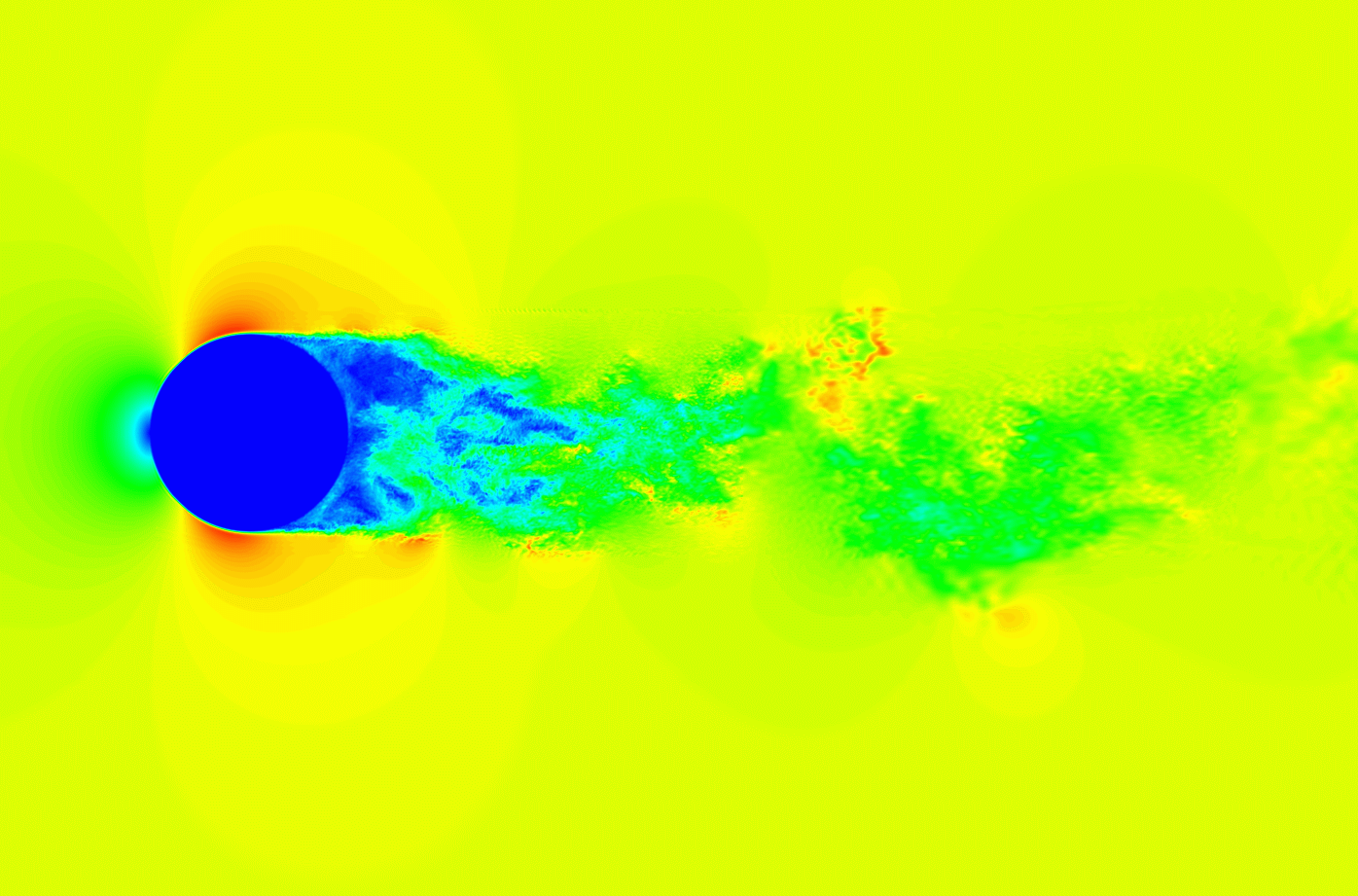}
   \subcaption{Instantaneous value for a dirty surface.}
   \label{fig:sphere_dirty_ins}
 \end{minipage}
 \begin{minipage}{0.4\hsize}
   \includegraphics[keepaspectratio,width=\textwidth]{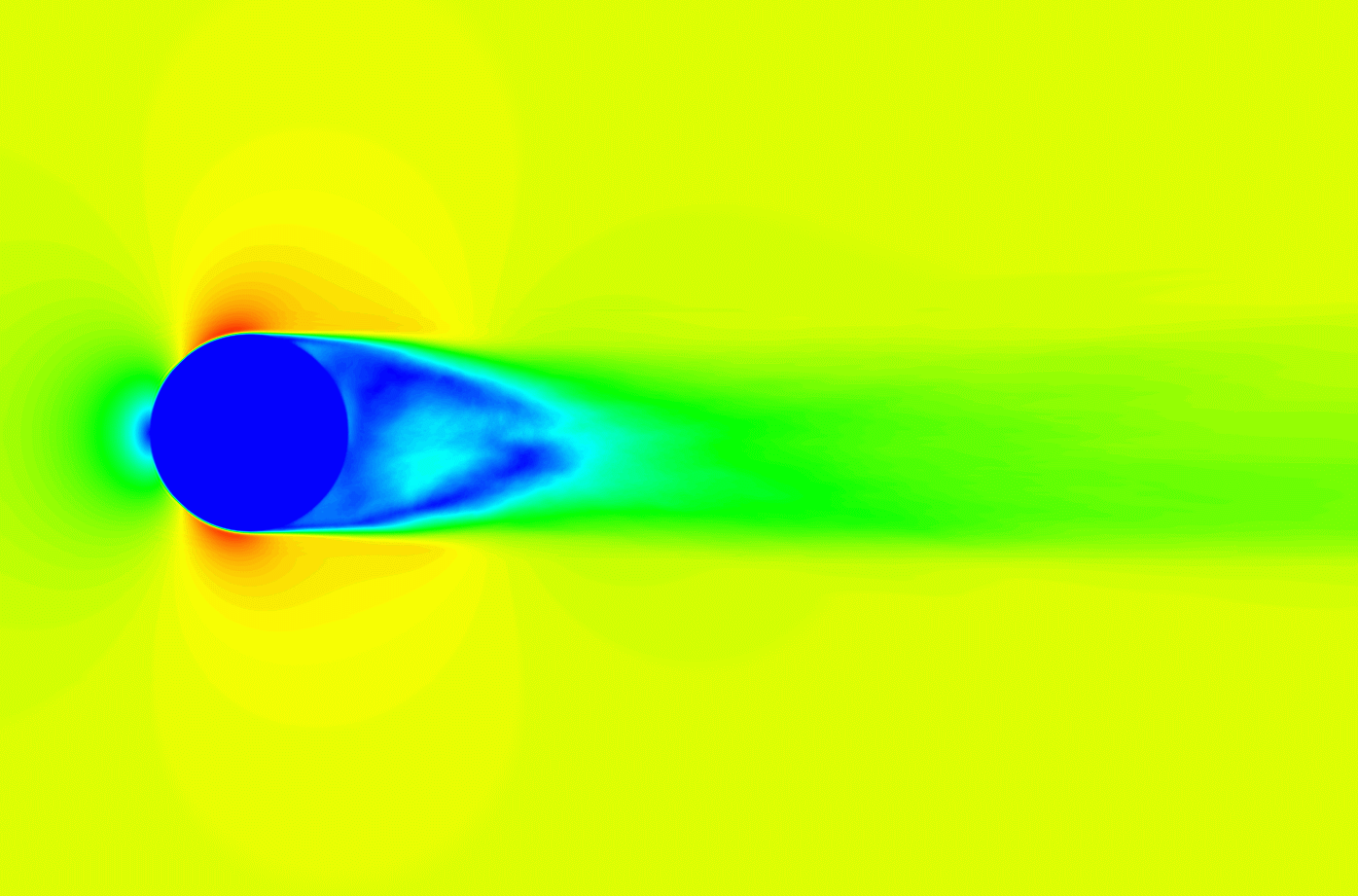}
   \subcaption{Time-averaged value for a dirty surface.}
   \label{fig:sphere_dirty_ave}
 \end{minipage}
 \begin{minipage}{0.4\hsize}
   \includegraphics[keepaspectratio,width=\textwidth]{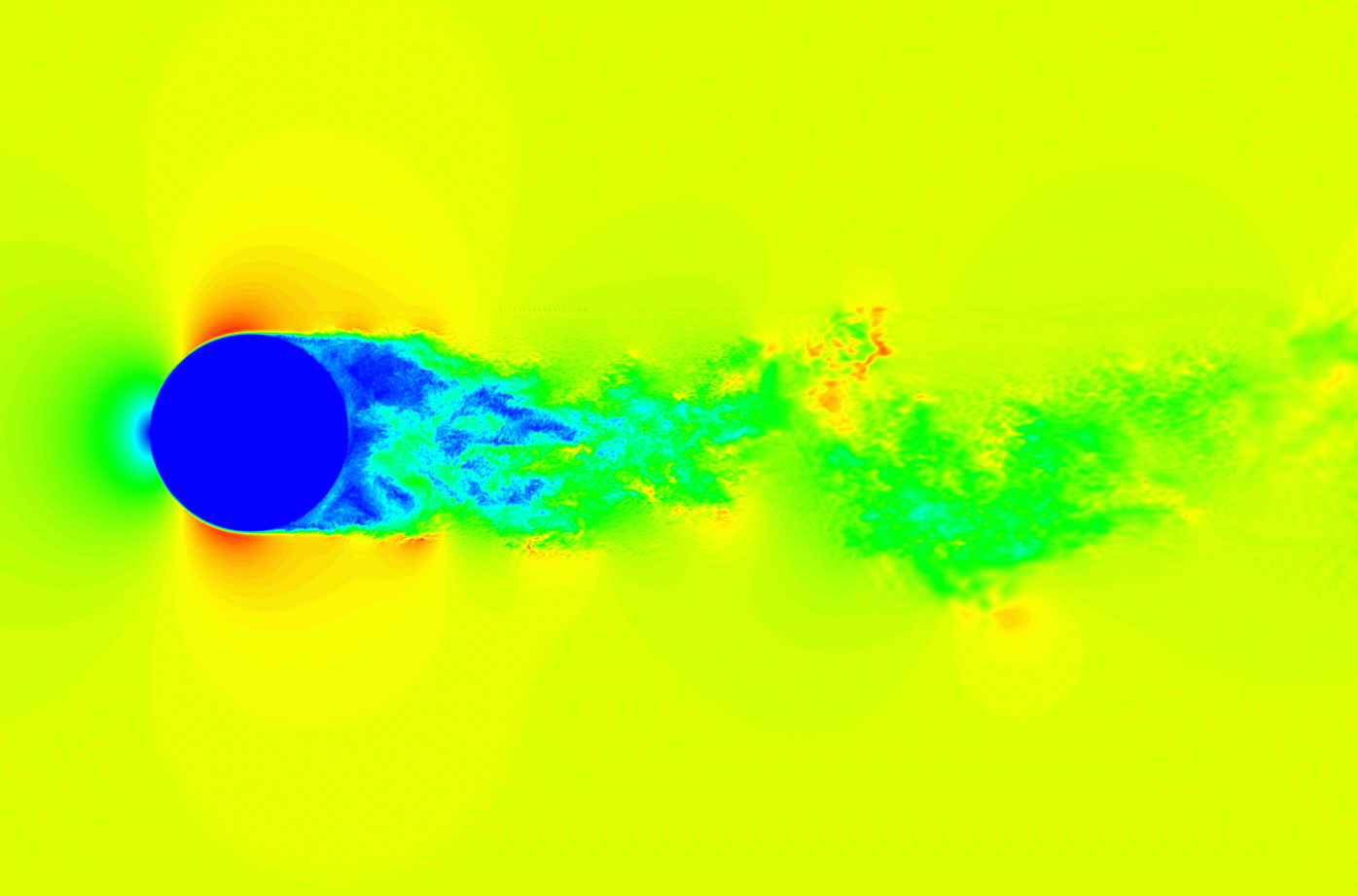}
   \subcaption{Instantaneous value for the frame and skin model.}
   \label{fig:sphere_frame_ins}
 \end{minipage}
 \begin{minipage}{0.4\hsize}
   \includegraphics[keepaspectratio,width=\textwidth]{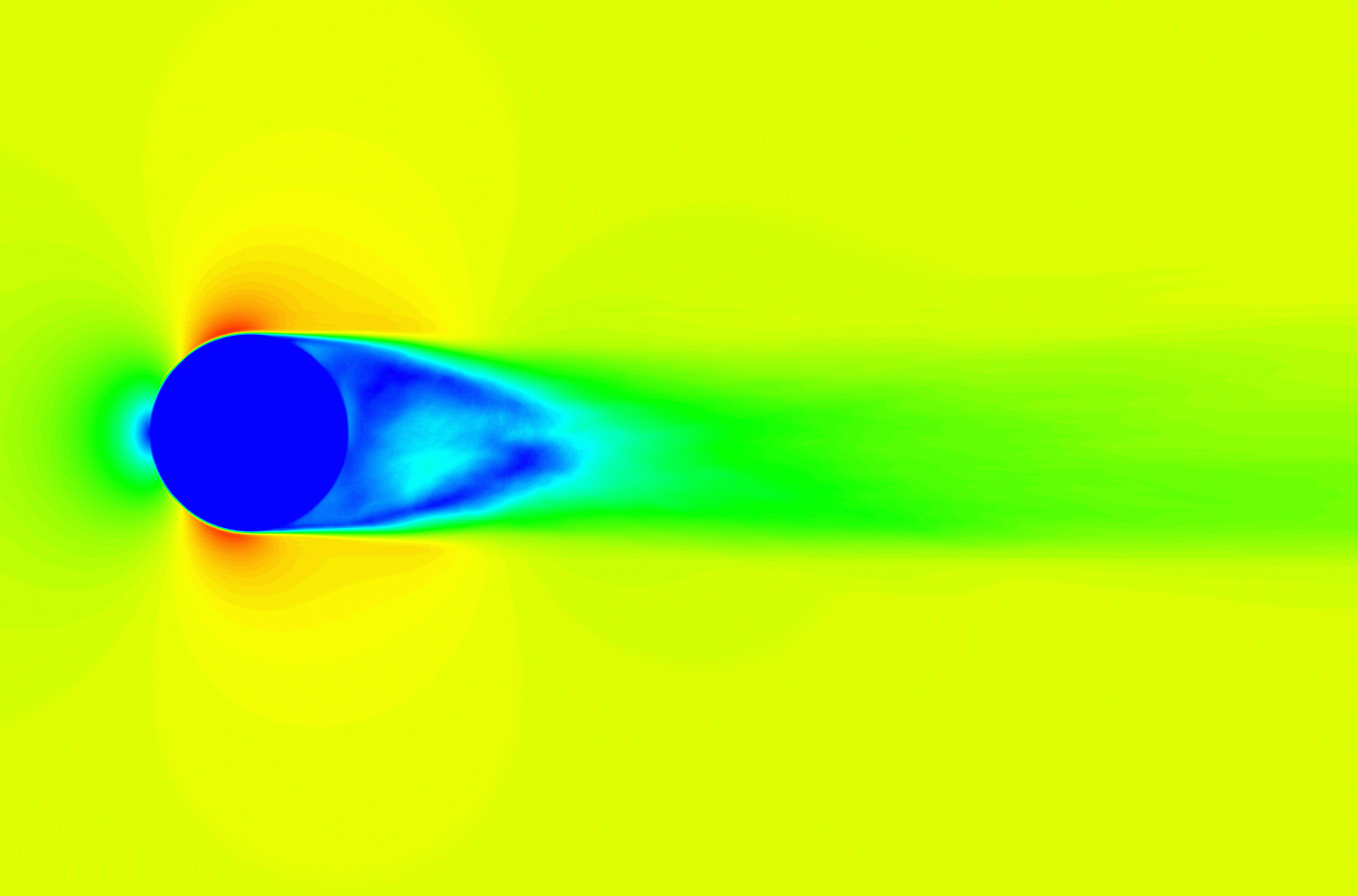}
   \subcaption{Time-averaged value for the frame and skin model.}
   \label{fig:sphere_frame_ave}
 \end{minipage}
 \caption{Non-dimensional velocity magnitude in the center section for $Re_{a} = 1.14 \times 10^{6}$ flows around a sphere.}
 \label{fig:sphere_highRe}
\end{figure}

Figures \ref{fig:sphere_moderateRe} and \ref{fig:sphere_highRe} show the instantaneous and time-averaged non-dimensional velocity magnitudes for each $Re$ in the center section of the sphere. Figure \ref{fig:sphere_highRe}\subref{fig:sphere_clean_ave} shows that for the case of a clean surface, there was laminar separation slightly downstream compared to Fig. \ref{fig:sphere_moderateRe}\subref{fig:sphere_moderate_ave}. Nevertheless, it was positioned upstream of the top-pole position of the sphere. Note that the flow was completely isolated at the outside and inside of the sphere. In Peskin's type IBM, unphysical flow may occur inside. Figure \ref{fig:sphere_highRe}\subref{fig:sphere_dirty_ave} and Figure \ref{fig:sphere_highRe}\subref{fig:sphere_frame_ave} show the results of the dirty surface and the frame and skin model respectively. Both show quantitatively the same flow field, and the same separation point, as a clean surface case. These are reasonable results, as the outermost surfaces have the same geometry.

\renewcommand{\thesubfigure}{\roman{subfigure}}

\begin{figure}[htbp]
 \centering
 \begin{minipage}{0.4\hsize}
   \includegraphics[keepaspectratio,width=\textwidth]{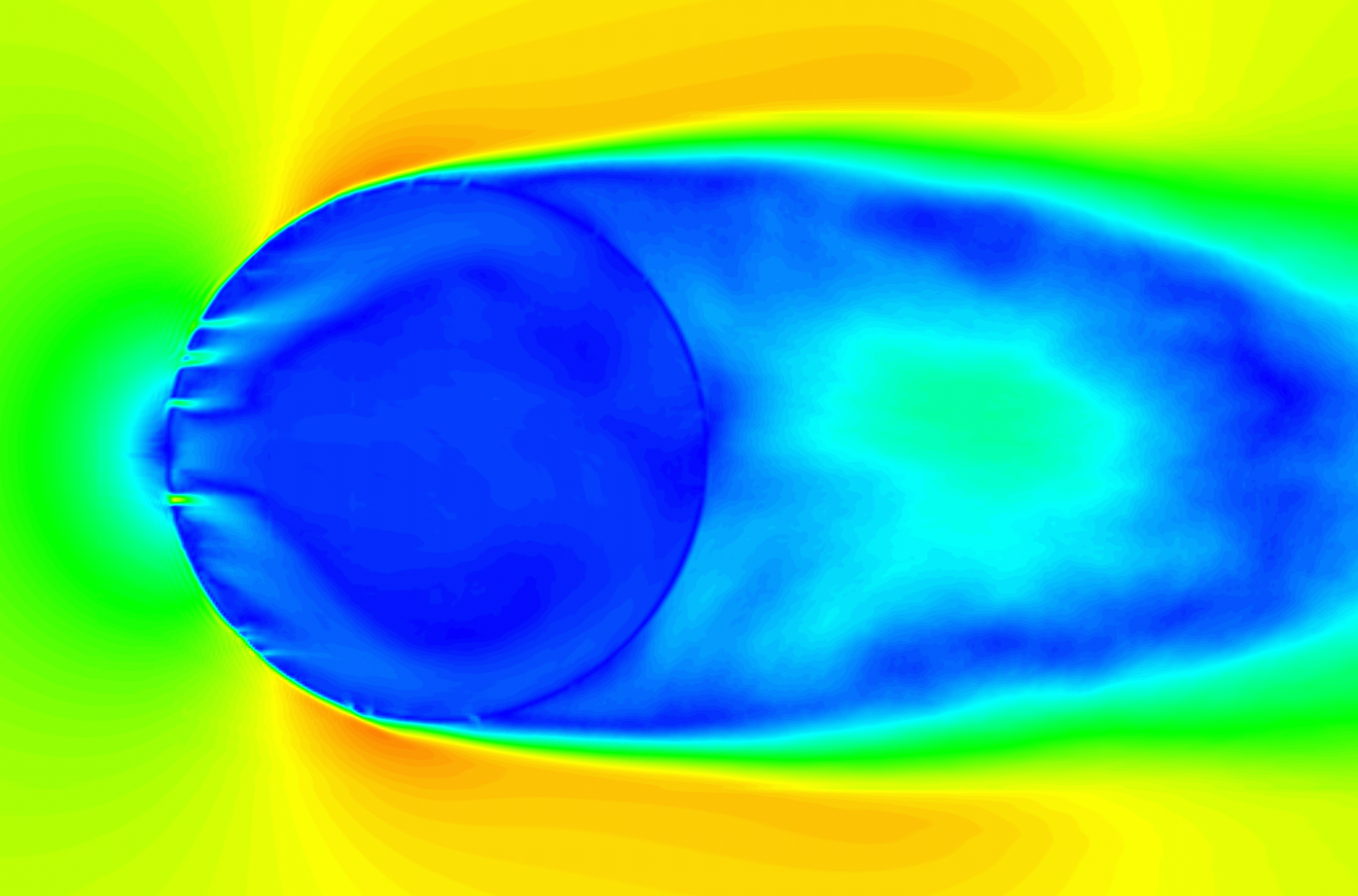}
   \subcaption{Velocity magnitude for the 10\% dirty surface.}
   \label{fig:sphere_dirty10_ave}
 \end{minipage}
 \begin{minipage}{0.4\hsize}
   \includegraphics[keepaspectratio,width=\textwidth]{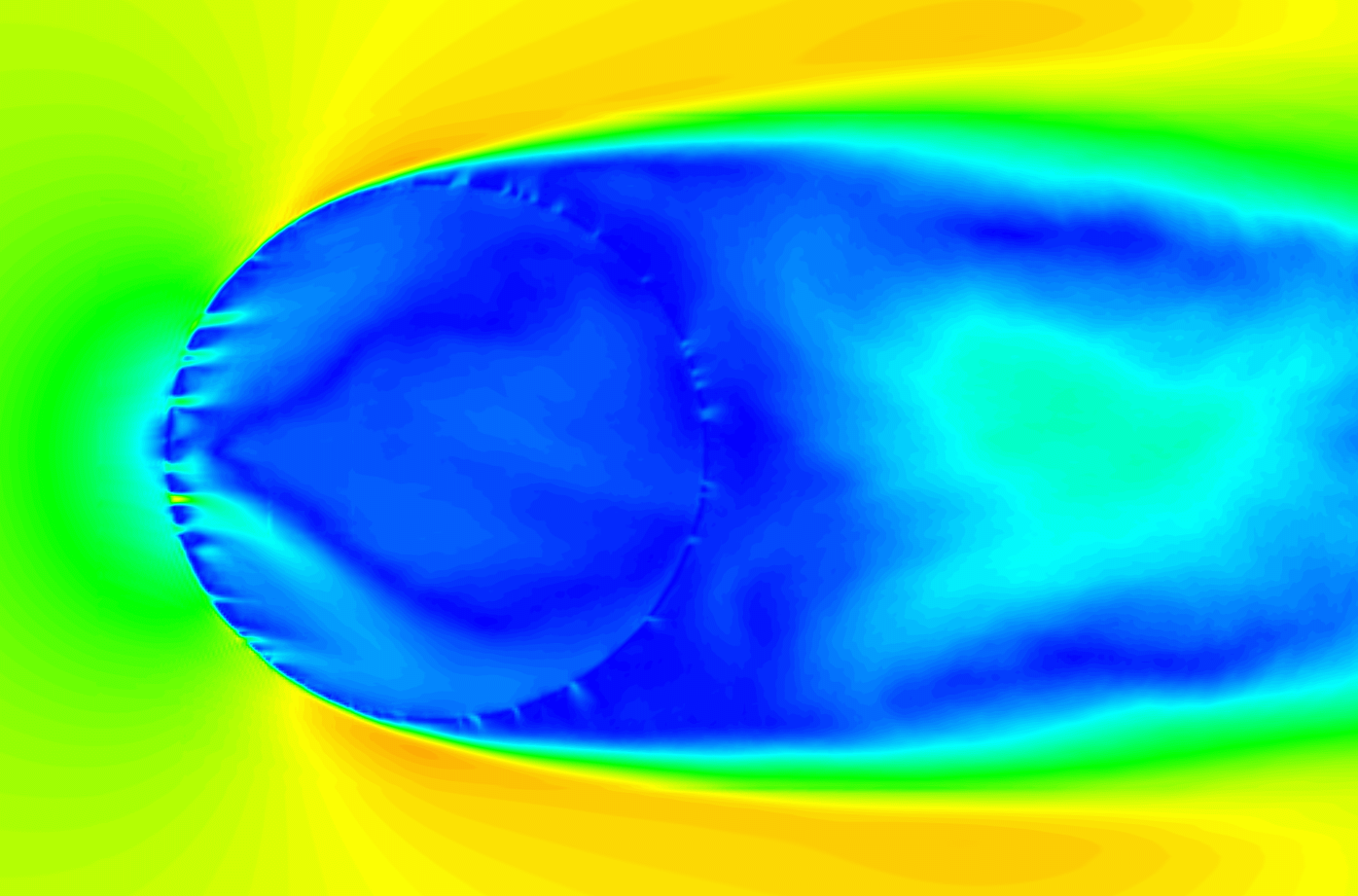}
   \subcaption{Velocity magnitude for the 20\% dirty surface.}
   \label{fig:sphere_dirty20_ave}
 \end{minipage}
 \begin{minipage}{0.4\hsize}
   \includegraphics[keepaspectratio,width=\textwidth]{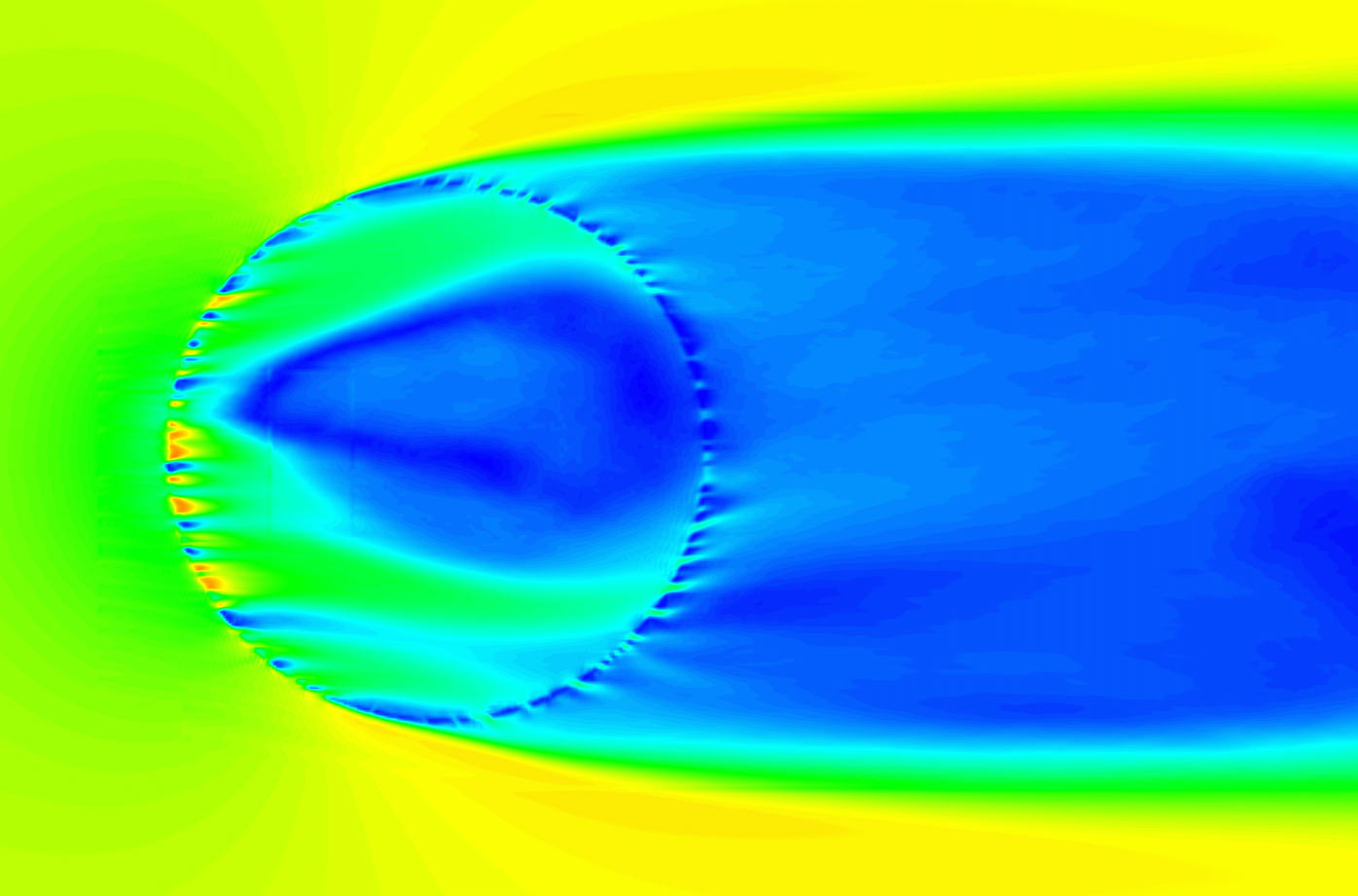}
   \subcaption{Velocity magnitude for the 50\% dirty surface.}
   \label{fig:sphere_dirty50_ave}
 \end{minipage}
 \begin{minipage}{0.4\hsize}
   \includegraphics[keepaspectratio,width=\textwidth]{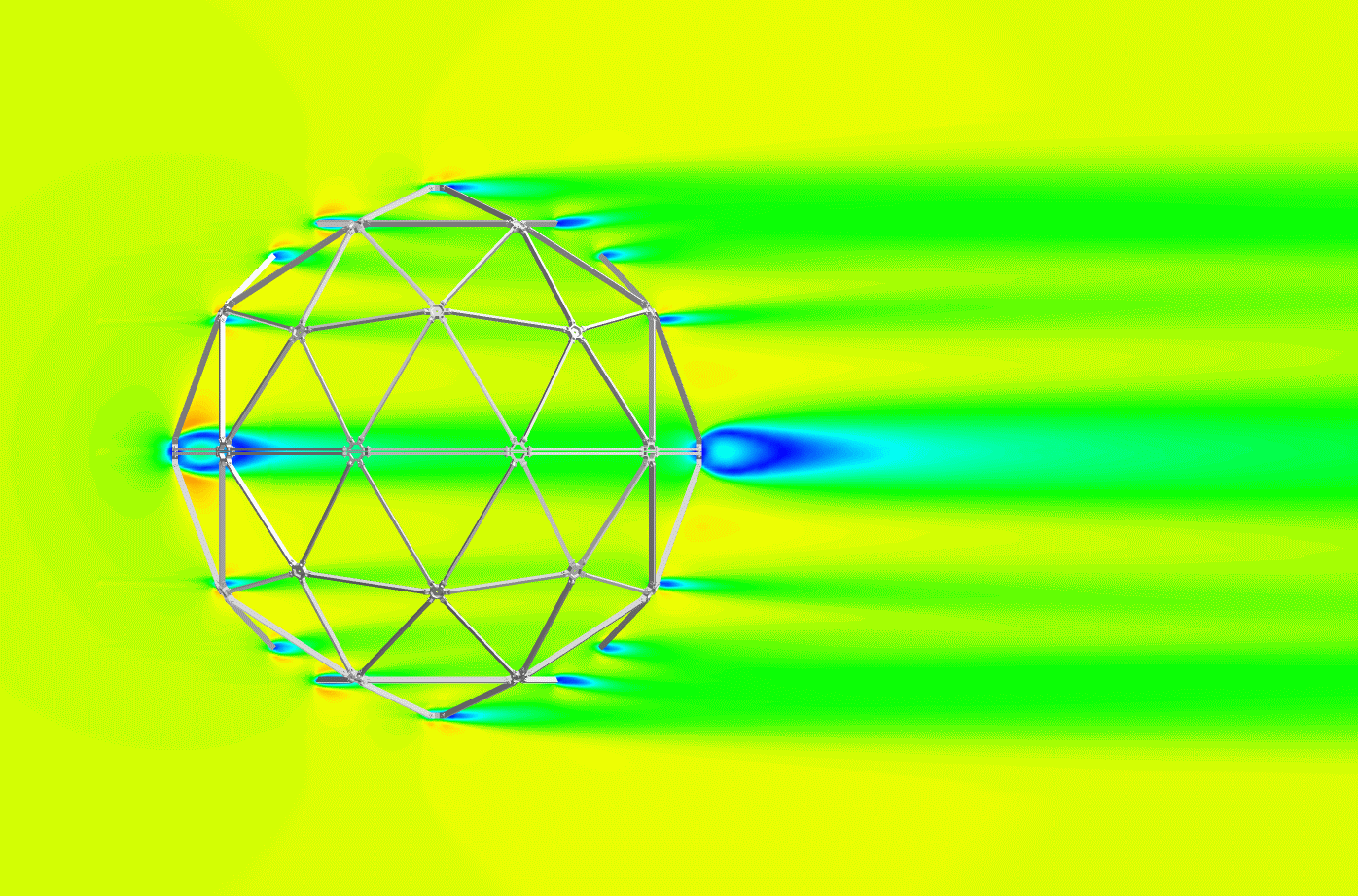}
   \subcaption{Velocity magnitude for the geodesic dome.}
   \label{fig:geo_dom_ave}
 \end{minipage}
 \begin{minipage}{0.4\hsize}
   \includegraphics[keepaspectratio,width=\textwidth]{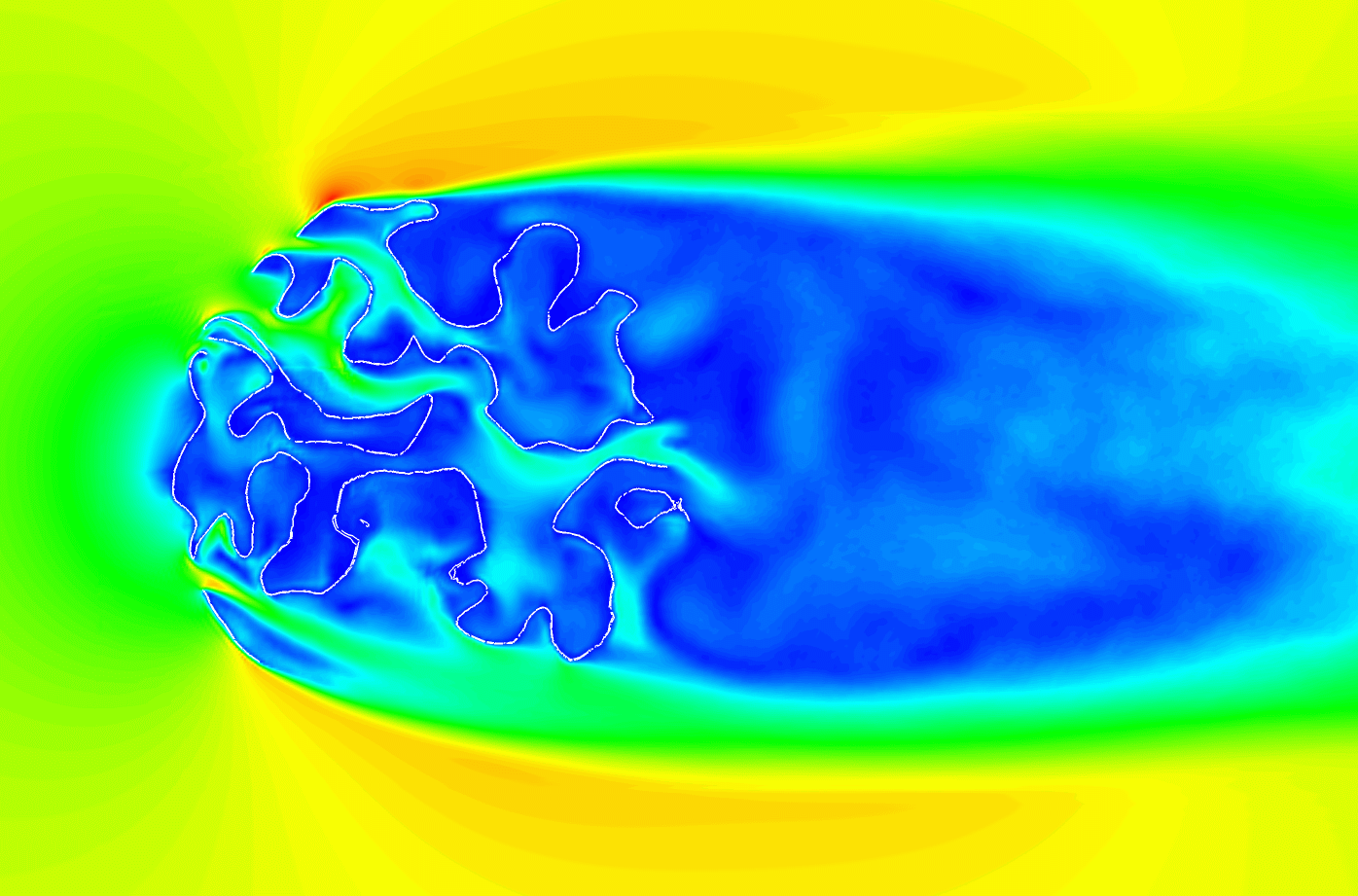}
   \subcaption{Velocity magnitude for the crumpled ball. The white line represents the shape of the central section.}
   \label{fig:crumpled_paper_ave}
 \end{minipage}
 \begin{minipage}{0.4\hsize}
   \includegraphics[keepaspectratio,width=\textwidth]{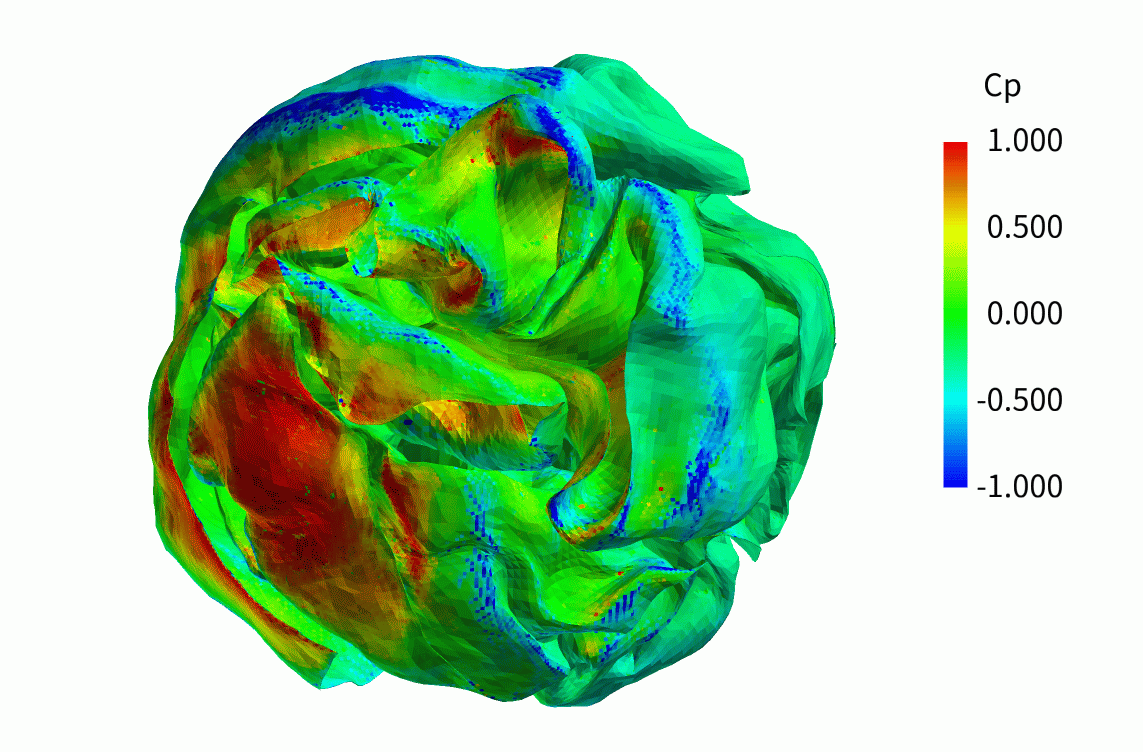}
   \subcaption{Surface pressure coefficient ($Cp$) for the crumpled ball.}
   \label{fig:crumpled_paper_Cp}
 \end{minipage}
 \caption{Time-averaged (non-dimensional) velocity magnitude in the center section and surface pressure for $Re_{a} = 1.14 \times 10^{6}$ flows around a dirty sphere.}
 \label{fig:sphere_dirty_flow}
\end{figure}

Figure \ref{fig:sphere_dirty_flow} shows the time-averaged velocity magnitudes in the center section of the dirty spheres. As the face reduction rate increased, the jet flow entered the sphere and then flowed out. The grid resolution was directly related to the formation of the jet: as the opening increased, the jet passing through the surface increased. This cannot be represented by Mittal's type IBM. In the geodesic dome case, the flow past the joint formed a separation bubble, and the flow past the frame created a shear layer. This shows that despite the errors shown in Fig. \ref{fig:geometry_error}\subref{fig:geodom_error}, the calculation can be performed without problems even for narrow channels that cannot be resolved by the grid. In the crumpled ball case, the flow penetrated the interior of the object along a complex flow path, and independent fluid rooms separated by the thin material were formed. This cannot be represented by both Mittal's type and Peskin's type of IBMs. In this way, it was possible to successfully capture the complex flow, which varied according to the details of the shape. In these cases, the same calculation grid was used. After preparing the first case, the execution of other geometry cases was automatically done simply by changing the input geometry file name.

\begin{table}[htb]
  \caption{Predicted drag coefficients of configurations.}
  \label{table:sphere_cd}
  \centering
  \begin{tabular}{lp{3cm}}
    \hline
    Case & $Cd$ \\
    \hline \hline
    Experiment (Schlichting \cite{schlichting1968boundary}) & 0.143 \\
    Simulation (Muto \cite{Muto2012}) & 0.180 \\
    Clean surface & 0.193 \\
    Dirty surface & 0.193 \\
    Dirty surface (10\%) & 0.447 \\
    Dirty surface (20\%) & 0.498 \\
    Dirty surface (50\%) & 0.551 \\
    Frame and skin & 0.191 \\
    Geodesic dome & 0.261 \\
    Crumpled ball & 0.462 \\
    \hline
  \end{tabular}
\end{table}

\begin{figure}[htbp]
  \centering
  \begin{minipage}{0.7\hsize}
    \includegraphics[keepaspectratio,width=\textwidth]{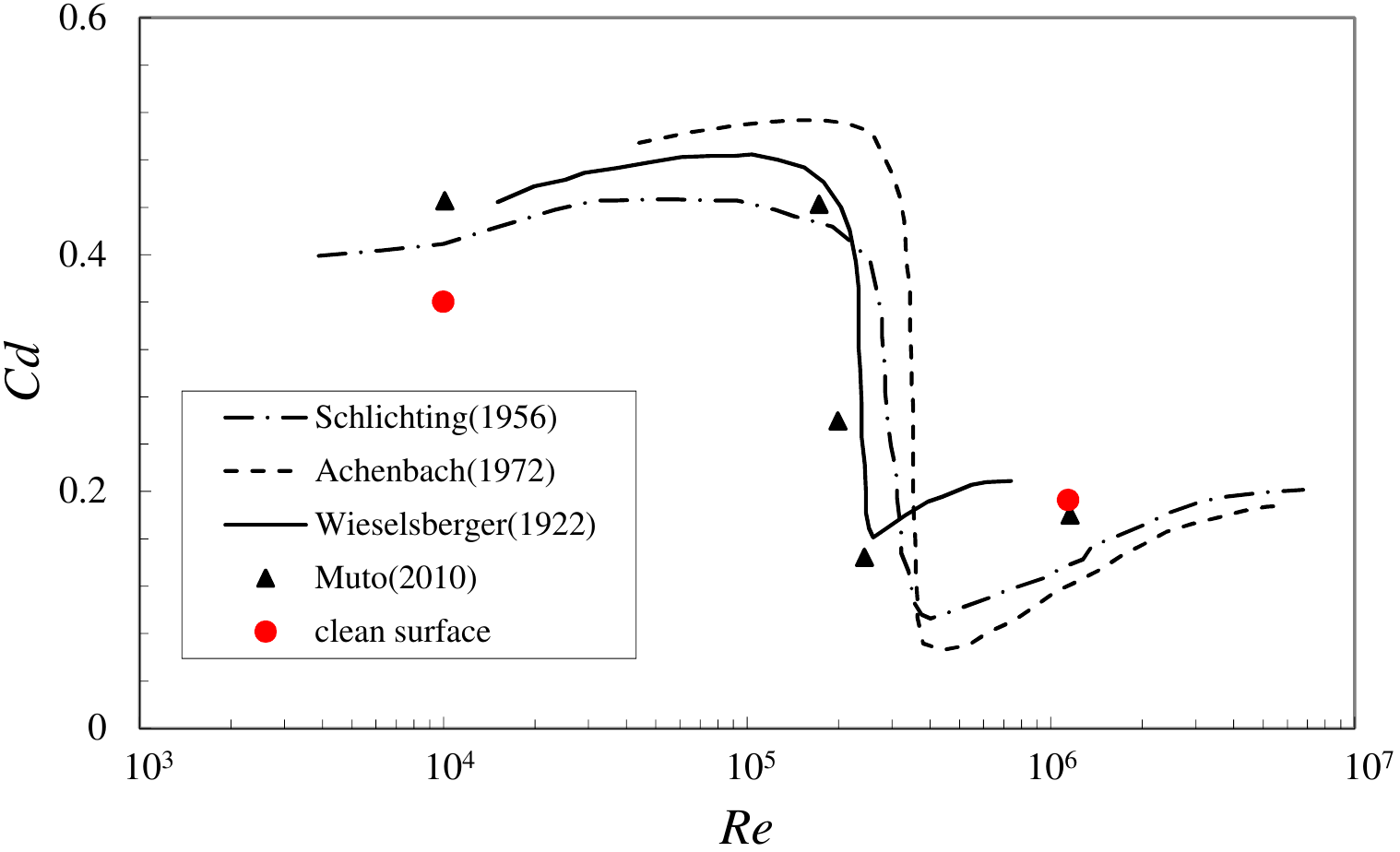}
    \caption{Comparison of the drag coefficient with results of past experimental and computational investigations of flow around a clean sphere, under moderate- and high-Re conditions. Black lines indicate past experimental results, while black symbols indicate past computational results. Red circles indicate the results of the present study.}
    \label{fig:sphere_cd}
  \end{minipage}
\end{figure}

Table \ref{table:sphere_cd} shows the predicted $Cd$. Figure \ref{fig:sphere_cd} shows the $Cd$ of the clean surface case, compared to the results of past studies. The $Cd$ is defined as:
\begin{align}
  \label{eq:cd_sphere}
 & Cd = f_x / (\frac{1}{2} \rho U_{0}^{2} \pi D^{2} / 4), &
\end{align}
where $f_x$ is a force integral in the flow direction. In all cases, the projected frontal area was the same. The force acting on the surface was integrated by employing Shepard's method, expressed by Eq. \eqref{eq:force}. However, here, the following filter was applied to prevent double-counting on the surface of the thin plate.

\begin{align}
  \label{eq:force_filter}
 & f(\bm{X}) = \frac{\sum_{k=1}^{m} \mathcal{S'} (\bm{X}) f_{k}}{\sum_{k=1}^{m} \mathcal{S'} (\bm{X})}, 
  \qquad {\rm where} \; \mathcal{S'}(\bm{X}) = \mathcal{Z}(\bm{X}) \otimes \mathcal{S}(\bm{X}), & \\
 & \mathcal{Z}(\bm{X}) = 
  \begin{cases}
    0, & \exists \bm{x} \in S( \mid \bm{X} - \bm{x} \mid + \mid \bm{x^{(k)}} - \bm{X^{(k)}} \mid \neq \Delta x ), \\
    1, & elsewhere,
  \end{cases}
  \qquad {\rm where} \; \mid \bm{x^{(k)}} - \bm{x} \mid = \Delta x . &
\end{align}

Here the filter function $\mathcal{Z} (\bm{X})$ was defined according to the distance between the surface node $\bm{X}$, the surrounding fluid grid point $\bm{x}^{(k)}$, and the surface node, which itself was defined on the surrounding fluid point $\bm{X^{(k)}}$. If the number of calculation points in the thin plate was less than one, the numerical value of the back surface (the inner surface of the thickness) was canceled. This treatment is important to avoid unexpected numerical errors when both front and back sides are involved, both for surface integral on thin-plate geometry. If this filter was not applied, it caused errors of 10-20\% in the surface integration.

\begin{figure}[htbp]
  \centering
  \begin{minipage}{0.7\hsize}
   \includegraphics[keepaspectratio,width=\textwidth]{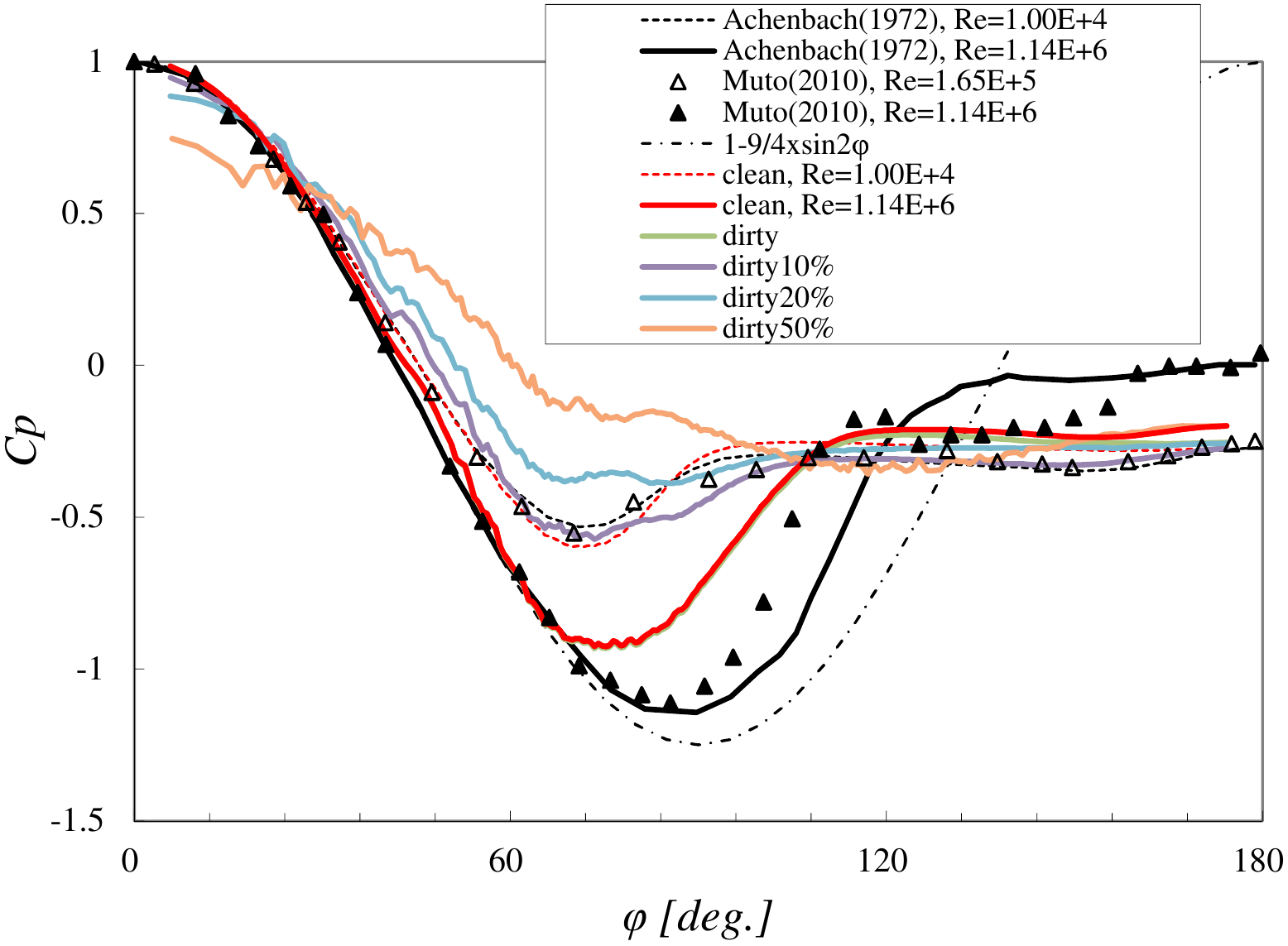}
    \caption{Profile of the surface pressure coefficient of flow passing a sphere under high-$Re$ conditions. Black lines indicate past experimental results and sinusoidal functions. Black symbols indicate past computational results. Red lines indicate the present results of a clean surface. Green, purple, blue, and orange lines show the results of dirty geometry models.}
    \label{fig:sphere_cp}
  \end{minipage}
\end{figure}

The $Cd$ value was time-averaged after the flow field became sufficiently stable. The value was underestimated for $Re_{b}$. The value was overestimated for $Re_{a}$, however, the case of the dirty surface showed the same value. The frame-skin case showed the same flow results, and the $Cd$ matched the clean surface case with about 1\% error, due to the filtering of Eq. \eqref{eq:force_filter}. A trend of increasing $Cd$ was observed as the surface reduction rate increased. Figure \ref{fig:sphere_cp} summarized the surface pressure coefficient ($Cp$) distribution obtained by averaging each meridian position. The $\varphi$ was the circumferential angle measured from the stagnation point. Here, the same equation of Eq. \eqref{eq:force_filter} was used for the evaluation of the surface pressure. As the equation formed the near-field profile, it could be applied for the surface evaluation. An example of the surface pressure is shown in Fig. \ref{fig:sphere_dirty_flow}\subref{fig:crumpled_paper_Cp}. The clean surface case had a typical laminar flow result at $Re_{b}$, and was in good agreement with the experimental result of Achenbach \cite{Achenbach1972} and the simulation result of Muto \cite{Muto2012}, which had a highly resolved unstructured grid. However, the result at $Re_{a}$ showed that the separation point shifted upstream, and then caused an overestimation of $Cd$. This indicates that excessive artificial viscosity was introduced around the separation point.

For the dirty geometries, as the face reduction rate increased, pressure recovery was observed near the top of the sphere. This is because, as was determined from the solution of the flow field, the boundary layer was separated forcibly by blowing out the jet from the hole resolved by the grid. In these cases, each hole typically had one to three cells. The result of the geodesic dome showed a relatively large $Cd$ value, despite the small projected frontal area, compared to the clean sphere. This is because frames orthogonally facing the flow direction received force effectively. This indicates that it is possible to have a larger drag force than a clean sphere in certain conditions, although there are no comparable measurement data.

In summary, the results for below critical $Re$ could be predicted well, and those for the above critical $Re$ showed relatively dissipative results. For shape data with a dirty topology, a stable and reasonable solution was obtained in all cases. The calculation conditions were set to implicit LES with appropriate viscosity, based on preliminary studies of the grid resolution and the blending ratio of the hybrid scheme. We concluded that the robust results of basic turbulent flow were captured in high-$Re$ conditions by appropriately selecting the artificial viscosity. However, the prediction accuracy was not sufficient. To increase the versatility of the method, we are planning to apply wall function modeling.

\renewcommand{\thesubfigure}{\alph{subfigure}}

\subsection{Complex geometries in real applications}

\begin{figure}[htbp]
 \centering
 \begin{minipage}{0.48\hsize}
   \includegraphics[keepaspectratio,width=\textwidth]{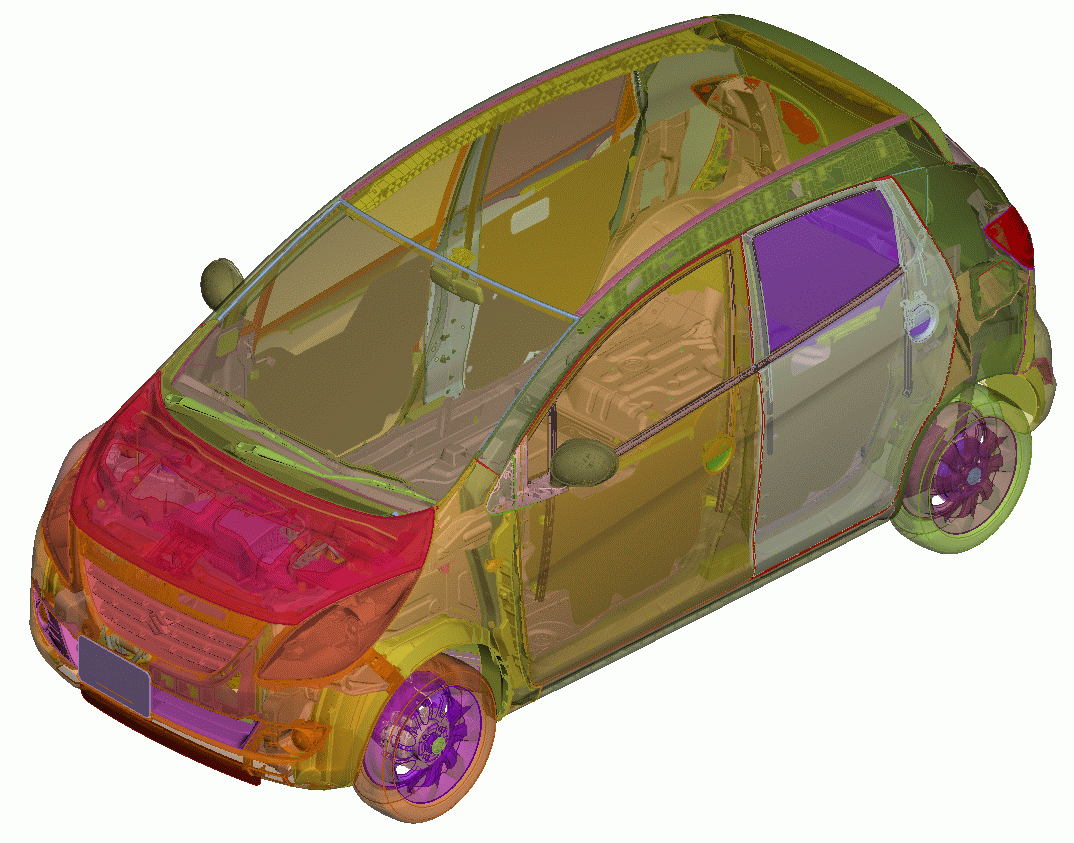}
  \subcaption{Geometric overview shown as transparent skin surfaces.}
  \label{fig:vehicle_geom}
 \end{minipage}
 \begin{minipage}{0.48\hsize}
   \includegraphics[keepaspectratio,width=\textwidth]{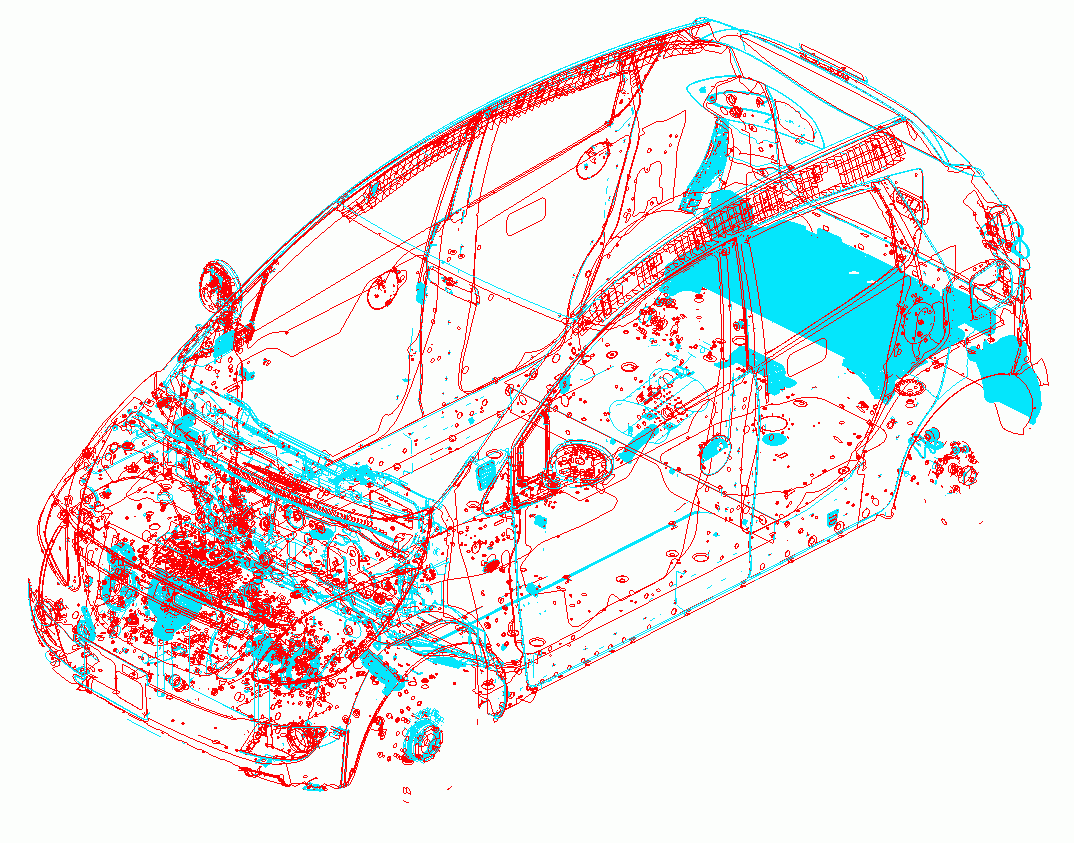}
  \subcaption{Problematic area of geometric data. Red lines indicate gaps where two faces are not connected; light blue lines indicate edges that are over-connected. There are about 500,000 error points.}
  \label{fig:vehicle_geom_error}
 \end{minipage}
 \caption{Snapshot of the full-vehicle CAD geometry, as an example of dirty geometry data.}
 \label{fig:vehicle_geom}
\end{figure}

Finally, as a test of high-$Re$ flow that has complex dirty CAD geometry, a full-vehicle aerodynamics case and a city area wind environment case were calculated. These have not been successfully reported using the existing IBM. The vehicle geometry is shown in Fig. \ref{fig:vehicle_geom}. The model has data for all vehicle components of the production model except for the interior; e.g., the model consists of the vehicle body, inner body frames, engine, transmission, suspension, brakes, heat exchangers, head/rear lamps, battery, pumps, auxiliaries, wire/harnesses of inside the engine bay, and bolts and nuts, among other components. The model is constructed with 9.2 million triangular elements with zero-thickness faces. The model has about half a million errors, and problematic face edges, as shown in Fig. \ref{fig:vehicle_geom}\subref{fig:vehicle_geom_error}. Figure \ref{fig:vehicle_geom}\subref{fig:vehicle_geom_error} shows that the red and light blue lines indicate the edges that the user needs to repair, including separated edges that should be connected, and over-connected edges that should be separated. Note that the geometry is not ''watertight'' and is not canonical in CFD. The geometric data cannot be used without pretreatment work, which would take approximately three to five days to complete in a well-organized manner in the conventional method \cite{furukawa2004application}. In fact, when we created unstructured grids for the same model in a previous study \cite{Yamazaki2012}, it took several weeks to complete the process, despite the use of commercial software with an excellent GUI and manual. In a further case, we performed a large-scale analysis, which took half a year to ensure a resolution of $1 mm$ around a vehicle's body. This clearly demonstrates the difficulty of the problem. Nevertheless, in this study, all the preliminary geometry-related work was eliminated, and the geometry data did not need to be simplified. Therefore, the primary factor of the nature of the flow causing perturbation (i.e., 'geometries') was fully retained. The CAD data file could be used as a direct input of the flow solver. This indicates that the countermeasure method based on the category classification discussed in section \ref{dirty_CAD_definition} is correct. Additionally, the calculation time of the grid generation using the workstation described in Tab. \ref{table:resource_WS} was only 10 minutes.

The calculation condition matched a full-scale vehicle running condition. The Reynolds number was $Re = 2.83 \times 10^{6}$. The computational grid consisted of 88 million cells, and the finest cell resolution placed around the vehicle body was $\Delta x = 6.10 mm$. The solution time was 1.45 seconds, with a time increment of $\Delta t = 1.0 \times 10^{-5}$ seconds. The computational resources used were 32 nodes of a supercomputer (see Tab. \ref{table:resource_OFP} for specs). The overall computational time was 23.9 hours.

However, in this case, there was a problem with the narrow flow path, i.e. issue (3) in section \ref{dirty_CAD_definition}, was not completely solved. Therefore, the following filter was applied to the external force term.

\begin{align}
  \label{eq:deadend_filter}
 & \mathcal{F'}(\bm{x},t) = \mathcal{B}(\bm{x}) \otimes \mathcal{F}(\bm{x},t), & \\
 & \mathcal{B}(\bm{x}) = 
  \begin{cases}
    0, & \sum_{k=1}^{6} \delta ( \bm{x^{(k)}} ) \geq 5, \\
    1, & elsewhere,
  \end{cases} &
\end{align}
where $\delta$ is the Dirac delta function defined in Eq. \eqref{eq:force}, $\mathcal{F}(\bm{x},t)$ is the discrete force term defined in Eq. \eqref{eq:discrete_forcing_term}, and $m = 6$ is the number of surrounding fluid points (the number of axial directions). If the wall-including cells were contained in more than five of the surrounding fluid cells, the IBM calculation was canceled. This is equivalent to terminating the calculation when the flow path was found in only one of the maximum of six directions.

\begin{figure}[htbp]
 \centering
 \begin{minipage}{0.48\hsize}
   \includegraphics[keepaspectratio,width=\textwidth]{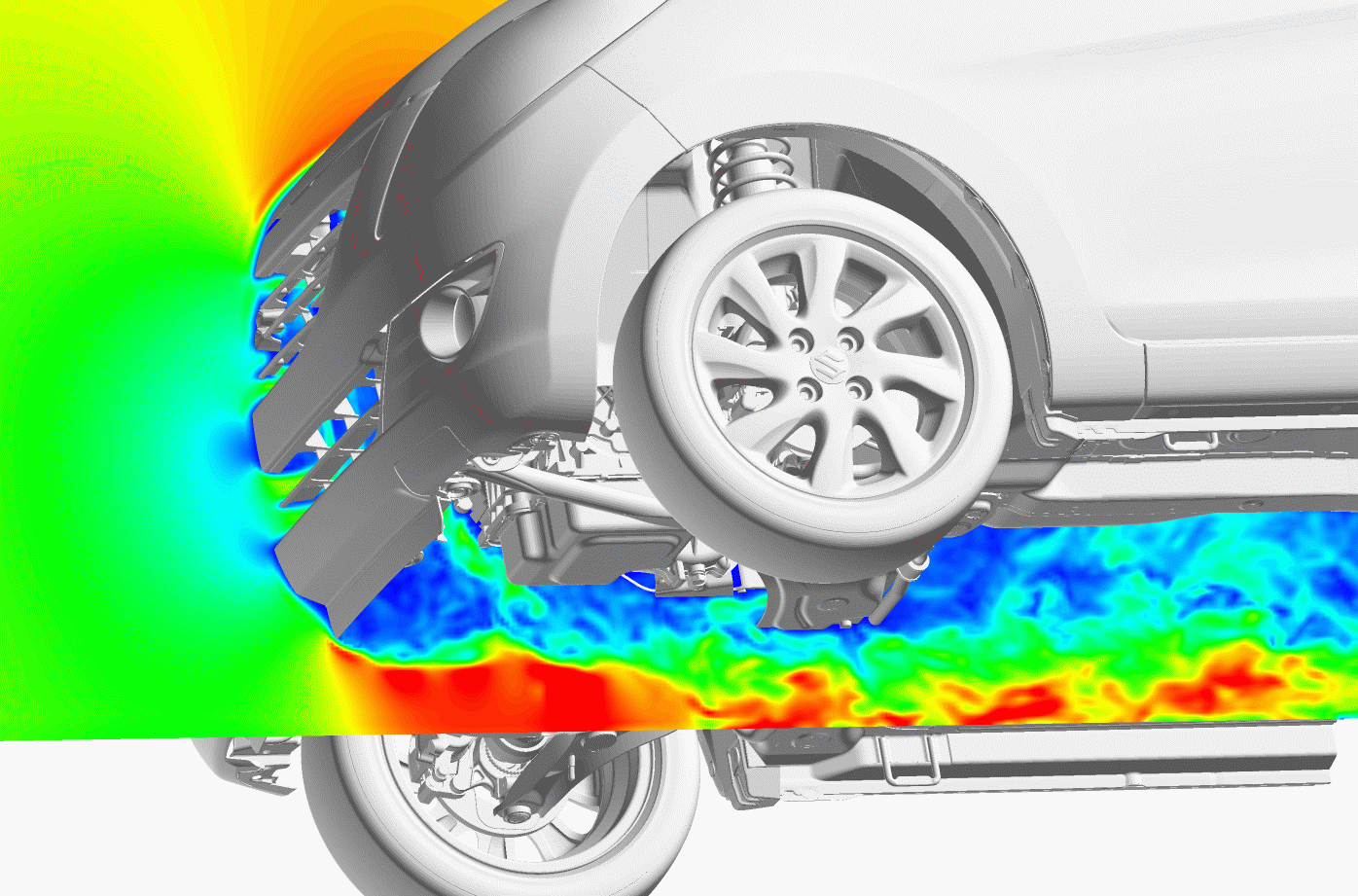}
  \subcaption{Magnified view of the instantaneous velocity magnitude of flow around the full-vehicle geometry.}
  \label{fig:vehicle_flow}
 \end{minipage}
 \begin{minipage}{0.48\hsize}
   \includegraphics[keepaspectratio,width=\textwidth]{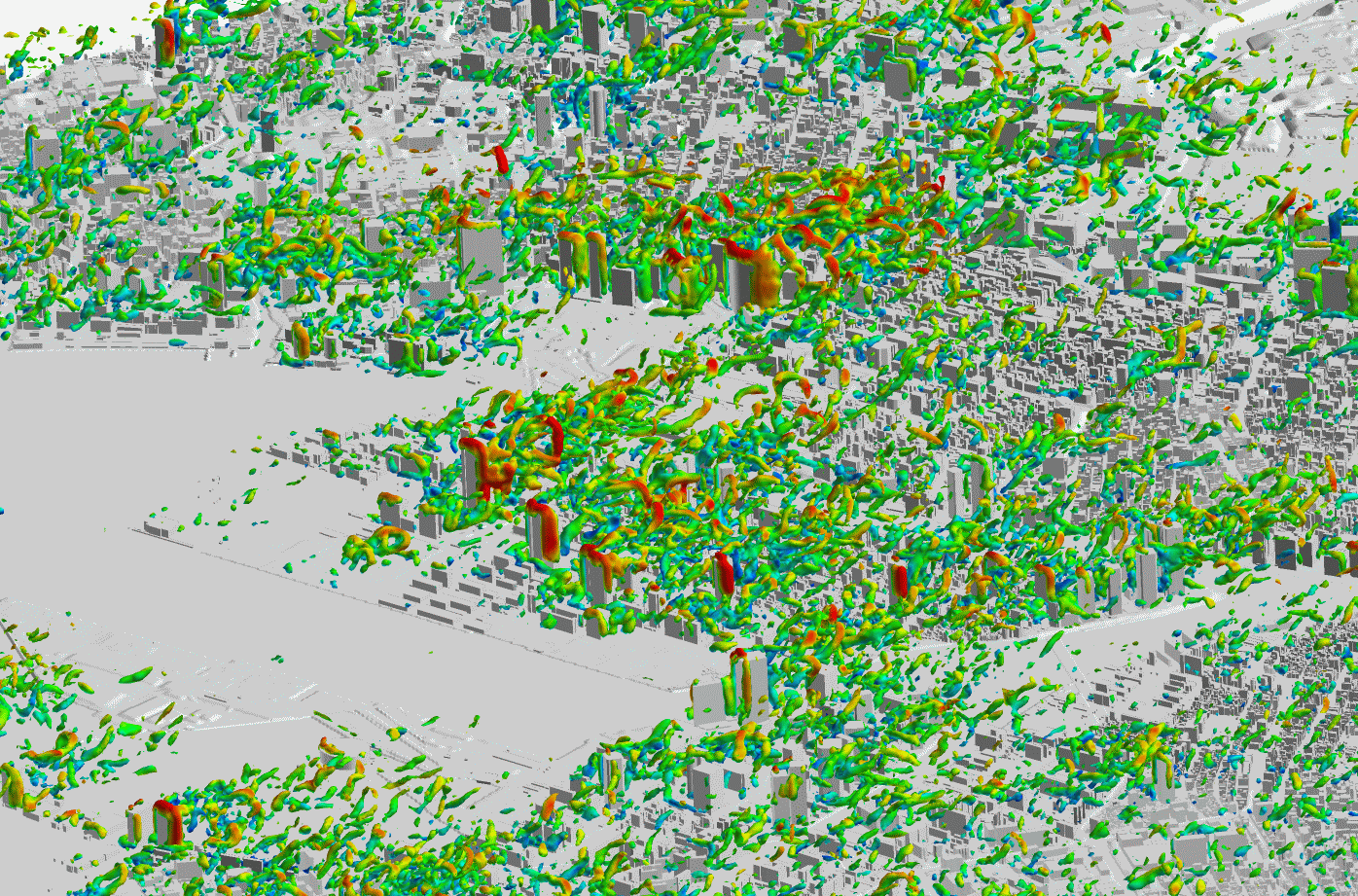}
  \subcaption{Overview of the vortex structure with Q-criterion, colored by the velocity magnitude of the wind environment simulation.}
  \label{fig:building_flow}
 \end{minipage}
 \caption{Test results of real applications of complex dirty geometry.}
 \label{fig:real_applications}
\end{figure}

Figure \ref{fig:real_applications}\subref{fig:vehicle_flow} shows the obtained result of the velocity magnitude of the center section. This geometry had an ''air-dam'' shape, which comprised a small thin air-spoiler attached to the bottom of the front bumper. This showed a reasonable flow change in the region of the underfloor. The case without the air-dam was also executed with the same calculation grid, which was performed automatically. The predicted $Cd$ differed by approximately 7\% to 9\% from the experimental result. Although the order of the error was the same as that of the conventional method, the solution of the flow field was dissipative. However, the delta of drag between the two configurations exhibited a better result than the other simulation results, which suggests that there was an advantage compared with using the unmodified geometry. The detailed results are reported in Onishi \cite{Onishi2018}, compared to the results of the conventional method. When using a finer or coarser resolution, an additional 10\% error was obtained. However, this depended on the resolution of the flow leakage into the vehicle interior, or the opening grill, or the heat exchanger, as well as the boundary layer. The current method depends on the given resolution and the numerical viscosity.

Figure \ref{fig:real_applications}\subref{fig:building_flow} shows another case of the different dirty topology, namely wind simulation for an urban city area. This case had two types of geometry data; the buildings and the terrain. The two geometry sets were overlaid. Each data point had a large number of defects of faces, however, the flow field showed reasonable characteristics in its vortex structure. Even in this case, the calculation could be performed almost automatically for cases where the wind direction had changed. This also indicates that the categorization in section \ref{dirty_CAD_definition} has been generalized well. Soon, several users of this software plan to report results in this field, and Cao \cite{Cao2019} has already reported the details of the fundamental study.

\section{Conclusions}

A topology-free boundary representation of the IBM, based on the BCM Cartesian grid approach, was proposed. This method is based on Mittal's ghost-cell boundary condition \cite{Mittal2007}, coupled with Peskin's \cite{Peskin1972} distributed forcing method, to handle gaps, overlaps, and defects of faces in dirty CAD data. The dummy cell definition was applied to avoid the search for fluid and solid regions by placing an arbitrary solid region that splits the regions into inner and outer regions locally. This also worked for the zero-thickness treatment, for the same reason. The formulation of the IBM was modified by projecting the interpolation onto the axial direction to avoid a failure when searching for interpolation points. This approach resulted in high affinity with the solution for avoiding the narrow flow path. The interpolation calculation was based on the ray-tracing algorithm; its parameters were adjusted to have an appropriate error. This method allows any type of three-dimensional complex geometrical topologies to be handled as is, without the need for any pretreatment of the geometry.

The verification analysis with the Taylor-Green vortex decay problem, in conjunction with the second-order central-difference scheme, resulted in first-order accuracy in the finer domain, though the coarser domain retained second-order accuracy.

To demonstrate the ability of the proposed method, three-dimensional flow around the Ahmed body was analyzed. Although the proposed method introduced artificial viscosity on the wall by using the first-order interpolation, the basic characteristics of the turbulent flow around a bluff body, including the fluctuating components, were captured by appropriately selecting the resolution of the computational grid.

Regarding the validation case of flow travelling past a flat plate for a wide range of angles of attack, the calculation results captured the tendencies of basic flow property of experiments, and quantitatively reproduced the state of the flow field around a thin plate.

Regarding the validation case of flow past a sphere for various geometries that have dirty topologies, the results for below critical $Re$ could be predicted well, and the results for above critical $Re$ showed relatively dissipative results. For shape data with a dirty topology, a stable and reasonable solution was obtained in all cases. The robust results of basic turbulent flow were captured in high-$Re$ conditions, by appropriately selecting the artificial viscosity. However, the prediction accuracy was not sufficient.

The method was applied to full-vehicle aerodynamic analysis and city area wind environment analysis to demonstrate its ability in simulating flows with complex geometries having dirty topology. All preparatory work for the geometry was eliminated, and no geometry simplification was required for these simulations. The CAD data was usable as a direct input to the solver, and robust and reasonable results were obtained. This indicates that the countermeasure method based on the category classification presented in this paper is correct.

The benefit of dealing with a detailed shape without modification is obvious, as is the automation of the process, especially in industrial CFD applications. We hope that this paper will contribute to the advancement of CFD technology by helping researchers to understand the geometry data used in industrial applications. However, under certain conditions, this method provided dissipative and sensitive results for a given resolution or viscosity. To improve the results and increase the versatility of the method, we are planning to enhance the order of accuracy around the wall and to apply wall function modeling in our future studies.

\section*{Acknowledgments}

I am grateful to my supervisors, Prof. K. Nakahashi and Prof. S. Obayashi of Tohoku University for their indispensable advice in the early stages of this research. We would like to thank Editage (www.editage.com) for English language editing. This research was supported by MEXT as ''Priority Issue on Post-K computer'' (Development of Innovative Design and Production Processes) and used computational resources of the HPCI System Research project (Project ID:hp150284, hp160232, hp170276, hp180192, hp190182 and hp150118, hp150206, hp160032).

\section*{References}

\bibliography{preprint}

\end{document}